\documentclass[10pt]{article}
\pdfoutput=1
\usepackage{jheppub}
\usepackage{amsmath,amssymb,amsfonts,mathbbol,graphicx,slashed,color,amsthm, mathtools, upgreek, enumerate, tensor}
\usepackage{subcaption}
\usepackage{setspace}
\usepackage[export]{adjustbox}
\usepackage{arydshln}
\usepackage{xcolor}
\usepackage{physics}
\usepackage[normalem]{ulem}

\graphicspath{{pics/}}

\allowdisplaybreaks

\colorlet{darkblue}{blue!70!black}
\colorlet{darkgreen}{green!50!black}
\colorlet{darkred}{red!50!black}

\newcommand{\ga}{\alpha}
\newcommand{\gb}{\varrho}
\newcommand{\gc}{k}
\newcommand{\gl}{\lambda}
\newcommand{\cN}{\mathcal{N}}

\newcommand{\mO}{\mathcal{O}}

\newcommand{\mfs}{\mathfrak{s}}
\newcommand{\mfc}{\mathfrak{c}}
\newcommand{\mfw}{\mathbb{w}}
\newcommand{\EF}{\vartheta}

\newcommand{\beq}{\begin{equation}}
\newcommand{\eeq}{\end{equation}}
\newcommand{\beqn}{\begin{eqnarray}}
\newcommand{\eeqn}{\end{eqnarray}}

\renewcommand{\thefootnote}{\textit{\alph{footnote}}} %temporarily uses letters instead

\title{Islands with Gravitating Baths: Towards ER = EPR}

 \author{Louise Anderson,\footnote{Now at Google, \href{mailto: louiseanderson@google.com}{louiseanderson@google.com}}}
 \author{Onkar Parrikar}
 \author{and Ronak M Soni}
 \affiliation{Stanford Institute for Theoretical Physics, 382 Via Pueblo, Stanford CA 94305}
 \vspace{5mm}

\vspace{1cm}

\emailAdd{louise.anderson@stanford.edu}
\emailAdd{parrikar@stanford.edu}
\emailAdd{ronakms@stanford.edu}

\abstract{
We study the Page curve and the island rule for black holes evaporating into gravitating baths, with an eye towards establishing a connection with the ER=EPR proposal. We consider several models of two entangled 2d black holes in Jackiw-Teitelboim (JT) gravity with negative cosmological constant.
  The first, ``doubled PSSY,'' model is one in which the black holes have end-of-the-world (ETW) branes with a flavour degree of freedom.
  We study highly entangled states of this flavour degree of freedom and find an entanglement-induced Hawking-Page-like transition from a geometry with two disconnected black holes to one with a pair of black holes connected by a wormhole, thus realising the ER = EPR proposal.
  The second model is a dynamical one in which the ETW branes do not have internal degrees of freedom but the JT gravity is coupled to a 2d CFT, and we entangle the black holes by coupling the two CFTs at the $AdS$ boundary and evolving for a long time.
  We study the entanglement entropy between the two black holes and find that the story is substantially similar to that with a non-gravitating thermal bath.
  In the third model, we couple the two ends of a two-sided eternal black hole and evolve for a long time.
  Finally, we discuss the possibility of a Hawking-Page-like transition induced by real-time evolution that realises the ER = EPR proposal in this dynamical setting.
}

\begin{document}

\maketitle
\renewcommand{\thefootnote}{\arabic{footnote}}
\setcounter{footnote}{0} 
\parskip=10pt

\section{Introduction}
Black holes have long been a subject of intense study, as they are expected to contain hints of non-perturbative quantum gravitational physics within semi-classical gravity.
Particular attention has been paid to the black hole information problem, which is an apparent tension between semi-classical gravity and unitary quantum evolution; we refer the reader to \cite{RajuReview,AMSTReview} for recent reviews.

While the naive `Hawking paradox' is not robust to small non-perturbative corrections, as argued for example in \cite{Maldacena:2003}, interest was reinvigorated by the discovery \cite{Mathur:2009,Almheiri:2013a,Almheiri:2013} of a version that is robust to these small corrections.
Consider a black hole that is formed by the collapse of some matter in a pure state; as it radiates away its mass, it gets entangled with the Hawking radiation.
Importantly, the Hawking radiation and the black hole are purified by each other.
A semi-classical calculation shows that the Hawking radiation is in a state very close to the thermal state.
This leads to a paradox for an old black hole when the apparent entropy of the Hawking radiation exceeds that of the remaining black hole, since Page's theorem guarantees that the entanglement entropy between two large subsystems in a generic state is given by the dimension of the smaller one, which in this case is the black hole.
The time at which this transition happens is known as the Page time $t_{Page}$ and the behaviour of the entropy --- initially increasing and then decreasing --- is known as the Page curve.

Suppose the calculation of the radiation's entropy gets small corrections that make it consistent with Page's theorem.
In this case, the black hole is maximally entangled with the `early radiation' up to a time $t$.
On the other hand, the requirement of a smooth horizon at the semi-classical level, that one might expect from the equivalence principle, means that the `late radiation' from time $t$ to time $t+\delta t$ is maximally entangled with the effective field theory degrees of freedom just behind the horizon.
But the EFT degrees of freedom, being part of the black hole, are already maximally entangled with the early radiation.
This is in conflict with the monogamy of entanglement.

This led to a flurry of work suggesting a variety of resolutions.
The one suggestion that is relevant to us is the following: that the above argument consists of a monogamy paradox only if we assume that the early radiation and the black hole interior are independent subsystems.
In other words, there is no paradox if the black hole interior degrees of freedom are \emph{encoded} in the early radiation, see for example \cite{Susskind:1993,Verlinde:2013,Papadodimas:2016,Nomura:2013,Susskind:2013}.
The most memorable version of this, riffing off the example of the duality between a thermofield double (TFD) state of two CFTs and a two-sided eternal black hole in $AdS$, is known as ER=EPR \cite{er-epr}.
Roughly, the suggestion is that all entanglement (EPR) constitutes a (possibly Planck-sized) non-traversable wormhole (ER) between the entangled degrees of freedom in quantum gravity.
While, as stated, this strong version of the proposal is clearly not verifiable with the tools at our disposal, we can nevertheless attempt to verify the ER=EPR proposal by ``condensing all the quantum wormholes'' into a classical wormhole by collapsing the radiation into another black hole \cite{er-epr-mvr}.

Recently, there has been significant progress on the black hole information problem \cite{Penington:2020,AEMM}, see also \cite{Almheiri:2019,Almheiri:2020,Almheiri:2020a,Almheiri:2020b,Rozali:2020,Chen:2020,Chen:2020jvn,Chen:2020b,Chen:2020c,Hernandez:2020,Hollowood:2020,Hollowood:2020a,Akal:2020twv,Geng:2020qvw, Balasubramanian:2020hfs, Balasubramanian:2020coy,Alishahiha:2020qza,Krishnan:2020fer,Krishnan:2020oun}, in $AdS$/CFT.\footnote{There have been calculations in other asymptotics \cite{islands-na-1,islands-na-2,islands-na-3,islands-na-4,islands-na-5,Dong:2020,Gautason:2020tmk,Sybesma:2020fxg,Geng:2021wcq,Caceres:2020jcn}, see also \cite{Laddha:2020} for a differing perspective. See also \cite{Mousatov:2020ics} for a possible string-theoretic origin of the encoding effect found with the island rule.}
These calculations involve coupling a holographic CFT in a black hole state to a (not necessarily gravitational) bath system that collects the radiation, and calculating the entanglement entropy (EE) using the quantum HRT formula \cite{RT,RT2,RT3,RT4}.
This formula states that the EE of a UV subsystem $A$ of a holographic system dual to general relativity (GR) coupled to an effective field theory is given by
\begin{equation}
  S_{E} (A) = \min \left\{ \underset{a\big|\partial a = X \cup A}{\text{ext}} S_{gen} (a) \right\}, \qquad S_{gen} (a) = \frac{\text{Area}(X)}{4 G_{N}} + S_{E,bulk} (a).
  \label{eqn:rt-formula}
\end{equation}
The bulk region $a$ whose generalised entropy $S_{gen}$ is the EE is called the \emph{entanglement wedge} (EW) and the corresponding surface $X$ is known as the \emph{HRT surface}.
$S_{E,bulk}(a)$ denotes the von Neumann entropy of the bulk fields within the region $a$.

We take $A$ to be the bath.
The important conceptual advance made in the above papers was the observation that in this case, the entanglement wedge (EW) $a$ can contain contractible disconnected components, or ``islands,'' deep in the bulk, e.g. in the black hole interior.
As the entanglement between the CFT and the bath increases, the minimal quantum extremal surface (QES), also known as the HRT surface, transitions from the empty surface to a surface near the horizon at the Page time.
This implies that, at late times, the interior of the black hole is in the entanglement wedge \cite{EWR} of the bath and therefore the radiation.
By entanglement wedge reconstruction \cite{EWR,EWR2,EWR3,EWR4,EWR5,Harlow:2016vwg}, this means that the interior is encoded in the radiation.

\begin{figure}[h]
  \centering
  \includegraphics[width=140mm]{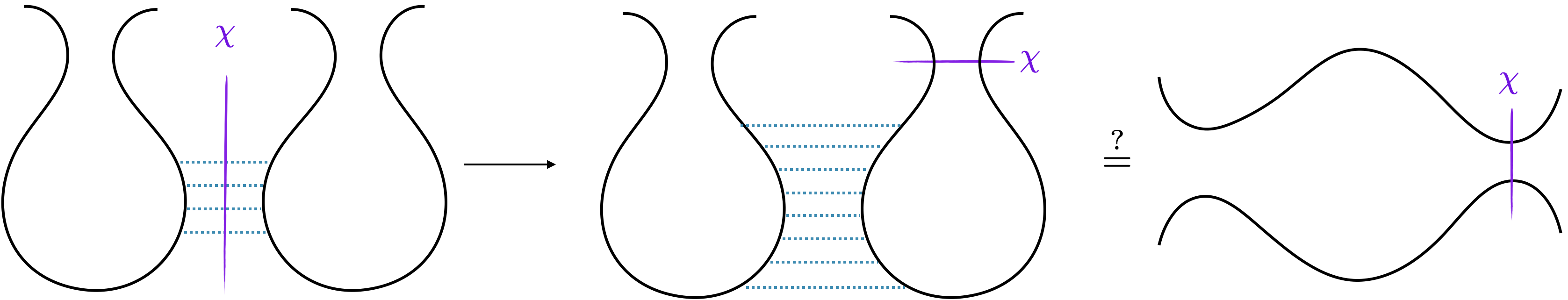}
  \caption{Schematic representation of the entanglement wedge transition in a model of two coupled black holes.
  The purple line is the HRT surface whose generalised entropy calculates the EE between the two boundary CFTs.
  At early times, each black hole interior is contained in the entanglement wedge of the corresponding boundary CFT.
  At late times, based on the island rule, we expect that the build-up of entanglement between the interiors causes one of the interiors to become part of the entanglement wedge of the \emph{other} CFT.
  In this paper, we ask if this is the same as the entanglement lines condensing into a geometric wormhole.}
  \label{fig:geoff-pics}
\end{figure}

The aim of our paper is to ask whether some of the recent conceptual and technical advances can be used to demonstrate the ER = EPR conjecture that the entanglement between the radiation and the black hole leads to the formation of a wormhole.
Consider the case in which the bath that collects the radiation is also a holographic CFT.
From the dual gravity perspective, we then have two black holes radiating into each other.
The expectation based on previous literature is sketched in the second line of figure \ref{fig:geoff-pics}, where one of the interiors become an island in the entanglement wedge of the other boundary CFT.
According to the ER = EPR proposal, however, we should find that after the Page time, a geometric (spatial) wormhole forms between the two black holes.
In a sense, then, this proposal states that the similarity between actual space and the ``entanglement lines'' in figure \ref{fig:geoff-pics} should be taken seriously, as in the case of holographic tensor networks, see e.g. \cite{Swingle:2012,Hayden:2016}.
If this is the case, the second picture in the second line of figure \ref{fig:geoff-pics} is equivalent to a connected geometry --- i.e. the entanglement between the interiors forms a ``bridge to the island.''\footnote{In this sense, this work can be seen as the spiritual converse of \cite{Bao:2019,Caputa:2020}, which try to build tensor networks from geometries.}

To explore the question of whether such a topology transition happens, we study various models in which the bath is a holographic system. (See also \cite{Geng:2020}, which previously studied a case with a gravitational bath and found similar results for the Page curve.)
As a cautionary remark, we remind the reader that the question of the topology of the spatial slice in the bulk dual of a state is a somewhat meaningless one, since there is no non-perturbatively diffeomorphism-invariant linear operator that can measure the `connectedness' of spacetime \cite{Marolf:2013,Papadodimas:2015,Jafferis:2017tiu}.
Nevertheless, we may simply calculate the geometry of the dominant saddle in various path integrals.
We believe that this is a sufficiently interesting exercise.

The examples we deal with are within the context of two-dimensional Jackiw-Teitelboim (JT) gravity \cite{J,T,AP,Maldacena:2016upp,Jensen:2016pah,Engelsoy:2016xyb}.
In two of the models, we take two pure-state black holes in two different asymptotically $AdS$ spacetimes and then entangle them.
Each pure state is one of the somewhat atypical Kourkoulou-Maldacena \cite{KM} states with end-of-the-world (ETW) branes behind the horizon, obtained by a projection acting on one half of a thermofield double.
In the first model, the doubled Penington-Shenker-Stanford-Yang (dPSSY) \cite{Penington:2019kki} model, we allow the ETW brane in both black hole interiors to have multiple flavours and entangle this flavour degree of freedom between the two black holes.
In the second model, we consider the JT gravity coupled to a 2d CFT in the bulk, following \cite{AEMM}.
At time $t=0$ we couple the two spacetimes (each with a black hole) by imposing transparent boundary conditions between them at the asymptotic boundaries, and then evolve for a long time.
The third model is similar to the second, except that the two black holes before the coupling are taken to be in a thermofield double state.

\iffalse
The specific quantity we would like to focus on to show the wormhole formation is as follows: consider the boundary path integral which calculates the norm of the state of the two coupled holographic systems described above.
The bulk saddle point that dominates the gravity path integral for this boundary condition is a natural candidate for the `bulk dual' geometry, and this is what we will mean by the bulk dual throughout this paper.
More generally, we can also calculate boundary correlation functions with a small number of light operators inserted in the boundary CFT.
\fi

The results in the three models of entanglement described above are different.
In the first model, at large entanglement, the bulk dual --- the dominant saddle in the path integral that calculates the norm of the state --- is a connected geometry, demonstrating the ER = EPR hypothesis.
The mechanism for this is unexpectedly simple.
Remember that even a factorised state of two boundaries generically has a non-zero $\mO(e^{-S})$ overlap with the thermofield double state.
From the dual gravity perspective in our setup, this is reproduced by a bulk geometry in which the ETW branes join up and create an ER bridge between the two (uncoupled and unentangled) boundaries \cite{Jafferis:2017tiu}.
Similarly, a connected geometry also contributes an exponentially sub-leading amount to the norm of the factorized (i.e. zero entanglement) state.
As we build up entanglement, the two contributions exchange dominance, as in the Hawking-Page transition \cite{Hawking:1982dh}, and we find that the leading contribution becomes the connected geometry.\footnote{There is a large body of work analysing dynamical topology-change in semi-classical gravity as well as string theory, see \cite{Wheeler:1957mu,Lavrelashvili:1987jg,Coleman:1988cy,Giddings:1987cg,Adams:2005rb,Saad2018,Saad:2019lba,Saad:2019pqd,Moitra:2021uiv} for a very incomplete set of references.}

Another way to think of the simplicity of the mechanism is to compare to the recent derivations of the island rule \cite{Penington:2019kki,Almheiri:2020b}.
In these papers, the essential contribution comes from a class of Euclidean wormholes, called replica wormholes.
These are Euclidean wormholes which turn up in a replica path integral computation with many `bra' and many `ket' copies of the state of interest (relevant for the R\'enyi entropy) and can connect all the copies to each other.
The ER bridge that forms between two unentangled `ket' boundaries in our study is analogous to these replica wormholes; indeed, it is a `ket-ket' wormhole.\footnote{There has been some discussion on another special case, called the `bra-ket' wormhole, see e.g. \cite{Chen:2020a,Anous:2020}.} 

In addition to the norm, we also study the entanglement entropy between the two coupled boundary quantum mechanics systems.
We find that the quantum extremal surface whose generalised entropy computes the entanglement entropy also lies in the connected geometry after the Page time.
A similar phenomenon was also found in \cite{Chen:2020a}.

In the second model, in which unentangled black holes are coupled in real time, the results are different.
As in the previous model, there is a second saddle with a connected geometry.
Further, the late time generalised entropy of a QES in this connected geometry is less than that in the naive disconnected geometry; however, a replica trick argument shows that the HRT surface remains in the disconnected geometry anyway.
We then explore whether any single-copy path integrals nevertheless exhibit a Hawking-Page-like transition, and find that some do. 

Finally, inspired by the alternate saddle above, we couple two black holes in a thermofield double state.
We find that the entanglement between the two black holes reduces till scrambling time, after which the story becomes substantially similar to the second model.

An outline of this paper is as follows: 
\begin{enumerate}
  \item In section \ref{sec:JT}, we review some crucial facts about Jackiw-Teitelboim gravity that we will use throughout. In particular, in section \ref{ssec:dPSSY-bulk} we construct the bulk geometries relevant to two coupled boundary CFTs, setting the stage for the calculations in the rest of the paper.
  \item In section \ref{sec:dPSSY}, we study the static doubled PSSY model. This is a simple toy model which illustrates our main results, without too much technical computation.  
  \item In section \ref{sec:evap}, we study the second model in which we allow two pure-state black holes to radiate into each other.
    Despite the relative difficulty of this problem, the main point is rather simple and is summarised with relatively few calculations in section \ref{ssec:evap-summary}.
    We show that the path integral corresponding to certain correlation functions exhibits an ER = EPR transition in section \ref{sssec:bulk-dual}.
  \item In section \ref{sec:conn-hist}, we study the third model in which we allow two black holes in a thermofield double state to radiate into each other.
  \item We end by discussing open questions and making some observations in section \ref{sec:conc}.
\end{enumerate}

%%%%%%%%%%%
\section{JT Gravity} \label{sec:JT}
The 2d gravity theory we work with throughout this paper is Jackiw-Teitelboim (JT) gravity with end-of-the-world (ETW) branes.
Its Euclidean action, without bulk matter, is
\begin{equation}
I[\phi, g] = -S_0 \chi  - \frac{1}{4\pi} \left[ \int_{M} \sqrt{g}\; \phi (R+2) + \int_{\partial M} \sqrt{h}\, \phi K \right] + \phi_r \mu \int_{\text{ETW brane}} ds,
\label{eqn:jt-action}
\end{equation}
where $S_0$ is the extremal entropy, $\chi$ is the Euler character of the Euclidean spacetime, $\mu$ is the tension of the ETW brane, and $\phi$ is the dilaton.
The model is further defined by two boundary conditions:
\begin{align}
  \text{Asymptotic $AdS$ boundary}: &\quad\quad \phi = \frac{\phi_{r}}{\epsilon}, \quad\quad du^2 \equiv \epsilon^2 ds^2|_{bd}, \nonumber\\
  \text{ETW brane boundary}: &\quad\quad n^{\alpha} \partial_{\alpha} \phi = \mu, \quad K = 0.
  \label{eqn:bd-conds}
\end{align}
We take the limit $\epsilon \to 0$ to recover $AdS_{2}$ physics.
The second equation in the first line is technically not a boundary condition but a definition of the UV time $u$.

Because of the dilaton equation of motion, the bulk is always $AdS_{2}$.
The dynamics of this theory reduces to the dynamics of a boundary particle, whose location is the boundary of the `cutout' of $AdS_2$.
Because of this simplicity, semi-classical Lorentzian JT gravity can be exactly solved by keeping track of three $SL(2,\mathbb{R})$ charges of the boundary; in appendix \ref{app:JT}, we review this and derive some results we will use in section \ref{sec:evap}.

Since the bulk is always $AdS_{2}$, we can use conventional coordinate systems.
In real time, they are
\begin{align}
    \text{Kruskal-Szekeres}: \qquad ds^2 &= \frac{4 d\mfw d \bar{\mfw}}{(1-\mfw\bar{\mfw})^2} \label{eqn:kruskal-coords} \\
    \text{Poincare}: \qquad ds^2 &= \frac{4 dx d \bar{x}}{(x+\bar{x})^2}, \qquad x = z+t_P, \bar{x} = z - t_P \label{eqn:poincare-coords} \\
    \text{Global}: \qquad ds^2 &= \frac{d\mfs d\bar{\mfs}}{\sin^2 \frac{\mfs + \bar{\mfs}}{2}} \qquad \mfs = \sigma + t_{gl}, \bar{\mfs} = \sigma - t_{gl} \label{eqn:global-coords}.
\end{align}
Additionally, we will also use two coordinate systems that do depend on the geometry
\begin{align}
    \text{Schwarzchild}: \qquad ds^2 &= -(r^2 - r_h^2) du^2 + \frac{dr^2}{r^2 - r_h^2}, \label{eqn:schw-coords} \\
    \text{UV}: \qquad ds^2 &= \frac{4 x'(y) \bar{x}'(\bar{y})}{[x(y) + \bar{x}(\bar{y})]^2} dy d\bar{y}, \qquad y = \rho + u, \bar{y} = \rho - u. \label{uv-coords}
\end{align}
The UV coordinate $y$ is so named because it is the light-cone extension of the UV time $u$, i.e., the intrinsic proper time of the boundary particle; this coordinate system only extends up to the causal horizons.
Note that we are using conventions in which one of the light-cone coordinates points backwards in time; this is useful for analytic continuation from Euclidean time.

The actual physical parameters of the geometry are encoded in the dilaton.
For a Schwarzchild black hole of temperature $T$, it takes the value
\begin{equation}
    \phi = \phi_r r = 2 \pi T \phi_r \frac{1 + \mfw \bar{\mfw}}{1 - \mfw \bar{\mfw}} = 2\phi_r \frac{1 + (\pi T)^2 x \bar{x}}{x+\bar{x} } = 2 \pi T \phi_r \frac{\cos t_{gl}}{\sin \sigma} = 2\pi T \phi_r \coth \left[ \pi T \rho \right].
    \label{eqn:bh-phi}
\end{equation}

Finally, we shall also need the analog of the HRT formula \eqref{eqn:rt-formula} in this theory.
We may in general couple multiple asymptotic boundaries and 2d systems, and define the boundary subregion to be a union $A = A_1 \cup A_2$ of some of the 1d asymptotic boundaries $A_1$ and a subset $A_2$ of a Cauchy slice of the 2d systems.
The HRT formula in this system is
\begin{equation}
    S_E (A) = \min \left\{ \underset{a \big| A_2 \subset a, \partial a = \left( \cup_i x_i \right) \cup A_1 \cup \partial A_2}{\text{ext}} S_{gen} (a) \right\}, \qquad S_{gen} (a) \equiv \sum_i (S_0 + \phi (x_i)) + S_{bulk} (a).
    \label{eqn:jt-rt}
\end{equation}

This formula has three different entropic quantities, all of which play a role in our discussion.
For clarity, we adopt a consistent notation, which we summarise in table \ref{tab:ents}, for these quantities.

\begin{table}[h]
    \centering
    \begin{tabular}{|c|c|} \hline
       Entropic Quantity  & Notation \\\hline\hline
       EE of 2d CFT & $S_{bulk}$ \\\hline
       Generalised Entropy of a General Region & $S_{gen,nE} = \sum_{endpts} (S_0 + \phi) + S_{bulk}$ \\ \hline
       Generalised Entropy of a Region Bounded by a QES & $S_{gen} = \text{ext}\; S_{gen,nE}$ \\ \hline
       EE between two QM systems/UV EE  & $S_E = \min S_{gen}$ \\ \hline
    \end{tabular}
    \caption{The various entropic objects we will deal with. We will stick with this notation throughout, for clarity.}
    \label{tab:ents}
\end{table}

\subsection{Saddle Point Geometries with ETW Branes} \label{ssec:dPSSY-bulk}
The gravitational set-up that we will analyse throughout this paper is the so-called Kourkoulou-Maldacena (KM) state \cite{KM} of a single boundary quantum mechanics (say, the SYK model).
It can be thought of as a projection operator acting on one end of the thermofield double (TFD) state of two SYKs, and is a useful toy model for a single-sided black hole. This is because from the dual gravity perspective, this projection looks like an ETW brane emanating from the position of the projection; the geometry ends at the location of this brane.
This state, then, is parametrised by two quantities; the mass $\mu$ of the ETW brane and the amount $\ell$ of Euclidean evolution involved in creating the state.

\begin{figure}[h]
    \centering
    \begin{subfigure}[t]{0.48\textwidth}
        \centering
        \includegraphics[height=30mm]{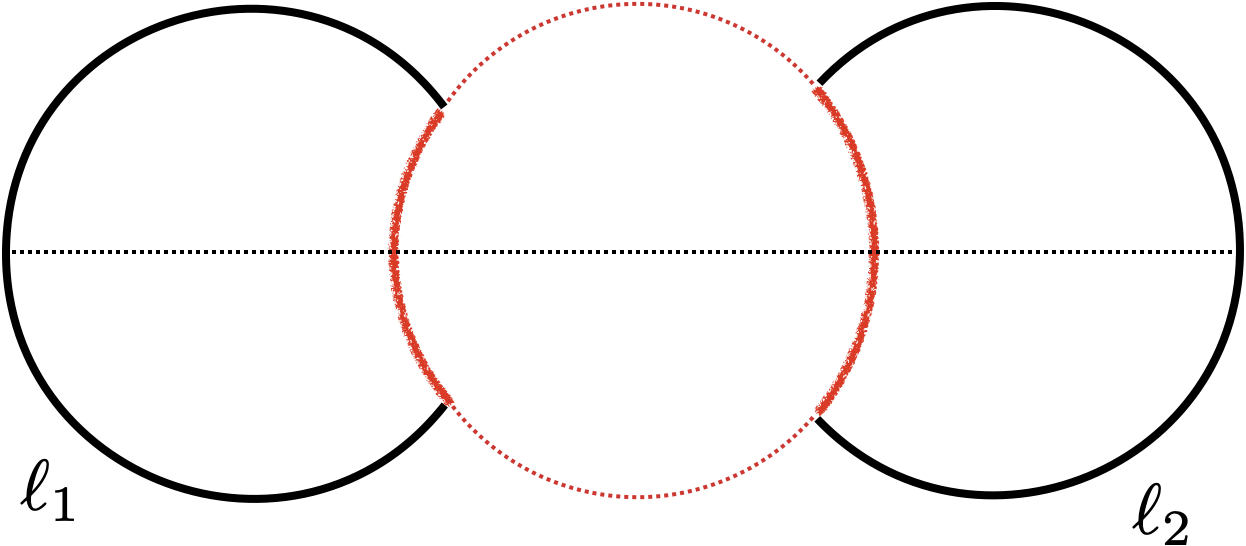}
        \caption{Disconnected geometry.}
	\label{fig:pssy-disconn}
    \end{subfigure}
    ~ 
    \begin{subfigure}[t]{0.48\textwidth}
        \centering
        \includegraphics[height=34mm]{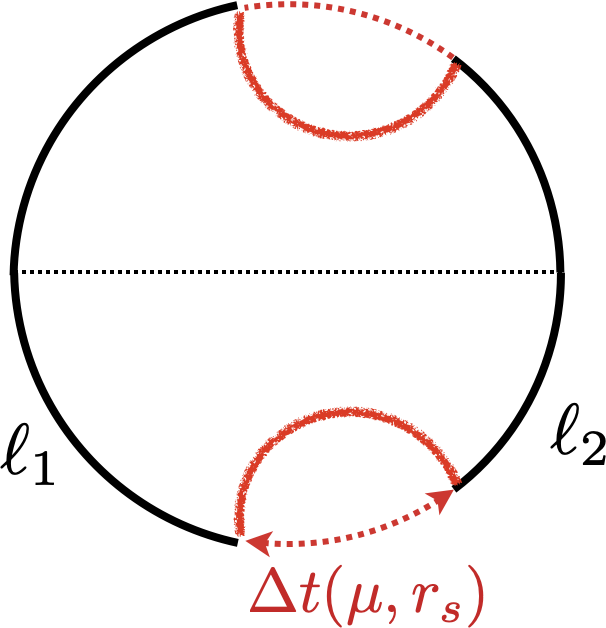}
        \caption{Connected geometry.}
	\label{fig:pssy-conn}
    \end{subfigure}
    \caption{The geometries contributing to \eqref{eq:dpssy-norm}. Here $\ell_1$ and $\ell_2$ are the lengths of one half (i.e., the ``ket'' part) of the Euclidean boundaries.}
    \label{fig:pssy-norm-gemoetries}
\end{figure}

In this paper, we will throughout consider two copies of this state.
Focusing on the norm path integral (i.e., the Euclidean path integral in the boundary corresponding to the norm of the state), there are two allowed bulk saddles, shown in figure \ref{fig:pssy-norm-gemoetries}.
We call these the disconnected and connected saddles respectively.
Notice that both of these have a Euclidean time-reflection symmetry, which we will use to analytically continue the geometries to real time in subsequent sections.

In this section, we calculate the saddle-point geometries for both of these topologies.
We highlight two important lessons.
First, that the Lorentzian continuation of the bulk geometry in the connected saddle is that of the eternal black hole.
Second, that the connected geometry is \emph{colder} than either disconnected one for all values of $\mu$, and consequently has a lower entropy.

To find the bulk geometry, we have to find, from among the geometries that fill in the boundary conditions given by the norm path integral, the one with minimal action.
Given the absence of matter, the set of bulk geometries we have to minimise the action over is that of Euclidean Schwarzchild black holes, see for example \cite{Harlow:2018tqv}.
The metric and dilaton are as in \eqref{eqn:schw-coords} and \eqref{eqn:bh-phi}.
So, the minimisation of the action is a minimisation over a single parameter, the horizon `radius' $r_{h}$.
We can repackage this into an inverse effective temperature,
\begin{equation}
    \beta_{eff} = \frac{2\pi}{r_{h}},
    \label{eqn:beta-eff}
\end{equation}
which is the size of an effective thermal circle, i.e. the periodicity of Schwarzchild time.
We will use $r_h$ and $\beta_{eff}$ interchangeably below.

The ETW brane lies on a geodesic, as required by the second boundary condition ($K=0$) in \eqref{eqn:bd-conds}.
The $AdS_{2}$-Schwarzchild black hole has a family of geodesics parametrised by the coordinate location $(t_{min}, \sqrt{r_h^2 + r_{min}^2})$ of its perihorizon, given explicitly by \cite{Harlow:2018tqv}
\begin{align}
    t(\lambda) &= t_{min} + \frac{1}{r_h} \tan^{-1} \left( \frac{r_h}{r_{\min}} \tanh \lambda \right) \nonumber\\
    r(\lambda) &= \sqrt{r_h^2 + r_{min}^2} \cosh \lambda, \qquad \lambda \in \mathbb{R}.
    \label{eqn:schw-geodesics}
\end{align}
Here, $\lambda$ is a proper length parameter.
Comparing the previous equation with \eqref{eqn:bd-conds}, we find that an ETW brane lies on a geodesic with
\begin{equation}
    r_{\min} = \mu.
    \label{eqn:r-min-mu-reln}
\end{equation}
Since $n^\alpha$ in \eqref{eqn:bd-conds} is the \emph{outward}-pointing normal, the sign of $\mu$ dictates which side of the geodesic is included.
We will restrict to $\mu > 0$, for which the outward-pointing normal points towards increasing $r$.\footnote{The reason to restrict to this case is that there is no connected saddle for $\mu < 0$.}
So the Schwarzchild time between the two ends of an ETW brane of mass $\mu$ in a background with Schwarzchild radius $r_h$ is
\begin{equation}
    \Delta t(\beta_{eff},\mu) = \frac{2}{r_{h}} \tan^{-1} \frac{r_h}{\mu} = \frac{\beta_{eff}}{\pi} \tan^{-1} \left( \frac{2\pi}{\beta_{eff}\mu} \right).
    \label{eqn:etw-brane-time}
\end{equation}

With these formulae in place, the strategy to find the saddle-point geometry will be as follows.
We add up all asymptotic boundary lengths and the time intervals in \eqref{eqn:etw-brane-time} and equate it to $\beta_{eff}$, giving an algebraic equation
\begin{equation}
    \beta_{eff} = \sum_{\text{bd components}} \ell_i + \sum_{\text{branes}} \Delta t_i(\beta_{eff}).
    \label{eqn:gen-eqn}
\end{equation}

First let us consider the disconnected geometry, created by a boundary of length $\ell_i$.
The norm has a boundary of length $2 \ell_i$, giving
\begin{equation}
    \beta_{eff} = 2 \ell_i + \frac{\beta_{eff}}{\pi} \tan^{-1} \left( \frac{2\pi}{\beta_{eff}\mu} \right).
    \label{eqn:disc-eqn}
\end{equation}
This is a transcendental equation and therefore not exactly solvable for arbitrary values of $\mu$.
In the limits $\mu \to 0,\infty$, however, the solutions are
\begin{align}
    \beta_{eff}  (\ell_i,\mu) = 
    \begin{cases}
    4\ell_i - \frac{16 \ell_i^2}{\pi^2} \mu + \mO(\mu^2)  \;,&\mu \rightarrow 0
    \\
     2 \ell_i + \frac{2}{\mu} + \mO(\mu^{-2}) \;, &\mu \rightarrow \infty
    \end{cases}.
    \label{eqn:disc-rh}
\end{align}
Further, solving \eqref{eqn:disc-eqn} for $\mu$ instead of $\beta_{eff}$ gives
\begin{equation}
    \mu = - r_h \tan (r_h \ell_i), \quad \mu \ge 0 \;\Rightarrow\; \beta_{eff} \le 4 \ell_i,
    \label{eqn:disc-mu}
\end{equation}
which shows that $\beta_{eff}$ is a monotonically decreasing function of $\mu$.
Thus, we get that the black hole entropy in this case is given by
\begin{equation}
    S_{BH} (\ell_i,\mu) = S_0 + \phi_r r_h(\mu,\ell_i) \in S_0 + \left( \frac{\pi}{2\ell_i} \phi_r, \frac{\pi}{\ell_i} \phi_r \right),
    \label{eqn:disc-ent}
\end{equation}
with the lower and upper limits given by $\mu=0,\infty$ respectively. 

Secondly, let us consider the connected geometry, whose norm path integral has two boundaries of lengths $\ell_1,\ell_2$ and two ETW branes, giving the equation
\begin{equation}
    \beta_{eff} = 2(\ell_1 + \ell_2) + \frac{2\beta_{eff}}{\pi} \tan^{-1} \left(\frac{2\pi}{\beta_{eff}\mu} \right).
    \label{eqn:conn-eqn}
\end{equation}
In the two limits, we find
\begin{align}
    \beta_{eff} =
    \begin{cases}
    \sqrt{\frac{2\pi^2}{\mu} (\ell_1 + \ell_2) } + \mO(\sqrt{\mu})  \;,&\mu \rightarrow 0
    \\
     2 (\ell_1 + \ell_2) + \frac{4}{\mu} \;, &\mu \rightarrow \infty
    \end{cases}
    \label{eqn:conn-rh}
\end{align}
The small $\mu$ divergence in $\beta_{eff}$ is related to the instability of the double trumpet geometry mentioned in \cite{Saad2018,Stanford2020}.
$\beta_{eff}$ is again a monotonically decreasing function of $\mu$.
So, we get for the black hole entropy in this case
\begin{equation}
    S_{BH} (\ell_1,\ell_2,\mu) = S_0 +  \phi_r r_h(\ell_1,\ell_2,\mu) \in S_0 + \left( 0, \frac{\pi}{\ell_1 + \ell_2} \phi_r \right),
    \label{eqn:conn-ent}
\end{equation}
where again the lower and upper limits correspond to $\mu = 0,\infty$ respectively.

While we will not need it, notice that this method can be straightforwardly extended to calculating the effective temperature of a path integral with $n > 2$ disconnected boundary components of lengths $2\ell_i$, $i = 1 \ldots n$.
In the two limits, the solution is
\begin{align}
    \beta_{eff} =
    \begin{cases}
    \frac{n-2}{n} \frac{\pi^2}{\mu} + \frac{4 \sum \ell_i}{n-2} + \mO(\mu)  \;,&\mu \rightarrow 0
    \\
     2 \sum \ell_i + \frac{2n}{\mu} + \mO(\mu^{-3}) \;, &\mu \rightarrow \infty
    \end{cases}
    \label{eqn:n-rh}
\end{align}

Coming back to the case of interest, we compare the limits of \eqref{eqn:disc-ent} with the corresponding limit of \eqref{eqn:conn-ent} to find that in both limits the connected solution has a lower entropy:
\begin{equation}
    S^{(\text{conn.})}_{BH} (\ell_1,\ell_2,\mu) < S^{(\text{dis.})}_{BH} (\ell_i, \mu), \quad i = 1,2.
    \label{eqn:conn-disc-ent-comparison}
\end{equation}
Further, it is easy to check numerically that this is the case for all $\mu > 0$.
This shows that the HRT surface, i.e. the extremal surface of minimal generalised entropy, is the bifurcation point of the connected geometry, as promised in the previous subsection.
The lower entropy can be traced to the fact that the connected geometry has a lower temperature, as can be seen by comparing \eqref{eqn:disc-rh} with \eqref{eqn:conn-rh}.

%%%%%%%%%%%%%%%%%%%%%%%%%%
\section{A Doubled PSSY Model} \label{sec:dPSSY}
We illustrate some of our main points in this section in a toy model based on that of \cite{Penington:2019kki}. We will introduce the setup of this toy model in section \ref{ssec:dPSSY-setup}.
We illustrate the main point of this work, i.e., the ER-EPR phenomenon, in section \ref{ssec:dPSSY-first-eg}.
Finally, we perform an exact analysis of the saddle point transition in a microcanonical version of our setup in appendix \ref{ssec:resolvent} using techniques from \cite{Yang:2018gdb,Penington:2019kki}.

\subsection{Setup} \label{ssec:dPSSY-setup}
The Pennington-Shenker-Stanford-Yang (PSSY) model consists of an asymptotically AdS black hole geometry in pure JT gravity, with an end-of-the-world (ETW) brane behind the horizon. The Euclidean action is given by \eqref{eqn:jt-action}. The ETW brane is taken to host some internal degrees of freedom, which will be labelled by the index $i$. We can think of these degrees of freedom as corresponding to the in-falling Hawking modes. The black hole plus ETW brane geometry is dual to a state in the boundary (ensemble averaged) quantum mechanics; we will denote this state by $|\ell, i\rangle$ and take it to be normalized to one. We can view this quantum mechanics state as being prepared at boundary Euclidean time $u=0$ by a Euclidean path integral over a Euclidean time segment of length $\ell$, with the boundary condition $i$ at the other end of the segment $u=-\ell$. In \cite{Penington:2019kki}, the process of black hole evaporation was modelled by considering an entangled state:
\beq
|\Psi\rangle = \frac{1}{\sqrt{k}}\sum_{i=1}^k |\ell, i\rangle_B \otimes |i\rangle_R,
\eeq
between the quantum mechanics dual to the black hole $B$ and a non-gravitational reference system $R$ which serves as a ``bath'' which absorbs the radiation. The parameter $k$ controls the amount of radiation which has been emitted by the black hole. As shown in \cite{Penington:2019kki}, when $k \ll e^{S_0}$, the entanglement entropy between the black hole and the reference system $R$ grows with $k$. The HRT surface is the empty surface, and the entire black hole is in the entanglement wedge of the boundary quantum mechanics $B$. However, when $k \gg e^{S_0}$, there is a new QES which takes over, namely the bifurcation point in the black hole geometry (which is an extremum for the dilaton). Beyond this, the entanglement entropy saturates to its extremal value $S_0 + \phi_r r_h$ (i.e., it is independent of $k$), and there is an ``island'' region in the black hole geometry which now moves over to the entanglement wedge of the radiation $R$.

In the present work, we are interested in considering a similar setup, but with one important difference --- we wish to take our reference system to also be gravitational. As a model for this, we will then consider an entangled state between two black holes in JT gravity with ETW branes behind their respective horizons, each hosting some internal degrees of freedom. We will call this the \emph{doubled PSSY model}. This seemingly minor modification has a dramatic effect on the physics --- after the Page time, i.e., when the logarithm of the brane entanglement rank $\log k$ exceeds the extremal entropy $S_{0}$, there is a \emph{new bulk geometry} which takes over, namely the eternal black hole geometry with a wormhole connecting the two boundaries. The new HRT surface is then the bifurcation point in this new geometry. This is to be contrasted with the original PSSY model, where the black hole geometry stays the same but there is a new QES which takes over after Page time. We now turn to the details of this model.

\subsection{ER = EPR: a first look} \label{ssec:dPSSY-first-eg}
\begin{figure}[h]
  \centering
  \includegraphics[width=40mm]{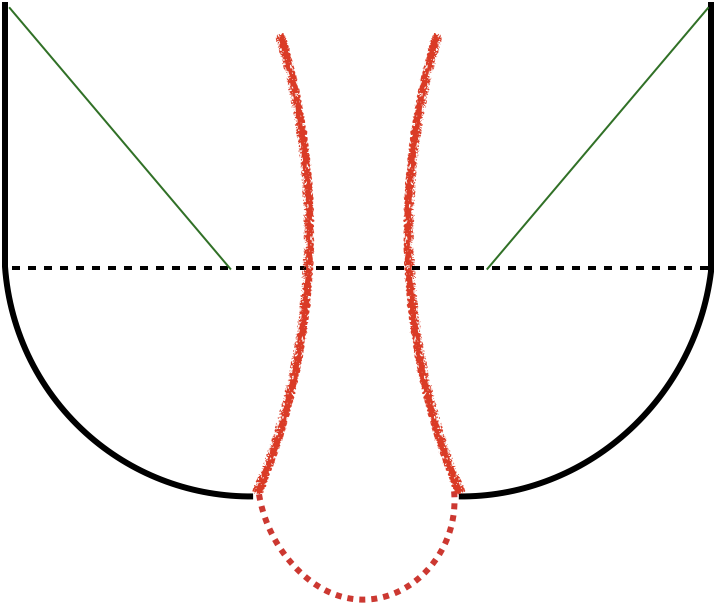}
  \caption{The naive bulk dual of $\ket{\psi}$.}
  \label{fig:dpssy-naive-lor}
\end{figure}

As discussed above, in the doubled PSSY model, we are interested in a state consisting of two entangled black holes, with asymptotic boundary lengths $l_1, l_2$:
\begin{align}
  \ket{\Psi} = \frac{1}{\sqrt{\cN}}\sum_{i,j=1}^{D} M_{ij} \ket{\ell_{1},i}_{1} \otimes \ket{\ell_{2},j}^*_{2}
  \label{eqn:dPSSY-state}
,\end{align}
where the superscript $*$ on the second factor denotes the state obtained after the action of the anti-unitary time-reflection operator, and we have introduced an arbitrary matrix $M_{ij}$ to model the various different patterns of entanglement between the black holes. The indices $i,j$ etc. will be taken to run over 1 to $D$. The role of the Page time parameter $k$ in the PSSY model will now be played, roughly speaking, by the rank of the matrix $M$; we will continue to use the symbol $k$ to denote this below. Pictorially, we could depict this state as:
\begin{align}
    |\Psi\rangle = \includegraphics[height=5em,valign=c]{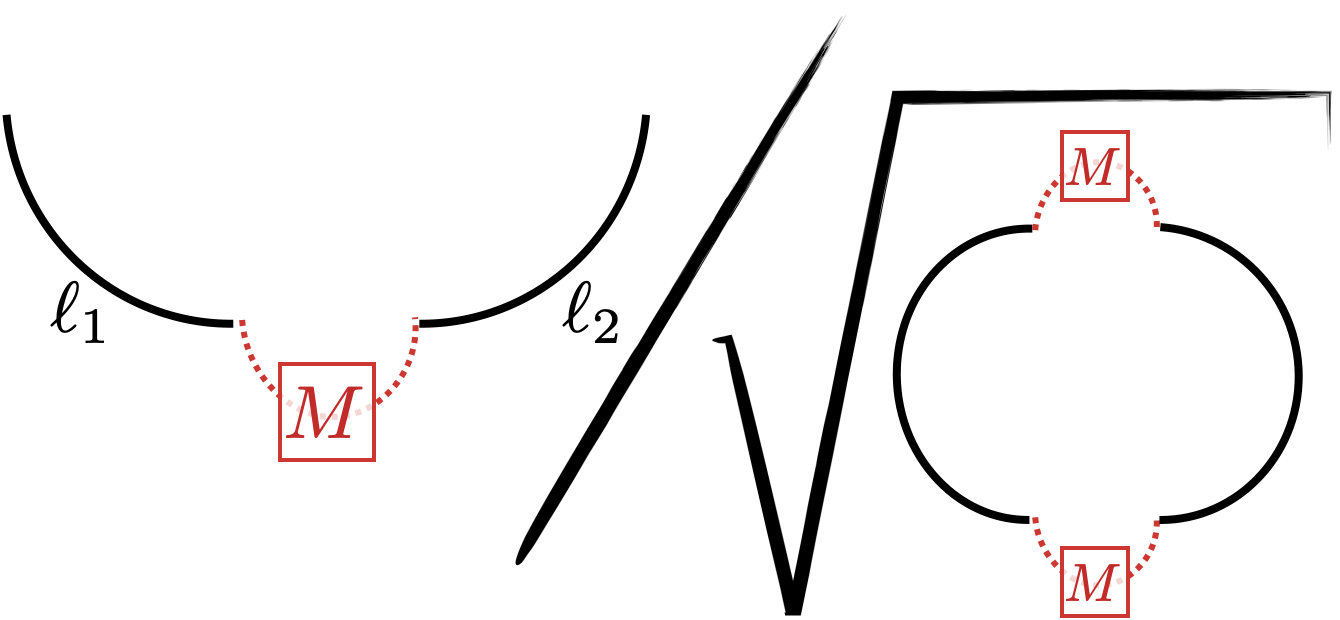}
    \label{fig:dpssy-state}
.\end{align}
For $k \ll e^{S_0}$, the bulk dual of this state consists of two $AdS_{2}$ black holes with ETW branes behind their respective horizons, as shown in figure \ref{fig:dpssy-naive-lor}.
In this case, the entanglement entropy between the two dual quantum mechanics systems is given by 
\beq
S_{E} = -\sum_i p_i\log\,p_i,
\eeq
where $p_i= \frac{\lambda_i}{\sum_j\lambda_j}$ with $\{\lambda_j\}$ being the eigenvalues of the positive matrix $M^{\dagger}M$. However, for $k \gg e^{S_{0}}$, this can exceed the UV dimension of the boundary quantum mechanical systems, giving rise to an information paradox. Following \cite{Penington:2019kki}, one could then look for new quantum extremal surfaces in this geometry. There are three extremal surfaces, namely the empty surface and the two bifurcation points of the individual black holes (see fig. \ref{fig:dpssy-naive-lor}).
The generalised entropy of either of the latter two surfaces is the coarse-grained black hole entropy 
\beq
S_{BH}(\ell_i,\mu) = S_{0} + \phi_{h}(\ell_i,\mu),
\eeq
while that of the empty extremal surface is $- \sum p_i \log p_i.$ Thus, a naive application of the HRT/island rule in this situation would lead us to conclude that 
\beq \label{eq:naive}
S_{E} \stackrel{?}{=} \text{min}\left(-\sum_i p_i\log\,p_i,\; S_{BH}(\ell_1,\mu),\; S_{BH}(\ell_2,\mu)\right).
\eeq
While this is enough to avoid the information paradox, we will now show that this formula is incorrect. In fact, what happens for $k \gg e^{S_0}$ is that a new gravitational saddle takes over, in analogy with the Hawking-Page transition \cite{Hawking:1982dh}. 

In order to see this, let us first study the norm of $\ket{\Psi}$; we will return to the computation of the entanglement entropy subsequently. From equation \eqref{eqn:dPSSY-state}, we find
\beq
\langle\Psi|\Psi\rangle = \frac{1}{\cN}\sum_{i,j,i',j'=1}^k M_{ij}M^*_{i'j'}\langle \ell_1,i'|\ell_1, i\rangle_1 \langle \ell_2,j | \ell_2, j'\rangle_2.
\eeq
We can now compute this overlap using the bulk gravity description.\footnote{More precisely, the gravity description computes the ensemble average $\overline{\langle \Psi|\Psi\rangle}$ \cite{Saad:2019lba}.} As shown in \eqref{eq:dpssy-norm}, there are two possibilities: (i) the disconnected geometry where the sum over the intrinsic brane degrees of freedom forms one loop, and (ii) a connected geometry where the sum over the intrinsic degrees of freedom gives two loops. 
\begin{align}
    \langle \Psi | \Psi \rangle  =\includegraphics[height=5em,valign=c]{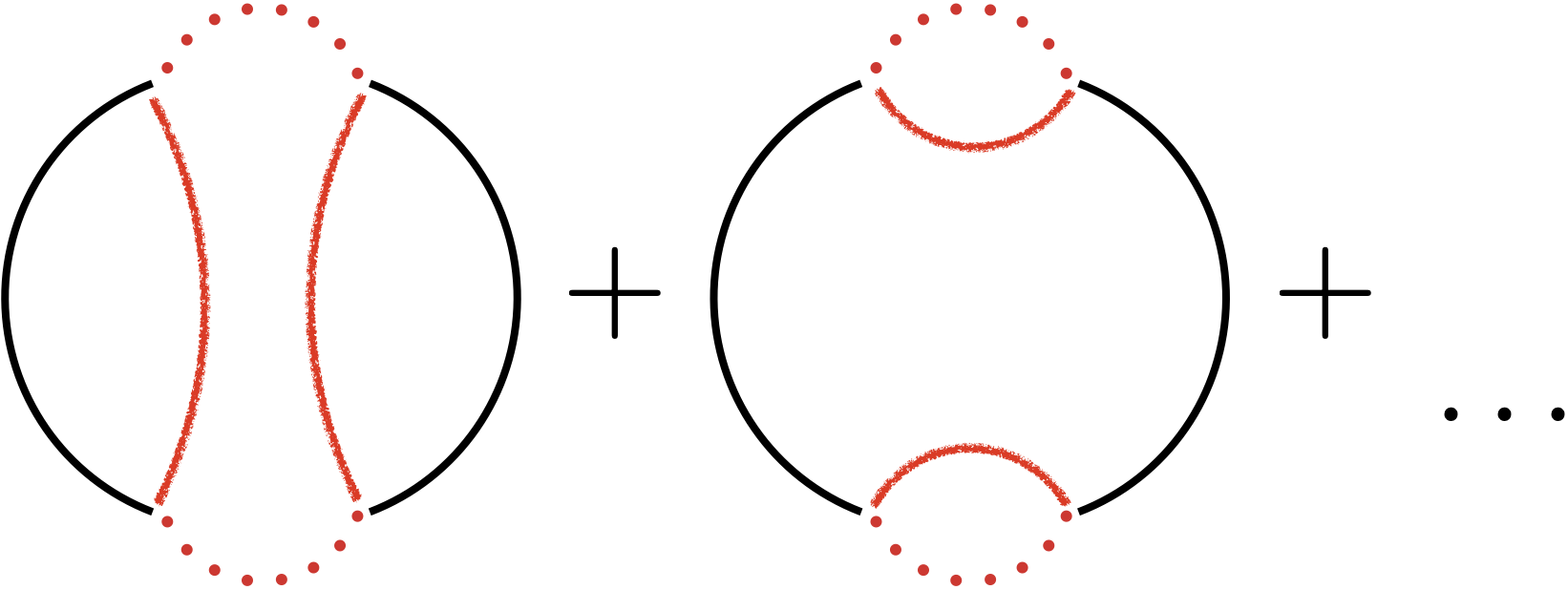} 
    \label{eq:dpssy-norm}
\end{align}
We have dropped non-planar as well as higher-genus contributions, which are further suppressed. Thus, we obtain
\beq\label{eq:norm}
\langle \Psi | \Psi \rangle =\frac{1}{\cN}\left( \mathrm{Tr}\,(M^{\dagger}M) e^{2S_0} Z_1(2\ell_1)Z_1(2\ell_2)+\mathrm{Tr}\,(M)\mathrm{Tr}\,(M^{\dagger}) e^{S_0}Z_2(2\ell_1,2\ell_2) +\dots\right),
\eeq
where $Z_1(\beta)$ is the gravity answer for the partition function with one asymptotic boundary of length $\beta$ in the disconnected geometry, and $Z_2(\beta_1,\beta_2)$ is the gravity partition function corresponding to the connected geometry with the boundary lengths $\beta_1$ and $\beta_2$ respectively. If we take $M$ to be the identity matrix for simplicity, then we see that for $k\gg e^{S_0}$, the connected geometry dominates. More generally, this transition happens when
\beq
\frac{\mathrm{Tr}\,(M)\,\mathrm{Tr}\,(M^{\dagger})}{\mathrm{Tr}\,(M^{\dagger}M)} \gg e^{S_0} \frac{Z_1(2\ell_1)Z_1(2\ell_2)}{Z_2(2\ell_1,2\ell_2)}. 
\eeq
If we cut open the connected geometry along the time reflection-symmetric slice in the bulk and use this as initial data to generate a Lorentzian spacetime, we get a two-sided eternal black hole geometry with a wormhole between the two boundaries, as shown in section \ref{ssec:dPSSY-bulk}. We thus find that when there is sufficient entanglement between the two systems, the dominant bulk geometry is connected, with a spatial wormhole joining the two black holes. We emphasize that the contribution of the connected geometry is enhanced here by a \emph{quantum} effect.  

Now we return to the question of entanglement entropy. In order to compute the entropy, we will study the R\'enyi entropy by computing the replica path integral for $\tr \rho_{1}^{n}$, and then analytically continue $n\to 1$. In order for the density matrix to be properly normalized, we must choose the normalization $\cN$ such that the right hand side of equation \eqref{eq:norm} is unity. The replica path integral is illustrated for the specific case $n=2$ in \eqref{eq:dpssy-repl}:\footnote{Since we can think of the ETW brane as the fixed point of a reflection symmetry \cite{KM}, we can double these two pictures.
After doubling, the ETW brane becomes just a massive particle, and the saddle-point geometries worked out in section \ref{ssec:dPSSY-bulk} are valid on each side of this massive particle.
The second picture then becomes the `type IV' replica wormhole from \cite{Balasubramanian:2021wgd}, with the difference that the bulk effective theory is massive rather than massless.
The second contribution in \eqref{eq:norm} is, after doubling, what \cite{Balasubramanian:2021wgd} call a `cylinder wormhole'.}
\begin{align}
    \tr \left(\rho_{1}^2\right) =  \includegraphics[height=10em,valign=c]{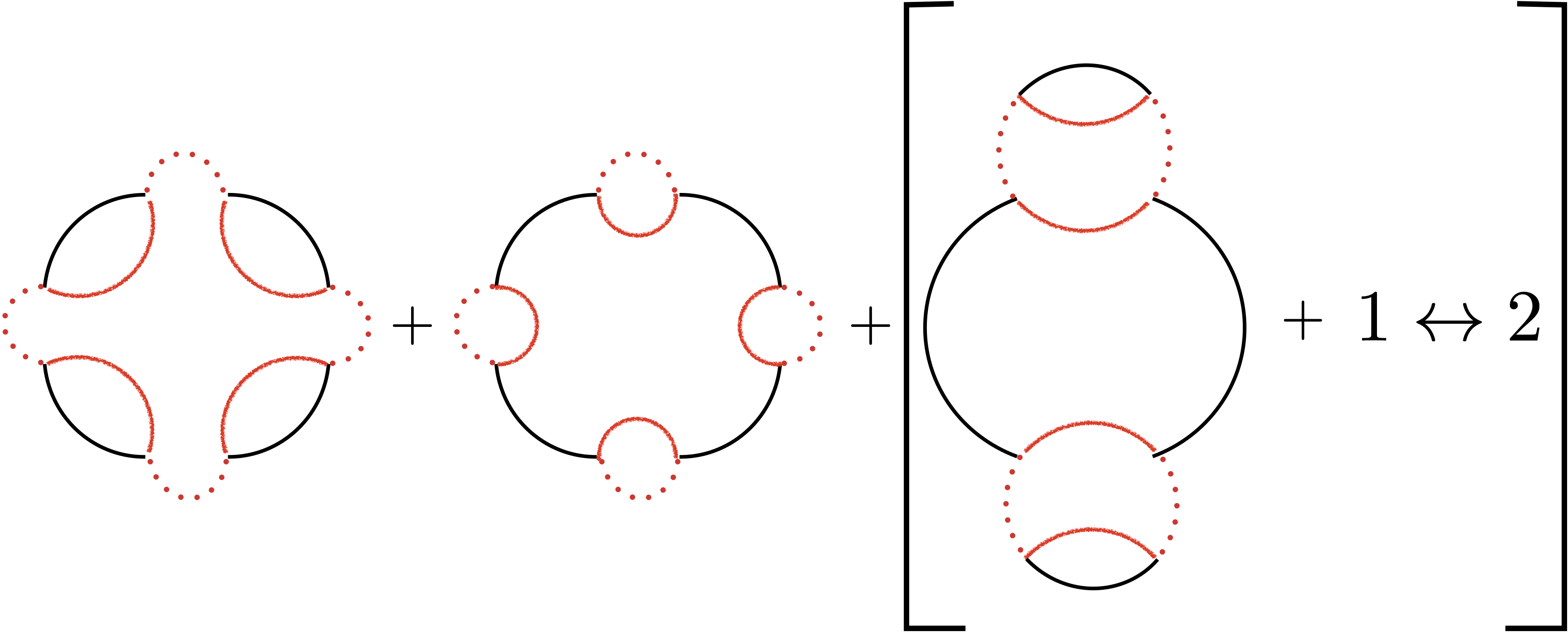}
    \label{eq:dpssy-repl}
\end{align}
In the third configuration above, all the boundaries in the connected component correspond to system 1 and the boundaries of the disconnected components all correspond to system 2; the fourth one is the same with systems 1 and 2 interchanged.
In \eqref{eq:dpssy-repl} we have only shown the replica-symmetric geometries;  there are also replica-symmetry-breaking geometries, but we will defer a discussion of these to Appendix \ref{ssec:resolvent}. Following \cite{Penington:2019kki}, we define $e^{S_{0}} Z_{r} (2\ell_1,\cdots,2\ell_r)$ as the partition function of Euclidean $AdS$ with $r$ disconnected boundary components of lengths $\{2\ell_i\}$ and ETW branes between adjacent boundary components.
At small $k$, the dominant contribution in the replica calculation is the one with most number of disconnected bulk components, as each of these components comes with a topological factor of $e^{S_0}$. In this limit, we get
\beq \label{eq:smallK}
\mathrm{Tr}\,(\rho_1^n) \sim \frac{\mathrm{Tr}\,[(M^{\dagger}M)^n]}{[\mathrm{Tr}\,(M^{\dagger}M)]^n},\;\;\;\;\cdots\;\;\; (k \ll e^{S_0}).
\eeq
At large $k$, the dominant contribution is the completely connected geometry as this has the most number of brane loops:
\beq\label{eq:largeK}
\mathrm{Tr}\,(\rho_1^n) \sim e^{(1-n)S_0} \frac{Z_n(2\ell_1,2\ell_2,\cdots,2\ell_1,2\ell_2)}{Z^n_2(2\ell_1,2\ell_2)},\;\;\;\;\cdots\;\;\; (k \gg e^{S_0}).
\eeq
Both of these are $\mathbb{Z}_{n}$-symmetric solutions, and thus the corresponding extremal surfaces are the $\mathbb{Z}_{n}$-symmetric points.
In the small $k$ limit, this procedure lands us on the empty surface as expected; all the dependence on the gravity path integrals in \eqref{eq:smallK} cancels out and we get after taking $n\to 1$: 
\beq
S_{E} = -\sum_{i} p_i\,\log\,p_i.
\eeq
On the other hand, the large $k$ limit holds a surprise, in that the QES is in a connected geometry, which, after continuation to Lorentzian signature, does \emph{not} give the geometry of figure \ref{fig:dpssy-naive-lor}. Indeed, the entanglement entropy obtained from equation \eqref{eq:largeK} is given by 
\beq
S_{E} = S_0 + \phi_h (\ell_1,\ell_2,\mu),
\eeq
where $\phi_h$ is the value of the dilaton at the extremal surface in the connected geometry. It was shown in section \ref{ssec:dPSSY-bulk}, specifically \eqref{eqn:conn-disc-ent-comparison}, that this is smaller than any of the generalised entropies that appear in \eqref{eq:naive}.
Thus, in the large $k$ limit, the correct entanglement entropy is given by the ``horizon area'' of a new gravitational saddle point, and the naive application of the island rule (equation \eqref{eq:naive}) fails. 

We regard this phenomenon --- the change in the dominant saddle geometry for these calculations --- as a realisation of the ER = EPR paradigm \cite{er-epr} and one of the central points of this paper. In the usual Hawking-Page transition, a new gravitational saddle dominates when some classical parameter controlling the boundary conditions, such as the temperature, is tuned. It is important to stress that in the present case, the transition to a connected geometry is a quantum effect; this transition is forced upon us by the large entanglement rank of the matrix $M^{\dagger}M$. In this sense, entanglement leads to a connected geometry. Another important point here is that the calculation above is unaffected by the precise structure of the entanglement between the two branes, i.e., the details of the matrix $M$, thus giving another example of classical gravity exhibiting averaging \cite{Saad:2019lba,Penington:2019kki,Faulkner:2020,Stanford2020,Belin:2020hea,Engelhardt:2020,Altland:2020ccq}.

%%%%%%%%%%%%%%%%%%%%%%%

\section{Coupling Two Pure-State Black Holes} \label{sec:evap}
In this section, we try to extend the discussion of section \ref{sec:dPSSY} to a state in which the entanglement is built up by coupling the two black holes in real time.
We first study the entanglement between the two black holes and find that there is no Hawking-Page-like transition that realises the ER = EPR proposal.
We then attempt to find such a transition for other quantities (such as correlation functions) in a single-copy path integral.

We consider a state of two 1D quantum mechanical (QM) systems obtained by beginning with two energetic pure states and evolving them with a coupled Hamiltonian for a real time $u$.
We attempt to calculate some path integrals, like the replica trick and simple correlation functions, semi-classically and find ourselves forced to consider a second saddle of connected topology.

The state we begin with is the Kourkoulou-Maldacena \cite{KM} state $\ket{\mu}$.
This state is a pure state of a 1D QM system that is dual to an ETW brane of mass $\mu$; we take the limit $\mu \to \infty$, which will simplify our calculations.
The state at time $ u = - i \delta$\footnote{We are choosing $u = 0$ to be the moment at which real-time evolution begins. As such, the ETW branes meet the boundary at $u = - i \ell - i \delta$} is two copies of $\mu$ evolved by a Euclidean time $\ell$,
\begin{equation}
  \ket{\psi(-i\delta)} = e^{- H_{L} \ell- H_{R} \ell } \ket{\mu}_{L} \otimes \ket{\mu}_{R}.
  \label{eqn:t-delta-state}
\end{equation}
We evolve with a coupled Hamiltonian to $u = 0$ to perform a joining quench regulated by $\delta$,
\begin{equation}
  \ket{\psi(0)} = e^{- (H_{L} + H_{R} + H_{int}) \delta} \ket{\psi(-i\delta)}.
  \label{eqn:t0-state}
\end{equation}
Here $\delta$ plays the role of a regulator, smoothing out the UV behaviour of the joining quench
We then evolve for a real time $u$ with the coupled Hamiltonian
to find the final state
\begin{equation}
  \ket{\psi_{f}} = e^{- i (H_{L} + H_{R} + H_{int}) u} \ket{\psi(0)}.
  \label{eqn:final-state}
\end{equation}

Let us now describe the bulk dual picture.
Each QM system is dual to a 2d system of JT gravity with negative cosmological constant coupled to a 2d CFT of central charge $c$.
We take the limit $1 \ll c \ll \phi_{r}/\ell$, and further assume that the 2d CFT is itself holographic with a 3d bulk dual whose effective theory is pure general relativity (GR) at leading order in $c$.
The trajectory of the QM system in time is conformal to the asymptotic boundary of the $AdS_{2}$.
The lack of exact conformal symmetry in one dimension is accounted for by the important role of an explicit cutoff $\epsilon$ in the bulk; all states in JT gravity are different cutouts of $AdS_{2}$, with the boundary having metric \eqref{eqn:bd-conds}.
As already mentioned, the state $\ket{\mu}$ is dual to an ETW brane of mass $\mu$.

Finally, we have to define the interaction Hamiltonian $H_{int}$.
We assume that it creates transparent boundary conditions for the bulk CFT in the limit $\epsilon \to 0$.
Away from $\epsilon = 0$, this specific boundary condition will prove somewhat inconvenient since it would require us to keep careful track of the shape of the cutouts in coupling the two bulk spacetimes.
As a way to pretend that the $AdS$ goes all the way to the asymptotic boundary, we attach a strip of flat space CFT to fill in the `gap', as it were.
The CFT strip has the metric
\begin{align}
  ds_{CFT}^{2} &= \frac{1}{\epsilon^{2}} dy d \bar{y}, \label{eqn:strip-metric} \\
  \frac{y - \bar{y}}{2} &= u.
  \label{eqn:y-defn}
\end{align}
Equation \eqref{eqn:y-defn} will define $y$ throughout this paper.
The $\epsilon^{2}$ in \eqref{eqn:strip-metric} allows for simple matching to the $AdS_{2}$ cutouts with boundary conditions \eqref{eqn:bd-conds}, when the gluing is purely along $u$ i.e. at constant $y + \bar{y}$.

We first give an overview of the calculation of entanglement entropy and a simple correlation function in section \ref{ssec:evap-summary}, and study the details in section \ref{ssec:disc-hist}.
We will also use some results that are derived in detail in section \ref{sec:conn-hist}.

\subsection{Overview of Results} \label{ssec:evap-summary}
\begin{figure}[h]
  \begin{subfigure}[c]{.5\textwidth}
    \centering
    \includegraphics[width=70mm]{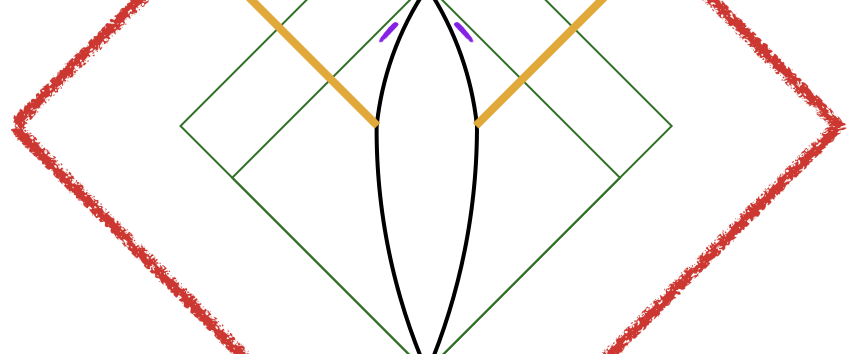}
    \caption{The disconnected history.}
    \label{fig:disc-hist}
  \end{subfigure}
  \begin{subfigure}[c]{.5\textwidth}
    \centering
    \includegraphics[width=70mm]{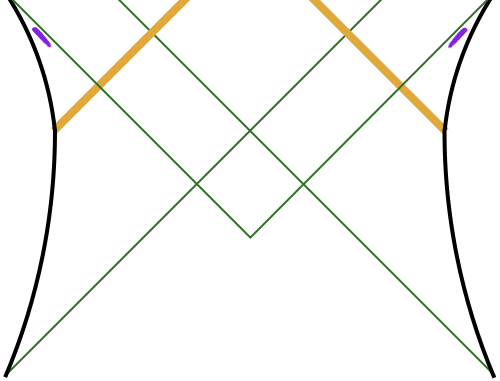}
    \caption{The connected history.}
    \label{fig:conn-hist}
  \end{subfigure}
  \caption{The disconnected and connected histories, without the splitting quench. The two boundaries are coupled at $u = 0$, which produces a symmetric pair of shocks (orange), kicking the boundary particle outward and moving the future horizon towards the boundary; all causal horizons are green lines. We have also indicated the rough position of the late time QESs.}
  \label{fig:hists}
\end{figure}

The naive bulk dual of the norm path integral is as follows.
At time $0$, we have two pure state black holes of inverse temperature $\beta_{0} = 2 \ell + \mO(1/\mu)$, see \eqref{eqn:disc-rh} for a derivation of the temperature.
Coupling them produces symmetric shocks that fall into the two black holes, heating them up to inverse temperature $\beta_{1} = \beta_{0} \left[ 1 - \mO(c \beta_{0}/\phi_{r}) \right]$.
These two systems are now in equilibrium and no more energy is exchanged, as we will show in section \ref{ssec:disc-hist}.
We thus have two identical coupled $AdS$-Vaidya black holes with ETW branes behind the horizons, as shown in figure \ref{fig:disc-hist}.

However, as with section \ref{sec:dPSSY}, this `disconnected history' is not the whole story.
There is an alternate saddle, a `connected history' that we show in figure \ref{fig:conn-hist}; the details of this saddle are studied in section \ref{sec:conn-hist}, in a different context.
We will study whether this saddle dominates over the one described above at late times, in any path integral.
We caution, given the difficulty of performing an exhaustive search over the space of semiclassical saddle points, that \emph{neither} of these saddles may be the leading saddle in any of these cases.

We can understand this connected saddle from the following perspective, which is beautifully explained in \cite{Jafferis:2017tiu}: even factorised states can have non-vanishing (but exponentially suppressed) overlap with the TFD state. In the present context, the realisation of this in bulk classical theory is a saddle in which the two ETW branes join up as in figure \ref{fig:pssy-conn}, and the geometry near the time-reflection symmetric slice is precisely that of a two-sided eternal black hole.
The inverse temperature of the black holes in this saddle is $\tilde{\beta_{0}} = 4 \ell + \mO(1/\mu) \approx 2 \beta_{0}$, see \eqref{eqn:conn-rh}.
Again, coupling the ends produces shocks but no more energy exchange thereafter, giving rise to a two-sided $AdS$-Vaidya geometry of inverse temperature $\tilde{\beta}_1 = \tilde{\beta}_0 \left[ 1 - \mO (c\beta_0/\phi_r) \right]$.
We summarise the main properties of the two histories in table \ref{tab:hists}.

\begin{table}[h]
    \centering
    \begin{tabular}[h!]{|c||c|c|}
      \hline
       Property & Disconnected History & Connected History \\
      \hline \hline
       Temperature before quench & $\beta_0 = T_0^{-1} = 2\ell$ & $\tilde{\beta}_0 = \tilde{T}_0^{-1} = 4\ell$ \\
      \hline
       CFT State Before Quench & Two copies of TFD state (Kruskal vacuum) & TFD state (Kruskal vacuum) \\
      \hline
      Location of ETW Brane(s) & Far behind horizon ($z \approx \mu$) & Nowhere \\
      \hline
    \end{tabular}
    \caption{A summary of the main properties of the two histories, shown in figure \ref{fig:hists}, that we have to consider in the dynamical situation.}
    \label{tab:hists}
\end{table}

We first study the replica trick path integral and look for the HRT surface that calculates the entanglement entropy between the two QM systems.
We then proceed to discuss the possibility of a similar phase transition for the bulk dual as well.

\subsubsection{The Entanglement Entropy} \label{sssec:summary-rt}
To calculate the entanglement entropy between the two QM systems, which we shall often refer to as the UV EE, we might try to use the 2d HRT formula \eqref{eqn:jt-rt}; as we shall see, it is too naive for our present purposes, but it is useful to calculate the generalised entropies that the HRT formula instructs us to minimise over.
To calculate the CFT EE $S_{bulk}$ we use standard CFT techniques for a CFT on the backreacted geometry, in sections \ref{sssec:disc-ent} and \ref{ssec:conn-ent}.

Before going ahead, we note that the symmetry of the setup will lead to the existence of pairs of degenerate QESs, which is somewhat non-standard.
According to the results of \cite{Murthy:2019,DW,MWW}, these provide an upper bound to the corresponding generalised entropies, with $\mO(\sqrt{\phi_{r}})$ corrections.
These corrections will not modify the main story.

At early times, the dominant QES is the empty surface and
\begin{equation}
  S_{E} (\text{early}) = S_{gen} (\varnothing) \approx  \frac{\pi}{3} c T_{1} u + \mO\left(\phi_r^0\right),
  \label{eqn:disc-ee-early}
\end{equation}
where $T_1$ is the temperature of the two black holes after the quench. 
This is the leading behaviour at $u \sim \phi_r$ and later, as we show in section \ref{sssec:disc-ent}.
We have ignored some transients at times much shorter than $\phi_{r}$.
Of course, as time goes on, this linear growth needs to be cut off at the Page time $u_{Page} \sim S_{0}/c$.
It is in fact cut off by a pair of symmetric QESs, shown by purple lines in figure \ref{fig:disc-hist}, giving an entropy
\begin{equation}
  S_{gen} (\text{disconnected,late}) \approx S_{0} + 2\pi T_{1} \phi_{r} - \mO(\sqrt{\phi_r}).
  \label{eqn:disc-ee-late}
\end{equation}

In the connected history as well there are two sets of QESs, that exchange dominance at a time $\sim \phi_{r}/c$.
There is a single QES that dominates at early times close to the bifurcate horizon with
\begin{equation}
  S_{gen} (\text{connected,early}) \approx S_{0} + 2\pi \tilde{T}_{0} \phi_{r} + \frac{\pi}{3} c \tilde{T}_1 u - \mO \left( \sqrt{c} \right),
  \label{eqn:conn-ee-early}
\end{equation}
where $\tilde{T}_0 \approx \frac{1}{2}T_0$ is the temperature of the eternal black hole in the connected saddle and $\tilde{T}_1$ is the post-quench temperature in this history.
This QES is not quite at the bifurcate horizon, but its position is symmetric between the two boundaries and to the future of the bifurcate horizon, as in \cite{Gao:2016bin}.
There is also a degenerate pair of surfaces at late times, right outside the late-time horizons, each of which is similar to that in \cite{Chen:2020jvn}, with
\begin{equation}
  S_{gen} (\text{connected,late}) = S_{0} + 2\pi \tilde{T}_{1} \phi_{r} - \mO(\sqrt{\phi_r}).
  \label{eqn:conn-ee-late}
\end{equation}

To find the minimum of the four generalised entropies above, we note a rather interesting fact: that before the quench, the black holes in the connected history are significantly cooler than those in the disconnected history, $\tilde{T}_{0} \approx \frac{1}{2} T_{0}$.
Since the stress energy carried by the shock does not scale with $\phi_{r}$ but the ADM energy of the pre-quench black holes does, we expect that $(T_{1} - T_{0})$ and  $(\tilde{T}_{1} - \tilde{T}_{0})$ are both $\mO (\phi_{r}^{-1})$.
This means that
\begin{equation}
  \tilde{T}_1 \approx \tilde{T}_0 = \frac{1}{2} T_0 \approx  \frac{1}{2} T_1,
  \label{eqn:temp-comp}
\end{equation}
Thus, the black holes in the connected geometry are colder and the shocks do not change this fact.
Just minimising these generalised entropies, then, would tell us that the UV EE saturates to the value \eqref{eqn:conn-ee-late} at late times and the HRT surface is in the connected history.

However, let us check this expectation more carefully.
We calculate the EE using the replica trick, in which we first calculate $\tr \rho^{n}$ and then analytically continue the R\'enyi entropy $S_{n} = \left( \log \tr \rho^{n} \right)/(1-n)$ in the index $n$ to $n = 1$.
$\tr \rho^n$ is written as $Z_{n}/Z_{1}^{n}$, where $Z_{n}$ is the path integral given by the following boundary condition: we take $n$ `ket' and $n$ time-reversed `bra' copies of the 1d path integral creating the state $\ket{\psi_f}$ in \eqref{eqn:final-state} and sew them up in the usual way corresponding to the replica path integral \cite{Calabrese:2009}.
Then, we have to find the dominant saddle consistent with these asymptotic boundary conditions.

Restricting to replica-symmetric saddles, we may take a $\mathbb{Z}_n$ quotient to get a geometry filling just one `ket' and one `bra' copy with a zero-dimensional ``twist brane,'' which has the following properties: (a) it has a `tension' $\sim \phi_r (1 - 1/n)$, (b) it acts as a twist operator for bulk quantum fields, and (c) it has to be homologous to the two QM systems\footnote{More clearly, this is a set of point sources on a Cauchy slice that splits the Cauchy slice into two regions, each of which contains one of the boundary QM systems.} \cite{Lewkowycz:2013,Dong:2016,Dong:2018}.
Following \cite{Almheiri:2020b}, we don't quotient the matter CFT so that the matter partition function on the quotiented manifold is effectively a R\'enyi partition function $Z_{CFT,n}$.
This path integral is thus computed by any bulk history consistent with the asymptotic boundary conditions corresponding to the norm of the QM state, but with one or more twist branes in the bulk.
The semiclassical partition function, keeping track only of terms that potentially scale with $S_0$, is
\begin{align}
  \log Z_{n} = \sum_{\text{semi-classical saddles}} &\left[  S_0 \left\{ n\; \chi_{quotient} + (1-n) \sum_{\text{twist branes}} 1 \right\} + \log Z_{CFT,n} + \dots \right],
  \label{eqn:semicl-action}
\end{align}
$\chi_{quotient}$ is the Euler characteristic of the quotient manifold, and the object in the braces is that of the unquotiented manifold \cite{Almheiri:2020b}.\footnote{This is easily proved by tiling the replicated manifold so that each twist brane is a vertex in the tiling and using the $V-E+F$ formula for $\chi$.}
In the limit $n\to 1$, the location of the twist brane is precisely the QES; thus, a QES in any history consistent with the asymptotic boundary conditions is a valid candidate for the true HRT surface.

We first notice that the disconnected history has $\chi_{quotient} = 2$ whereas the connected one has $\chi_{quotient} = 1$, and at late times both have one twist brane in the bulk, and so the connected history is suppressed because of the topological term in the action, as in section \ref{sec:dPSSY}.
In section \ref{sec:dPSSY}, however, the analog of $\log Z_{CFT,n}$ competes with this, giving a factor of $k^{n}$.
In our case, this is a CFT partition function on the background that includes the twist brane as well as its backreaction.
Since the mass of the brane vanishes as $n \to 1$ (though it is still $\mO(\phi_r)$) its backreaction is generically $\mO( (n-1) \phi_r^0)$.\footnote{We have to maintain the order of limits $\phi_r^{-1} \ll n-1$; otherwise one can get wrong results near phase transitions.}
Thus, in this limit, we can approximate the background (at leading order in $n-1$) as the saddle-point of the norm path integral; meaning that $Z_{CFT,n}$ is simply related to the twist-operator correlation function in this geometry, as $Z_{CFT,n} = Z_{CFT,1}^n \left\langle \prod_i \sigma_n (x_i) \right\rangle$.
Because of the unitarity of time evolution, $Z_{CFT,1}$ is independent of time in our set-up.

But we have seen that in both histories the twist operator correlation function saturates to some $\mO(1)$ constant at late times, meaning that $Z_{CFT,n}$ does so as well.
Therefore, the HRT surface is the degenerate pair in the \emph{disconnected} history and the UV EE is
\begin{equation}
  S_{E}(u) = \min \left\{ S_{gen} (\varnothing), S_{gen} (\text{disconnected,late}) \right\}.
  \label{eqn:final-ee}
\end{equation}
In this case with multiple bulk saddles of different topology, the HRT surface is \emph{not} the QES of minimal generalised entropy.

\subsubsection{A Transition of The Bulk Dual?} \label{sssec:bulk-dual}
The existence of this alternate history is tantalising, however, and we are led to ask whether there are \emph{any} boundary path integrals for which the dominant saddle becomes the connected history.
The norm path integral, on its own, is insufficient.
Without any insertions at late time, the norm path integral is insensitive to the amount of real time evolution, since real time evolution is unitary.
In bulk semiclassical theory, this is realised by the fact that the on-shell semiclassical action is real on each Lorentzian sheet and so it cancels between forward and backward time evolution.

Thus, we need to do something at time $u$, like averaging over time or inserting an operator.
We insert a single primary operator $O$ of dimension $\Delta$ in the central CFT strip, at the position $y = - \bar{y} = u$, and explore the possibility of a transition to the connected history. Presently, we do not have a complete picture for such a transition, but we will make preliminary observations.

The unnormalised partition function with the insertion, then, is
\begin{equation}
  Z_{full} = e^{- I_{grav,disc}} Z_{CFT,disc} \langle O(u) \rangle_{disc} + e^{- I_{grav,conn}} Z_{CFT,conn} \langle O(u) \rangle_{conn},
  \label{eqn:opf-Z}
\end{equation}
where we have summed over both histories.
Here, $\langle O(u) \rangle$ is the usual one-point function in the respective history.
Because of unitarity, the one-point function is the only factor in either term that depends on $u$.

Thus, the dominant bulk dual is the one with the lower value of
\begin{equation}
  I = - S_{0} \chi - \log \langle O(u) \rangle + \dots,
  \label{eqn:action}
\end{equation}
where $\chi$ is the Euler characteristic of the bulk geometry.
The $\dots$ terms are $u$-independent terms, including the Schwarzian action\footnote{The solutions are real, and so this cancels between the sheets} and the CFT partition function, that don't matter at leading order since they do not scale with $S_{0}$.
The disconnected history has $\chi = 2$ and the connected history has $\chi = 1$, leading to the disconnected history being dominant at early times.
We will show that, at late times, the two terms compete with each other.

To calculate the one-point function, we use the rule that
\begin{equation}
  \langle O(u) \rangle_{g} = e^{-\Delta\; \Omega(u)} \langle O(u) \rangle_{\hat{g}}, \qquad g_{\mu\nu} = e^{2 \Omega} \hat{g}_{\mu\nu}.
  \label{eqn:opf-trans-gen}
\end{equation}
In this section, we choose convenient fiducial metrics in the two histories and show that the quantity
\begin{equation}
  \tilde{I} = - S_{0} \chi + \Delta\; \Omega (u)
  \label{eqn:red-action}
\end{equation}
becomes smaller in the connected history than in the disconnected, and then argue that this implies a change in the dominance of the saddle.

\begin{figure}[h!]
  \centering
  \includegraphics[width=100mm]{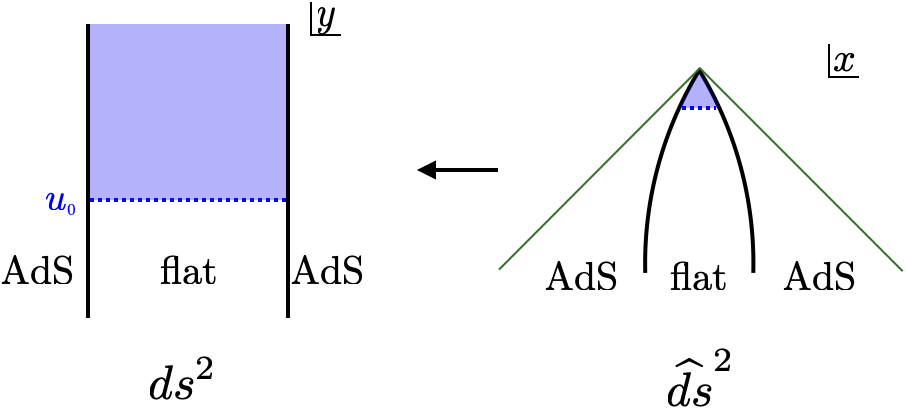}
  \caption{On the left, the flat strip is a rectangle in $y$ coordinates. An infinite rectangle given by $u > u_{0}$ is squeezed into a small finite volume in Poincare $x$ coordinates. This is the source of the large conformal factor in \eqref{eqn:logZ-disc}. A similar squeezing occurs in the connected history, except that we consider global rather than Poincare coordinates. }
  \label{fig:squeezing}
\end{figure}

In the disconnected history, we take the fiducial metric to be the flat metric in Poincare coordinates,
\begin{equation}
  \hat{ds}_{disc}^{2} = dx d\bar{x}, \quad x = z + t_{P}.
  \label{eqn:ghat-disc}
\end{equation}
We will justify this choice in section \ref{ssec:disc-hist}.
Since we are in the central CFT strip, the physical metric is
\begin{equation}
  ds^{2} = \frac{d y d \bar{y}}{\epsilon^{2}} = \frac{1}{\epsilon^{2} x'(y) \bar{x}'(\bar{y})} dx d \bar{x}, \qquad x(y) = \frac{1}{\pi T_{1}} \tanh (\pi T_{1} y).
  \label{eqn:g-disc}
\end{equation}
We have here substituted the future-of-the-shock coordinate transformation in the $AdS$-Vaidya geometry derived in appendix \ref{app:JT}, see \eqref{eqn:tP-vaidya-soln}.
The conformal factor is, then,
\begin{align}
  \Omega &\sim \log \frac{1}{\epsilon t_{P}'(u)} = - \log \epsilon + \log \cosh^{2} \left( \pi T_{1} u \right) \nonumber\\
  &\approx 2 \pi T_{1} u
  \label{eqn:logZ-disc}
\end{align}
The source of this growth is clear when we draw out the $y$ and $x$ coordinate systems, as in figure \ref{fig:squeezing}.
So, we have for the reduced action \eqref{eqn:red-action}
\begin{equation}
  \tilde{I}_{disc} = - 2 S_{0} + 2 \Delta\; \pi T_{1} u.
  \label{eqn:I-disc}
\end{equation}

The convenient fiducial metric in the connected history is the flat metric in \emph{global} coordinates,
\begin{equation}
  \hat{ds}_{conn}^{2} = d \mfs d \bar{\mfs}, \quad \mfs = \sigma + t_{gl}.
  \label{eqn:ghat-conn}
\end{equation}
Again, we will justify this in section \ref{sec:conn-hist}.
The physical metric is
\begin{align}
  ds^{2} = \frac{1}{\epsilon^{2}} d y d \bar{y} &= \frac{1}{\epsilon^{2} \mfs'(y) \bar{\mfs}' (\bar{y})} d\mfs d \bar{\mfs}, \nonumber\\
  &\qquad\mfs(y) = 2 \tan^{-1} \frac{\tanh 2\pi \tilde{T}_{1} u}{\sqrt{1+2\alpha}}, \qquad \tilde{T}_1 = \tilde{T}_0 \sqrt{1+2\alpha}.
  \label{eqn:g-conn}
\end{align}
The coordinate transformation is derived in appendix \ref{app:JT}, see \eqref{eqn:t-gl-vaidya-soln}.
The conformal factor we now find to be
\begin{align}
  \Omega &=  \log \frac{1}{\epsilon t_{gl}'(u)} = - \log \left( \epsilon  2\pi \tilde{T}_{1} \right) + \log \cosh 2\pi \tilde{T}_{1} u \nonumber\\
  &\approx  2\pi\tilde{T}_{1} u.
  \label{eqn:logZ-conn}
\end{align}
So, the reduced action \eqref{eqn:red-action} in this case is
\begin{equation}
  \tilde{I}_{conn} = - S_{0} + 2\Delta\; \pi \tilde{T}_{1} u+ \cdots.
  \label{eqn:I-conn}
\end{equation}

The reduced action \eqref{eqn:red-action} becomes smaller in the connected history when
\begin{align}
  \tilde{I}_{disc} - \tilde{I}_{conn} &> 0 \nonumber\\
  \Rightarrow u &> \frac{1}{2 \Delta\; \pi \left( T_{1} - \tilde{T}_{1} \right)} S_{0} \approx \frac{1}{\Delta\; \pi T_{0}} S_{0}.
  \label{eqn:conn-dom}
\end{align}
This is sufficient to show that the connected history dominates this path integral only if we assume that $\langle O (u) \rangle_{\hat{g}}$ neither diverges nor vanishes at large u.
This is not an innocuous assumption, since the conformal factors are much the same in the decoupled case; in this case, $\langle O(u) \rangle_{\hat{g}}$ diverges exponentially so as to restore Schwarzchild time-translation symmetry.
Roughly speaking, this divergence is a result of the operator going exponentially close to the boundary, at $z = (x+\bar{x})/2 = 0$ or $\sigma = (\mfs + \bar{\mfs})/2 = 0$, in the $\hat{g}$ metric at late times.
However, with the coupling between the black holes, the CFT does not see a boundary there and so we do not find this divergence.
We show this quantitatively in sections \ref{sssec:disc-bulk} and \ref{ssec:conn-bulk} for the disconnected and connected histories respectively.
Thus, we find that the path integral with a single insertion of a primary operator is dominated by the connected history at late times.

We now make some observations about this transition.
\begin{enumerate}
\item It is clear from the answer \eqref{eqn:conn-dom} that different operators transition at different times.
This goes back to the fact mentioned in the introduction that the idea of a particular `bulk dual' is somewhat meaningless, and the best we can say is that different saddles dominate different calculations.

\item In \eqref{eqn:opf-Z} and throughout this section, we considered the \emph{unnormalised} partition function with an insertion. The one-point function is the ratio of this object with the partition function sans insertion. However, the normalisation factor does not transition even when \eqref{eqn:opf-Z} does, leading to a somewhat non-standard answer for this correlation function.
In particular, this discrepancy means that the full one-point function is $\sim e^{-S_0} \langle O(u) \rangle_{conn}$.

\item The CFT strip is entirely unimportant here; we could have as well measured a one-point function of a QM operator dual to a primary operator in the bulk.
This is consistent with its introduction as a regulator.

\item The same effect is \emph{not} visible in higher-point functions of a set of operators that have identity in their OPE.
The reason is simple: if we place (say) two operators at an $\mO(1)$ distance in the $y$-plane, they end up at an $\mO(e^{-2\pi T u})$ distance in the $x$-plane or $\mfs$-cylinder, so that while the disconnected part dies off the connected part (i.e. expectation value of the identity term in the OPE) does not because that latter distance cancels out the conformal factor.

Suppose we try to place the two operators far enough apart so that this cancellation is avoided, say at $y_1 + \bar{y}_1 \sim e^{2\pi \tilde{T}_1 u}, y_2 + \bar{y}_2 = 0$.
The operator at $y_1$ is deep in the bulk and needs to be written as an HKLL operator on the boundary.
Then, in terms of boundary operators, this correlation function involves operators separated by a time $\Delta u \sim S_0$, large enough that there might be other saddles, like the baby-universe-emitting saddle of \cite{Saad:2019pqd}, that dominate the calculation.\footnote{Another problem is that this is deeper than a distance $\sim \beta \log \phi_r$ in the bulk and so one expects that the HKLL reconstruction is too oscillatory to be useful, see for example \cite{Rey:2014dpa}.}

\item The transition \emph{is} visible in any correlation function of operators that do not have identity in the OPE.

\item The lack of a transition of R\'enyi entropies is because they behave like a CFT two-point function rather than one-point function at late times.
On the other hand, it behaves like a two-point function because of the very same replica wormholes that give the island rule and the right Page curve; meaning that the same saddles that are required to reproduce the Page curve prevent the realisation of ER = EPR in this set-up.
\end{enumerate}

\paragraph{A Different Transition: Projector onto the Ground State}
Finally, we can also consider the question of what happens when we act on the product state $e^{- \ell (H_L + H_R)} |\mu\rangle_L\otimes |\mu\rangle_R$ with a projector onto the vacuum state of an interacting Hamiltonian $(H_L + H_R + H_{int})$, this time without any Lorentzian time evolution.
One way to accomplish this is to consider the state
\beq
|\psi(\delta)\rangle =\frac{1}{Z} e^{-U(H_L + H_R + H_{int})}|\mu\rangle_L\otimes |\mu\rangle_R,
\eeq
in the limit $U \to \infty$. We take $H_{int}$, as above, to be such that it creates transparent boundary conditions for the bulk 2d CFT across the asymptotic boundaries.
Now consider the bulk calculation of either the norm of $|\psi(\delta)\rangle$, or some correlation function with a small number of light operators.
It was shown in \cite{MQa} that this state corresponds to the eternal traversable wormhole, which is a connected geometry.

More concretely, we could as before consider either the disconnected or the connected geometry in the bulk.
We observe that in the disconnected geometry, the CFT lives on a background which is conformally equivalent to the half-plane, while in the connected geometry (which is suppressed by $e^{-S_0}$), the CFT lives on a background which is conformally equivalent to a cylinder of length proportional to $U$.
In the $U \to \infty$ limit, the connected geometry therefore receives an enhancement factor of $e^{\alpha c \delta}$ coming from the Casimir energy on the cylinder, where $\alpha$ is an $O(1)$ constant.
When $U \sim S_0/c$, this enhancement overcomes the suppression from the entropy and the connected geometry dominates the norm/correlation function.
In this way, we see that at sufficiently large $U$, the connected geometry dominates in the bulk.
This is again a quantum-entanglement-induced phase transition --- the operator $e^{-U (H_L + H_R + H_{int})}$ projects onto a highly entangled state of the two boundary quantum systems, and this large entanglement causes a Hawking-Page-like transition in the bulk geometry.

\subsection{The Disconnected History} \label{ssec:disc-hist}
Now that we've summarised the main story, we present the derivation of the geometry and the QESs in the disconnected history.
In this history, at time $-i\delta$, we have two black holes of inverse temperature $\beta = 2 \ell + \mO(\mu^{-1})$.
In the limit $\mu \to \infty$, the ETW branes are at $z = \mu \to \infty$.
The first step is to calculate the trajectory of the boundary particle after the joining quench.
Then, we study the conformal factor at late time, and also find all the QESs in this history.

The state of the 2d CFT in each black hole before the quench is the \emph{Kruskal vacuum}, i.e., the state at the time-reflection symmetric slice created by the 2d CFT Euclidean path integral on the ``lower half'' Euclidean black hole.
A useful fact about this state is that it is also the vacuum in the related flat metric $d\mfw d \bar{\mfw}$.
This can be seen by calculating the conformal anomaly contribution to the non-trace components of the stress tensor from stripping off the $1/(1-\mfw\bar{\mfw})^2$ conformal factor and finding that it vanishes, see e.g. \cite{AEMM}.
Because Kruskal coordinates and Poincare coordinates are related by an $SL(2,\mathbb{R})$ transformation,
\begin{equation}
    x = \frac{1+\mfw}{1-\mfw}, \qquad \{\mfw \bar{\mfw} = 1\} \mapsto \{x + \bar{x} = 0 \}.
    \label{eqn:kruskal-to-poincare}
\end{equation}
we find that the state is actually in the vacuum in the metric $dx d \bar{x}$ as well; we will therefore refer to it as the \emph{Poincare vacuum}.

\begin{figure}[h]
  \centering
  \includegraphics[width=0.98\textwidth]{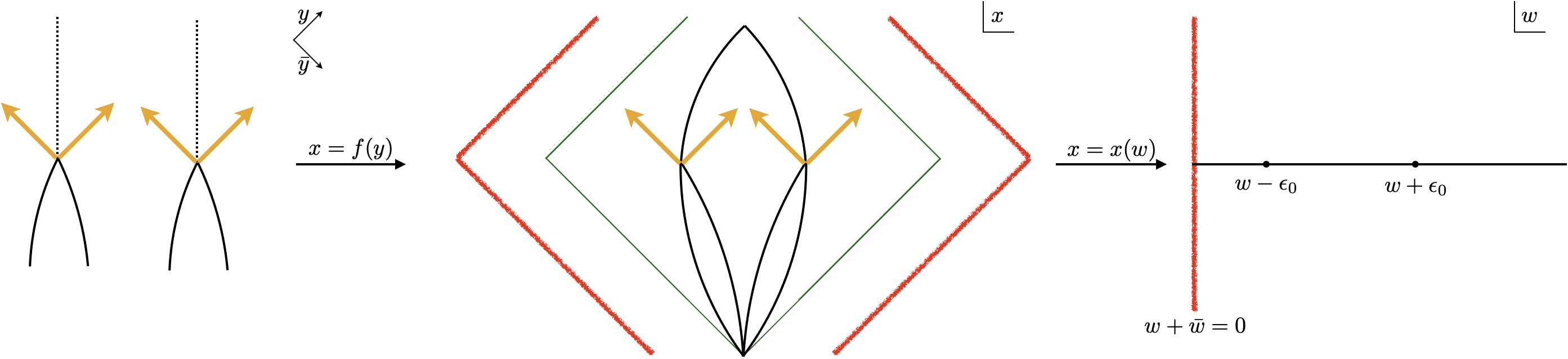}
  \caption{The conformal transformations for the disconnected history. Red lines, as above, are ETW branes, green lines are the causal horizons \emph{before} gluing, and orange lines are positive energy. All three pictures are in real time. The left-most figure is the $y$-plane, in which time is the UV time $u$. The middle figure is the Poincare plane; it is the only Penrose diagram here. The right-most picture is the $w$-plane, which is just half of $\mathbb{R}^2$ ($w + \bar{w}\ge 0$); in this plane, the state is the half-line vacuum for all time.}
  \label{fig:disc-conf}
\end{figure}

\subsubsection{The Bulk Geometry} \label{sssec:disc-bulk}
The gluing is done in physical time, i.e. by matching the proper times $u$ of the boundary particles, as shown in the first panel of figure \ref{fig:disc-conf}.
Also shown in the figure is the fact that the gluing results in the release of energy; our job is to calculate this energy and therefore the bulk metric.
The CFT is in a thermal state in the two $AdS$ spacetimes; we can rewrite the bulk metrics in Poincare coordinates,
\begin{align}
  ds_{i}^{2} &= \frac{4 dx_{i} d \bar{x}_{i}}{(x_{i}+\bar{x}_{i})^{2}} = \frac{4 x_{i}'(y) \bar{x}_{i}'(\bar{y}) dy d \bar{y}}{\left[ x_{i}(y) + \bar{x}_{i}(\bar{y}) \right]^{2}}, \qquad i = L,R, \nonumber\\
  ds_{C}^{2} &= \frac{dy d\bar{y}}{\epsilon^{2}}.
  \label{eqn:ads-coords}
\end{align}
With the boundary condition \eqref{eqn:bd-conds} the strip can be smoothly glued to the bulks at constant $y + \bar{y}$.

By stripping off the conformal factors, we get three \emph{different} conformally related flat metrics: the $y$, $x_{L}$, $x_{R}$ planes (where we have appropriately extended the ranges of these coordinates to cover both Poincare patches as well as the central strip).
We denote the conformal transformations between the three planes as
\begin{equation}
  x_{R} = x_{R} (y), \qquad x_{L} = - x_{L} (y) \qquad x_{R} = -x_{R} \circ x_{L}^{-1} (x_{L}).
  \label{eqn:conf-trans}
\end{equation}
The signs take care of the fact that the left bulk is reflected.
A crucial simplification follows from symmetry,
\begin{equation}
  -x_{L} = x_{R} = x.
  \label{eqn:symm}
\end{equation}
So, we have in fact only two planes, the $y$ and $x$ planes.
In the $y$ plane, the CFT in the $i^{th}$ bulk, spacelike to the joining quench, is in a thermal state; whereas in the $x_i$ plane, it is in the vacuum.

Of course, even in the $x$ plane it is not in the vacuum \emph{after} the quench.
However, as noted in \cite{CCQuench,AEMM}, the state after the quench is a descendant of the strip vacuum, i.e. there is a conformal transformation $x \to w(x)$ to a half-plane in which there is no quench.\footnote{In the Euclidean language, the quench geometry corresponds to a plane (labelled by $x$) with two slits along the imaginary axis placed in a time-reflection symmetric configuration. This plane can then be conformally mapped to a half-plane (labelled by $w$), where the slits get mapped to the boundary of the half plane.}
On the slice where Poincare time $t_{P} = 0$, or $x = \bar{x}$, we may write the conformal transformation to the line where $w = \bar{w}$ as
\begin{align}
  w = \frac{w_{0}^{2}}{w_{0} - x} \theta(-x - \epsilon) + (w_0 + x) \theta(x-\epsilon) + y \theta(\epsilon-x)\theta(x+\epsilon)
  \label{eqn:disc-conf-1}
\end{align}
at leading order in $\epsilon$.
Here, we have used the fact that the Poincare coordinate is cut off at $z = \epsilon f'(u)$ and $f'(0) = 1$.
The reason for the asymmetric treatment of the two bulks is that in a CFT, infinity is a single point and so, if both bulks extended to $|w| = \infty$ the quench would have involved joining the two bulks \emph{at both ends}.
Equation \eqref{eqn:disc-conf-1} has the property that $w'$ is continuous all the way across, and that it is an $SL(2,\mathbb{R})$ transformation within each of the three regions, so that it reproduces the correct stress tensor expectation values at points spacelike separated from the quench. 

Another important point is that we are interested in the limit $w_0 \to 0$, since we need correlation functions on this time-slice to factorise as $\left\langle \prod_i O_i(x<0) \prod_j O_j(x>0) \right\rangle = \left\langle \vphantom{\prod_j O_j} \prod_i O_{i}(x<0) \right\rangle \left\langle \prod_j O_j(x>0) \right\rangle$.
In the limit $w_0 \to 0$, operators in the $x>0$ region are much further from operators in the $x<0$ region than each other and the boundary, and so we find the required factorisation \cite{AEMM}.\footnote{ In \cite{CCQuench}, the map \eqref{eqn:disc-conf-1} is written slightly differently as $w = \left(x + \sqrt{x^2 + \delta^2} \right)/2$. In the limit $\delta \to 0$, this coincides with \eqref{eqn:disc-conf-1} with the identification $w_0 = \delta/2$.}

Since $T_{ww}$\footnote{We are, as always, using the convention that
\begin{equation*}
  T_{x^\alpha x^\beta} = \langle T_{\alpha\beta} \rangle_{ds^2 = \eta_{\alpha\beta} dx^\alpha dx^\beta}.
\end{equation*}
} vanishes by definition (i.e., the $w$-plane is chosen such that the CFT is in the vacuum on it), we find that at $\epsilon = 0$,
\begin{align}
  T_{xx} &= (w')^{2} \,T_{ww} - \frac{c}{24\pi} \{ w,x \}\nonumber\\
  &\approx \frac{c}{12\pi w_{0}} \delta(x).
  \label{eqn:disc-Tzz}
\end{align}
This and a similar calculation for $T_{\bar{w}\bar{w}}$ shows that the joining quench produces two shocks with no further energy exchange, and thus the backreacted geometry of each black hole is an AdS-Vaidya spacetime with a single shock.\footnote{In the non-symmetric case, one has to solve a differential equation, but here the answer is clear.}
The CFT strip which we inserted between the two black holes to facilitate the coupling only resolves the above shock to two closely-spaced shocks; to the future of these, this correction is not so important.

As mentioned above, it is thus easy to take into account the backreaction of this energy distribution.
The post-coupling state is just two $AdS$-Vaidya black holes of temperature $T_{1} > T_{0} = \frac{1}{2\ell} $, and the boundary particle trajectory is
\begin{equation}
  t_P(u) = \frac{1}{\pi T_{0}} \tanh ( \pi T_0 u ) \theta(-u) + \frac{1}{\pi T_{1}} \tanh ( \pi T_1 u ) \theta(u), \qquad \beta_{1} = \beta_{0} - \mO(c \beta_{0}/\phi_{r}) < \beta_0.
  \label{eqn:disc-trajectory}
\end{equation}
The temperature difference scales with the small parameter $c \beta_{0}/\phi_r$ because the black hole has ADM energy $(\pi T_0)^2 \phi_r$, see appendix \ref{app:JT}, and the shock adds an amount of energy $\sim c/w_0$ that does not scale with $\phi_r$; so that we have $\Delta E/E \sim \Delta \beta/\beta \sim c \beta/\phi_r$. 

A quantity of interest is the conformal factor between the CFT strip and the vacuum plane:
\begin{align}
  ds^{2} = \frac{1}{\epsilon^{2}} dy d \bar{y} &= e^{2 \Omega(y)} dw d \bar{w}, \nonumber\\
  e^{2 \Omega(y)} &= \frac{1}{\epsilon^{2}} \left| \frac{x'(w)}{x' (y)} \right|^{2} \nonumber\\
  &\xrightarrow{y = -\bar{y} = u} \frac{1}{\epsilon^{2}} \cosh^{4} (\pi T_{1} u) \nonumber\\
  \Rightarrow \quad \Omega(u) &\xrightarrow{u \to \infty} 2\pi T_1 u.
  \label{eqn:conf-factor}
\end{align}
This was the main input in \eqref{eqn:logZ-disc}, which went into determining the bulk dual.

The other important input in \eqref{eqn:logZ-disc} was the choice of the flat metric in Poincare coordinates as a fiducial metric.
Since the state in the $w$ plane is the vacuum, in reality it is $dw d \bar{w}$ that should be the fiducial metric.
However, the growing part of $w'(y)$ is $x'(y)$,
\begin{equation}
  w'(y) = w'(x) x'(y) \approx \frac{w_{0}^{2}}{x^{2}} x'(y) \xrightarrow{y = - \bar{y} = u \to \infty} (\pi T_{1} w_{0})^{2} x'(y).
  \label{eqn:real-conf-factor}
\end{equation}
So we see that \eqref{eqn:logZ-disc} is the right answer at leading order. 

We can also use these results to calculate the one-point function of a primary operator at $y = - \bar{y} = u$.
\begin{equation}
    \langle O(u) \rangle = \left\langle O \left( w = t_P(u), \bar{w} = \frac{w_0^2}{t_P(u)} \right) \right\rangle_{dw d\bar{w}} e^{- \Delta \Omega} \propto \frac{1}{t_P(u)^\Delta} e^{-2 \pi T_1 \Delta u}.
    \label{eqn:disc-opf}
\end{equation}
This validates the assertion around \eqref{eqn:opf-trans-gen} that only the conformal factor competes with the $S_0$ term.

\subsubsection{Entropies} \label{sssec:disc-ent}
For a point at the center of the strip and arbitrary time $y = -\bar{y} = u$, the calculation of quantum extremal surfaces is the same in most particulars as that in the ``thermal equilibrium'' calculations in \cite{Chen:2020jvn}, in which they couple an $AdS$ black hole to a \emph{non-gravitational} bath at temperature $T_{1}$.

The reason for this agreement between these two seemingly distinct cases can be understood as follows.
Under a Weyl transformation the CFT EE transforms as
\begin{equation}
    S_{bulk,g} = S_{bulk,\hat{g}} + \frac{c}{6} \sum_{endpts} \Omega, \qquad g = e^{2\Omega} \hat{g}.
    \label{eqn:ee-conf}
\end{equation}
We will take $\hat{ds}^2 = dw d \bar{w}$, as above.
$S_{bulk,\hat{g}}$ is a completely fixed function of the $w,\bar{w}$ coordinates, the CFT EE on the half-line vacuum; for a region with either one or two endpoints it is
\begin{align}
    S_{bulk,\hat{g},1pt} \left( (w,\bar{w}) \right) &= \frac{c}{6} \log (w + \bar{w}) + \log g \nonumber\\
    S_{bulk,\hat{g},2pt} \left( (w_1,\bar{w}_1),(w_2,\bar{w}_2) \right) &= \frac{c}{6} \log \left[(w_1 - \bar{w}_1)(w_2 - \bar{w}_2) \eta\right] + \log G(\eta), \qquad \eta = \frac{(w_1 + \bar{w}_1)(w_2 + \bar{w}_2)}{(w_1 + \bar{w}_2)(w_2 + \bar{w}_1)}.
\end{align}
$G(\eta)$ is a theory-dependent function that has a particularly simple form when the 2d CFT is itself holographic \cite{Chen:2020jvn}.
Taking the simple case of $g = 1$ in a holographic CFT, the bulk entropy is
\begin{align}
    S_{bulk,\hat{g},2pt} \left( (w_1,\bar{w}_1),(w_2,\bar{w}_2) \right) = \begin{cases} \frac{c}{6} \log \left[ (w_1 - w_2) (\bar{w}_1 - \bar{w}_2) \right] & \eta > \frac{1}{2} \\ S_{bulk,\hat{g},1pt} \left( (w_1,\bar{w}_1) \right) + S_{bulk,\hat{g},1pt} \left( (w_2,\bar{w}_2) \right) & \eta < \frac{1}{2} \end{cases}.
\end{align}
This simple form corresponds to an exchange of dominance of HRT surfaces.
The details of the physical situation, then, enter in two ways: in the placement of the end-points in the $w$ plane, and in the conformal factors.

We are interested in calculating the CFT entropy when one end-point is in the central strip and any other possible end-points are, say, in the $x>0$ bulk region.
The map from the $y = - \bar{y}$ line to the $w$ plane is clearly given by $w(y)$, and the conformal factor is $\epsilon^2 w'(y) \bar{w}'(\bar{y})$.
For any other points, they are given by $w(x)$ and $(x+\bar{x})^2 w'(x) \bar{w}'(x)$ respectively.
$w(x)$ is fixed entirely by the fact that the $x>0$ bulk is in the vacuum in the Poincare conformal frame.
The dependence on the actual nature of the bath, then, is encoded in the function $x(y)$.
But $x(y)$ is the map between the conformal frame in which the gluing happens (the $dy d \bar{y}$ frame) and the map in which the $x>0$ bulk is in the vacuum before the quench (the $dx d\bar{x}$ frame).
Since the gluing happens at $u>0$, it only cares about the temperature of the $x<0$ region \emph{after} the quench.
Thus, it doesn't matter whether the $x<0$ region \emph{started out} at temperature $T_1$ or got there by backreaction of the joining quench; the actual inputs into the CFT entropy calculation only care that it is in fact at that temperature after the quench. 

One important difference is that \cite{Chen:2020jvn} considers not a pure state black hole but a thermofield double in the $x>0$ region, meaning that they do not have the option of the empty surface as an entangling surface.
Using the transformation \eqref{eqn:disc-conf-1} to do the CFT calculation, we find that the generalised entropy of the empty surface is
\begin{equation}
  S_{gen} (\varnothing) = S_{bulk} (AdS_{L}) = \frac{c}{6} \log \frac{\sinh^{2} (\pi T_{1} u)}{ \pi^{2} T_{1}^{2} \epsilon w_{0} } \xrightarrow{u \to \infty} \frac{\pi}{3} c T_{1} u.
  \label{eqn:empty-gen-ent}
\end{equation}

\begin{figure}[h]
  \centering
  \includegraphics[width=.97\textwidth]{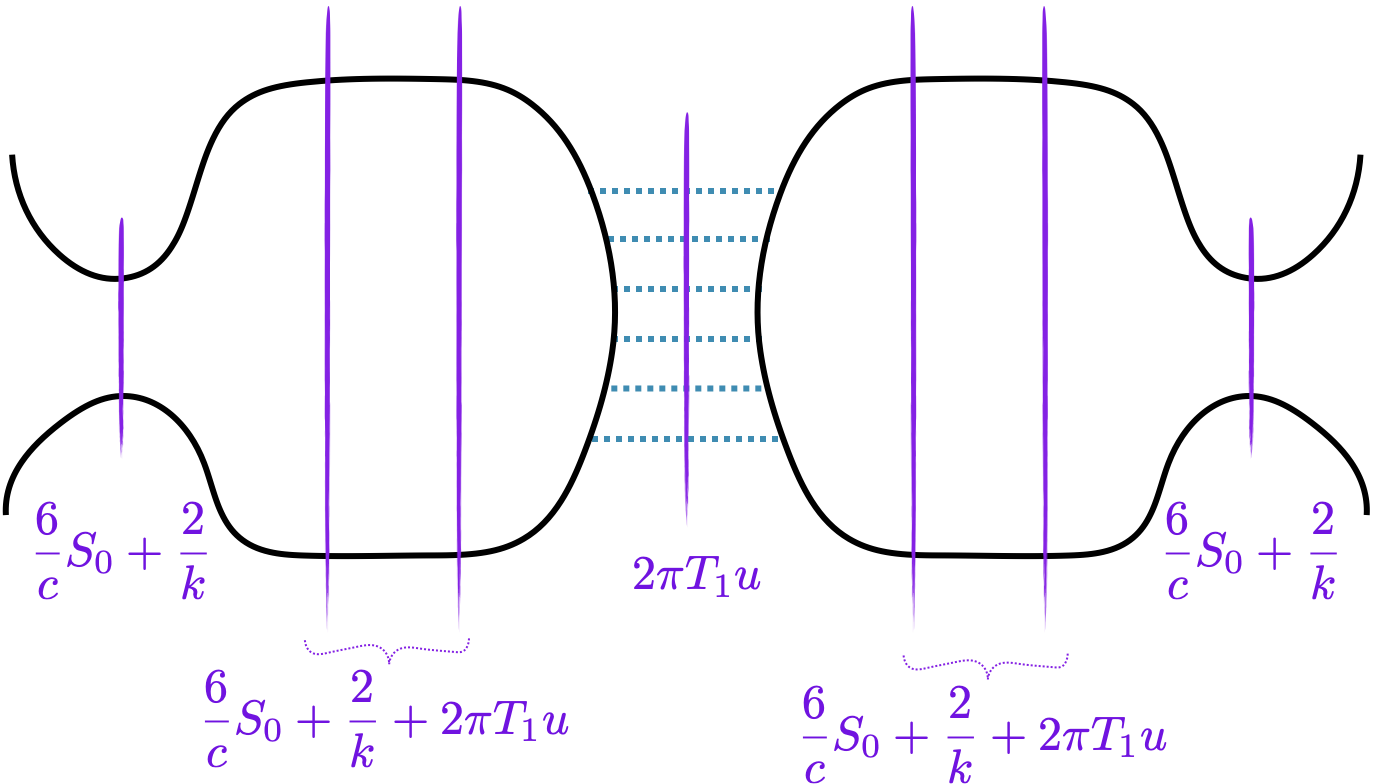}
  \caption{All the QESs in the disconnected history. Dashed blue lines denote entanglement DOFs, and purple lines mark the location of QESs, and they are labelled by the value of $\frac{6}{c} S_{gen}$. $u \gg u_{Page}$. \newline The QESs marked with the same generalised entropies aren't actually degenerate, but our analysis is insufficient to distinguish them. \newline We have defined the quantity $k = \frac{c}{6\pi \phi_{r} T_{0}}$.}
  \label{fig:disc-lunch}
\end{figure}

Another important difference between our set-up and \cite{Chen:2020jvn} is that our set-up is symmetric, and so non-empty QESs come in degenerate pairs.
\cite{Chen:2020jvn,Chen:2020} find three (pairs of) candidate QESs in this set-up.
Because all the non-empty QESs have entropy at least $S_{0}$ and the final late-time QES dominates in \cite{Chen:2020jvn} at time $u \sim \phi_{r}/(c\beta_{1})$, only the empty surface and the late time steady-state extremal surface need be considered for the calculation of the entanglement.
However, the other two are needed for the calculation of the spacetime's lunch structure in figure \ref{fig:disc-lunch}.

Since the original bifurcate horizons in each bulk are classical extremal surfaces, there is always a pair of QESs close to them.
This QES is dealt with explicitly in \cite{Chen:2020jvn} and so we simply copy the generalised entropy, which at late enough times is
\begin{equation}
    S_{gen} (\text{disconnected,bifurcate}) = S_0 + 2\pi T_0 \phi_r + \frac{\pi}{3} c T_1 u - \mO \left( \sqrt{\phi_r} \right).
    \label{eqn:disc-bif-gen-ent}
\end{equation}

There is also another QES only slightly to the past of the shock and almost null-separated from the boundary endpoint.
The generalised entropy of a general point in the right $AdS$ ($x + \bar{x} > 0$) to the past of the shock ($\bar{x} > 0$) and space-like separated from the UV end-point $x = - \bar{x} = t_P(u)$ ($t < x$) is
\begin{equation}
    S_{gen,nE} (x = t_P(u) + \delta x, \bar{x}) = 2 \phi_r \frac{1 + (\pi T_0)^2 x \bar{x}}{x + \bar{x}} + \frac{c}{6} \log \left[ \frac{2}{\epsilon w_0} \frac{t_P(u)}{t_P'(u)} \frac{\bar{x} \delta x}{x+\bar{x}} \right], \qquad 0 < \delta x, \bar{x} \ll 1.
    \label{eqn:disc-connected-rt-gen-ent}
\end{equation}
Here, we've used that $\eta \approx 1$ in this regime.
Assuming $\bar{x},\delta x = \mO(1/\phi_r)$, we find for the derivatives of the generalised entropy
\begin{align}
    \frac{6}{c} \partial_x S_{gen,nE} &= - \frac{12\phi_r}{c} \frac{1}{(x+\bar{x})^2} + \frac{1}{\delta x} + \mO \left( \phi_r^0 \right) \nonumber\\
    \frac{6}{c} \partial_{\bar{x}} S_{gen,nE} &= - \frac{12\phi_r}{c} \frac{1 - (\pi T_0)^2 x^2}{(x+\bar{x})^2} + \frac{1}{\bar{x}} + \mO \left( \phi_r^0 \right).
    \label{eqn:disc-qes3-eqns}
\end{align}
Setting both of these to $0$, we find that the QES is at
\begin{align}
    x_* &= t_P(u) + \frac{c}{12 \phi_r} t_P(u)^2 + \mO \left( \phi_r^{-2} \right), \nonumber\\
    \bar{x}_* &= \frac{c}{12\phi_r} \frac{t_P(u)^2}{1 - (\pi T_0 t_P(u))^2} + \mO \left( \phi_r^{-2} \right) \qquad \xrightarrow{u \to \infty} \frac{c}{24\phi_r} \frac{1}{1 - \frac{T_0}{T_1}} \sim w_0^2.
    \label{eqn:disc-qes3-pos}
\end{align}
We have used that $T_1 = T_0 + \mO(c/(\phi_r w_0^2))$; one can check that, because of the $w_0$ factor, there is no order of limits issue and this is the position of the QES at arbitrarily late time.
Finally, the generalised entropy at late times is
\begin{equation}
    S_{gen} (\text{disconnected,at-shock}) = S_0 + 2 \pi T_1 \phi_r + \frac{c}{6} \log \frac{(c/12\phi_r)^2}{1 - \left( T_0/T_1 \right)^2} + \frac{\pi}{3} c T_1 u - \mO \left( \sqrt{\phi_r} \right).
    \label{eqn:disc-qes3-ent}
\end{equation}
This QES is more entropic than the one near the bifurcate horizon by an $\mO(1)$ amount --- but this difference is dominated by the error due to this QES being one of a degenerate pair.

Finally, we can use the results in \cite{Chen:2020jvn} for the position and generalised entropy of the late-time surface.
In the boundary-adapted $y,\bar{y}$ coordinates in which the steady-state nature is manifested as an invariance under $u$ translations, the extremal surface that calculates the entropy at time $u$ is
\begin{align}
  u_{QES}(u) &= u \nonumber\\
  \frac{y_{QES} + \bar{y}_{QES}}{2} &= \pm \frac{\beta_{1}}{2\pi} \log \frac{24 \pi \phi_{r}}{c \beta_{1}} + \mO \left( \frac{c \beta_{1}}{\phi_{r}} \right).
  \label{eqn:disc-late-rt-pos}
\end{align}
The $\pm$ reflects the fact that this is actually a degenerate pair of HRT surfaces.
Its generalised entropy, and therefore the late-time EE between the two QM systems, is
\begin{equation}
  S_{gen} (\text{disconnected,late-time}) = S_{0} + \frac{2\pi}{\beta_{1}} \phi_{r} - \mO \left( \sqrt{\phi_{r}} \right).
  \label{eqn:disc-late-gen-ent}
\end{equation}
The $\mO(\sqrt{\phi_{r}})$ terms above denote corrections to the holographic EE near phase transitions \cite{Murthy:2019,DW,MWW}.
Comparing with \eqref{eqn:empty-gen-ent}, we see that there is an exchange of dominance between the empty surface and this late time QES in the disconnected saddle at
\begin{equation}
  u_{Page} = \frac{3}{\pi} \frac{S_{0}}{c T_{1}} + \frac{6 \phi_{r}}{c} - \mO \left( \sqrt{\phi_{r}} \right).
  \label{eqn:disc-page}
\end{equation}

\section{Coupling Two Black Holes Connected by a Wormhole} \label{sec:conn-hist}
In this section, we consider a two-sided eternal black hole. We couple the two sides at some initial time and let them radiate into each other.
We will find that, apart from a pre-scrambling-time transient, the UV EE behaves much like the previous case.
Apart from being an interesting problem in its own right, this calculation is also relevant as the connected history of section \ref{sec:evap}.
The Lorentzian geometry is shown in figure \ref{fig:conn-hist}.

At time $-i \delta$, we take a two-sided black hole of inverse temperature $\tilde{\beta}_{0}$ coupled to a holographic 2d CFT.
We then do a joining quench, similar to the one in section \ref{sec:evap}, of the two ends to couple the two black holes.

Working out the bulk geometry is a harder exercise in this case, and involves a 2d CFT calculation that has not been dealt with in the previous literature.
We deal with this part of the problem in section \ref{ssec:conn-bulk}, and use the lessons learnt here to calculate the Page curve in section \ref{ssec:conn-ent}

\subsection{The Bulk Geometry} \label{ssec:conn-bulk}
The state of the CFT before the quench is in a descendant of the Kruskal vacuum, or the thermofield double (TFD).
It is not quite in the TFD state because of the presence of the ETW brane; however, in the limit $\mu \to \infty$, we can approximate it as the TFD state, since the ETW brane shrinks to $0$ size in this limit, see \eqref{eqn:schw-geodesics} and \eqref{eqn:r-min-mu-reln}.
Calling the Kruskal-Szekeres coordinates $\mfw$, the boundary of $AdS$ is at $\mfw \bar{\mfw} = 1$, which is a hyperbola in real time.
For this reason, despite being natural, these are inconvenient coordinates to study the joining quench in.

\begin{figure}[h]
  \centering
  \includegraphics[width=0.98\textwidth]{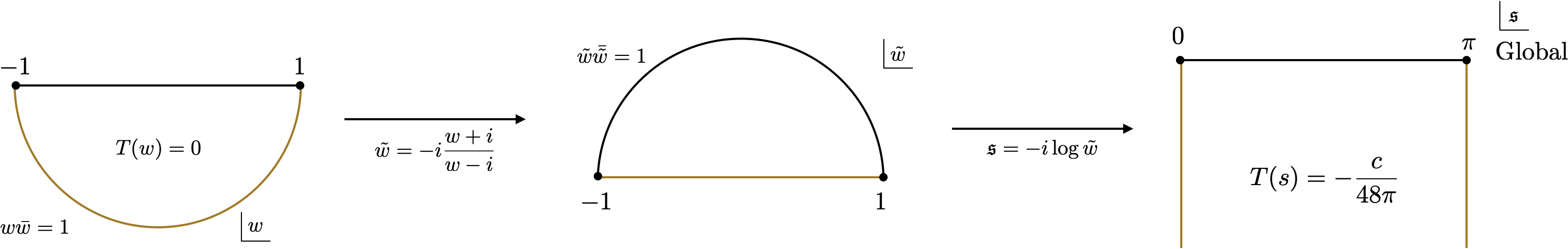}
  \caption{The conformal transformations relating the Kruskal and global vacuums. In this figure the Kruskal-Szekeres coordinate is denoted by $w$ rather than $\mfw$.}
  \label{fig:conn-ket}
\end{figure}

\begin{figure}[h]
  \centering
  \includegraphics[width=0.98\textwidth]{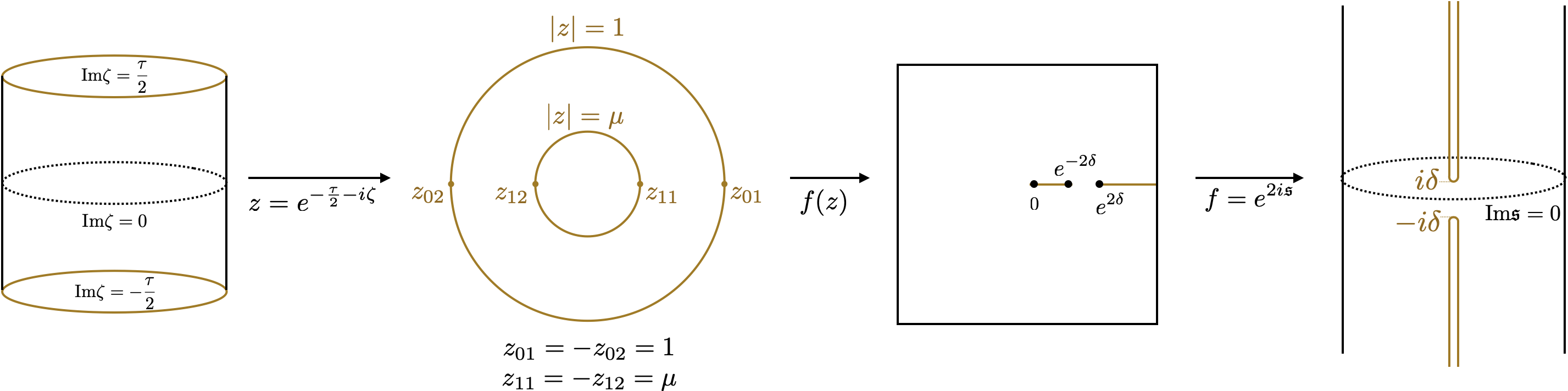}
  \caption{The conformal transformation from the $U(1)$ symmetric $\zeta$ cylinder to the $\mfs$ cylinder in which the quench happens. There are two slits in the $\mfs$ cylinder because we need to use this manifold for expectation values, and so need both a `ket' as well as a `bra' copy of the state. This figure is in Euclidean space whereas figure \ref{fig:disc-conf} in the previous subsection is in Lorentzian space.}
  \label{fig:conn-ket-bra}
\end{figure}

The useful set of conventional $AdS$ coordinates are global coordinates, $\mfs \equiv \sigma + t_{gl}$, related to the Kruskal-Szekeres coordinates by
\begin{equation}
    \mfs = - i \log \frac{1 - i \mfw}{\mfw - i}, \qquad \{\mfw \bar{\mfw} = 1\} \mapsto \{\sin (\mfs + \bar{\mfs}) = 0\},
    \label{eqn:kruskal-to-global}
\end{equation}
see figure \ref{fig:conn-ket}.
While this is merely a coordinate transformation in $AdS_{2}$, this is a non-trivial conformal transformation from the metric $d\mfw d \bar{\mfw}$ to the metric $d\mfs d \bar{\mfs}$.
The stress tensors in these two flat metrics are
\begin{equation}
    T(\mfw) = 0, \qquad T(\mfs) = \frac{\{\mfs,\mfw\}}{\mfs'(\mfw)^2} = - \frac{c}{24\pi} \frac{1}{2}.
    \label{eqn:conn-T}
\end{equation}
The first equation is justified in the beginning of section \ref{ssec:disc-hist}.
As in the previous section, the map from $\mfw$ to $\mfs$ is merely a coordinate transformation in $AdS_2$ but a conformal transformation in flat space; including the contribution from the $AdS_2$ Weyl factor $\sin^2 \frac{\mfs + \bar{\mfs}}{2}$ gives $\langle T(\mfs) \rangle_{AdS} = 0$, see eg \cite{MQa}.

The metric in the $AdS$ and the gluing strip regions respectively is
\begin{align}
  ds_{AdS}^{2} &= \frac{d\mfs d \bar{\mfs} }{\sin^{2} \frac{\mfs + \bar{\mfs}}{2}} = \frac{\mfs'(y) \bar{\mfs}' (\bar{y}) dy d \bar{y}}{\sin^{2} \frac{\mfs(y) + \bar{\mfs} (\bar{y})}{2}} \nonumber\\
  ds_{C}^{2} &= \frac{d y d \bar{y}}{\epsilon^{2}}.
  \label{eqn:ads-coords-2}
\end{align}
The gluing, as before, is along the $u$ coordinate.
The symmetry between the two QM systems is here given by $y \leftrightarrow - \bar{y}$ and $\mfs \leftrightarrow \pi - \bar{\mfs}$.
Similarly to the previous section, the symmetry dictates that the relation between $\mfs$ and $y$ is continous across the gluing and the effect of the joining quench is just a symmetric pair of shocks.
We will now consider the CFT in the flat metric $d\mfs d \bar{\mfs}$.

To calculate the stress energy and EEs after the quench, we would like to, following \cite{CCQuench}, calculate correlation functions on the Euclidean $\mfs$ `cylinder' in figure \ref{fig:conn-ket-bra} and analytically continue the position of the insertions to real time.
This manifold is obtained by taking one `ket' and one `bra' copy of the state, so that we can calculate correlation functions on it.
We can easily see that this manifold has the topology of a cylinder, i.e. its Euler characteristic is $\chi = 0$.\footnote{This quench has not, to our knowledge, been discussed in the CFT literature, apart from a calculation of the overlap with the vacuumn \cite{Dubail:2011} that is calculated by a genus $0$ manifold.}
This means that this case is more complicated than the previous case, since the state after the quench is a descendant not of the $S^{1}$ vacuum but of a Cardy state $\ket{B}$ corresponding to the boundary conditions at the asymptotic $AdS$ boundary.

Thus, the simplest state that our state of interest on the $\mfs$ cylinder is conformally related to is not the vacuum but a $U(1)$-symmetric excited state of the form $e^{-\frac{\tau}{2} H} \ket{B}$.
Adding the bra copy of this state, we get a cylinder of circumference $2\pi$ and length $\tau$.
The relevant sequence of conformal transformations that takes this cylinder to the $\mfs$ cylinder is illustrated in figure \ref{fig:conn-ket-bra}.
The  quantity $\tau$ is a conformal invariant and will play an important role below; it is related to the so-called `modulus' of the annulus by
\begin{equation}
    \mu = e^{- \tau}.
    \label{eqn:mu-defn}
\end{equation}
As an aid to understanding, we compare the roles of coordinates used in the two histories in table \ref{tab:coords}

\begin{table}[h]
    \centering
    \begin{tabular}[h!]{|p{60mm}||c|c|}
      \hline
       Role of coordinate & Disconnected History & Connected History \\
      \hline \hline
       Coordinate in which the state \newline is spatial-translation-invariant & $w$ & $\zeta$ \\
      \hline
       Conventional $AdS$ coordinate & $x$ (Poincare) & $\mfs$ (global) \\
      \hline
       UV coordinate & $y$ & $y$ \\
      \hline
    \end{tabular}
    \caption{A comparison of the roles various coordinates play in section \ref{ssec:disc-hist} and this section.}
    \label{tab:coords}
\end{table}

The crucial function $f(z)$ in figure \ref{fig:conn-ket-bra} is a so-called doubly connected Schwarz-Christoffel map \cite{dep-sc,driscoll2002schwarz}, and it is given by 
\begin{align}
  f(z) &= e^{2\delta} + C \int_{1}^{z} \frac{\Theta \left(\mu, \frac{\mu \tilde{z}}{z_{11}} \right) \Theta \left(\mu, \frac{\mu \tilde{z}}{z_{12}} \right) \Theta \left(\mu, \frac{\tilde{z}}{\mu z_{01}} \right)}{\left[ \Theta \left(\mu, \frac{\tilde{z}}{\mu z_{02} } \right) \right]^{3}} d\tilde{z} \nonumber\\
  & \quad \Theta (\mu,z) = \prod_{j=1}^{\infty} \left( 1 - \mu^{2j-1} z \right) \left( 1 - \frac{\mu^{2j-1}}{z} \right) \nonumber\\
  & \quad f(z_{01}) = e^{2\delta}, \quad f(z_{02}) = \infty , \quad f(z_{11}) = e^{-2\delta} , \quad f(z_{12}) = 0.
  \label{eqn:dsc-map}
\end{align}
Here, $C$ is an integration constant.
By symmetry, the `pre-vertices' $z_{ij}$ are at
\begin{equation}
  z_{01} = - z_{02} = 1, \quad z_{11} = - z_{12} = \mu.
  \label{eqn:dsc-prevertices}
\end{equation}

The modulus $\mu$ is hard to calculate analytically in general, but it turns out that in the limit $\delta \to 0$, $\mu=e^{-\tau}$ increases towards one (i.e. the $\zeta$ cylinder gets more `squashed').
We confirm this by analysing \eqref{eqn:dsc-map} in the $\tau \to 0$ limit in appendix \ref{app:dcSC}, finding
\begin{equation}
  \tau \approx \frac{\pi^2}{\log \frac{4}{\delta}}
  , \quad C \approx \frac{4\pi}{\tau}e^{-\frac{\pi^2}{2\tau}+\frac{\tau}{2}}.
  \label{eqn:hu-mu0}
\end{equation}
We can check this using the numerical package \cite{hu1998algorithm};\footnote{We thank Chenglie Hu for correspondence.} for the smallest value of $\delta$ with which the algorithm converges we find
\begin{equation}
  \delta = 5\times 10^{-6}, \quad \mu \approx .4838, \quad C \approx .0278,
  \label{eqn:hu-mu}
\end{equation}
which matches with \eqref{eqn:hu-mu0}. All code used in this section is available online \cite{conn-code}.

\begin{figure}[h]
  \centering
  \includegraphics[width=100mm]{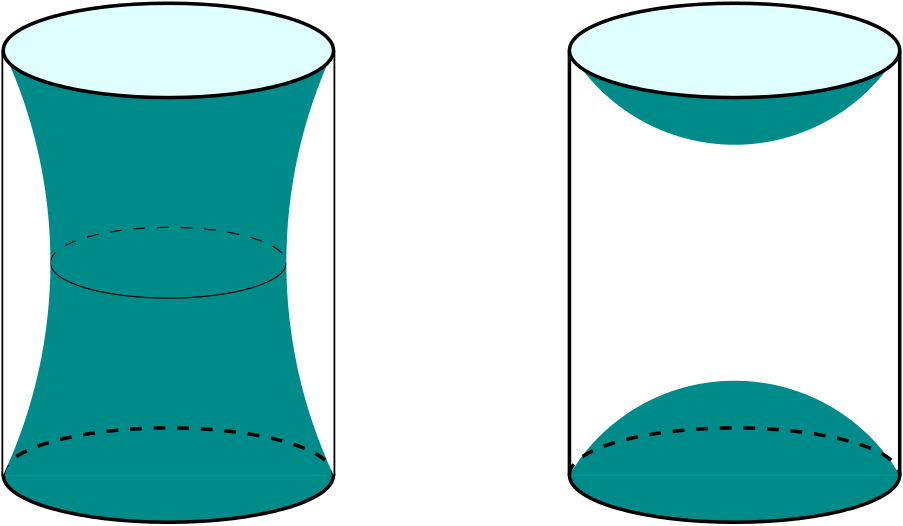}
  \caption{Phases of the 3d ETW brane in the $\zeta$ cylinder.}
  \label{fig:3d-brane-phases}
\end{figure}

The importance of the $\zeta$ cylinder is that it provides to us a $U(1)$-symmetric state that is related to the post-quench state by a conformal transformation.
However, even this simpler state is complicated in a general CFT.
As mentioned previously, the 2d CFT is holographic, dual at leading order to pure 3d general relativity with negative cosmological constant.
In that case, the boundary condition translates to a 3d ETW brane with a specified tension $\mathbb{T} \in (-1,1)$  \cite{Takayanagi:2011,Fujita:2011}, and the bulk geometry dual to the $\zeta$ cylinder is one of the two geometries in figure \ref{fig:3d-brane-phases}, as discussed in detail in \cite{Cooper:2018cmb}.
The one in which the two branes don't meet has the geometry of vacuum $AdS_{3}$ in the bulk whereas the one in which they do has the geometry of a black hole.
This will be enough to calculate the entanglement of the 2d CFT and therefore the positions of the quantum extremal surfaces in this history.

The first question is which of the two phases in figure \ref{fig:3d-brane-phases} dominates.
Because the contribution of the conformal transformation to the CFT partition function (i.e. the 3d on-shell action) is independent of state, we only need to check which phase dominates on the $\zeta$ cylinder.
Using formulas in \cite{Cooper:2018cmb}, we find that the black hole phase dominates when the brane tension $\mathbb{T}$ is small enough so that
\begin{equation}
  \tanh^{-1} \mathbb{T} < \frac{\pi^{2}}{4 \tau} - \frac{\tau}{4}.
  \label{eqn:bh-domination-condition}
\end{equation}
In the limit $\tau \to 0$, we find that the black hole phase always dominates.
This is the expected answer physically, since we know that the post-quench 3d geometry has to have an ETW brane falling away from the 2d boundary as in \cite{Shimaji:2018czt}; an easy way to see this is that correlation functions spacelike to the quench have to take the strip vacuum values and so the corresponding 3d geodesics have to end on an ETW brane.
Since the analytic continuation to real time happens at a slice of time-reflection symmetry, and the brane does not pass through this slice in the vacuum phase on the right of figure \ref{fig:3d-brane-phases}, we conclude that the correct phase is the black hole phase on the left.

This phase has been studied extensively in \cite{Cooper:2018cmb,Almheiri:2018} and it will be sufficient to use their formulas without modification.
The bulk geometry (dual to the $\zeta$ cylinder) in this phase is given by a 3d BTZ black hole,
\begin{equation}
  ds_{3d}^{2} = -(r^{2} - r_{h}^{2}) dt^{2} + \frac{dr^{2}}{r^{2} - r_{h}^{2}} + r^{2} d\phi^{2}, \quad \zeta = \phi + t, \bar{\zeta} = \phi - t.
  \label{eqn:3d-g}
\end{equation}
Here, the black hole radius is given by
\begin{equation}
  r_{h} = \frac{\pi}{\tau}.
  \label{eqn:3d-rh}
\end{equation}
The trajectory of the ETW brane in real time is
\begin{equation}
  \cosh (r_{h} t) \sqrt{\frac{r^{2}}{r_{h}^{2}} - 1} = \frac{\mathbb{T}}{\sqrt{1 - \mathbb{T}^{2}}}.
  \label{eqn:3d-brane-pos}
\end{equation}
For $\mathbb{T} > 0$ the ETW brane is in a second exterior  connected by a wormhole, similarly to the $\mu > 0$ case in section \ref{ssec:dPSSY-bulk}.

Finally, we will also need the stress tensor on the $\zeta$ cylinder, $T(\zeta)$.
Since we have taken the limit $\tau \to 0$, it is a very thin cylinder.
The conformal transformation
\begin{equation}
  \tilde{\zeta} = \frac{\pi}{\tau} i \zeta
  \label{eqn:zeta-vac-trans}
\end{equation}
is a modular transformation that gives a cylinder of width $\pi$ and circumference $\frac{2\pi^{2}}{\tau}$.
In the limit $\tau \to 0$, we find the state at $\Im \tilde{\zeta} = const$ is the vacuum on a strip of width $\pi$.
This gives
\begin{equation}
  T(\tilde{\zeta}) = - \frac{c}{24\pi} \frac{1}{2} \qquad \Rightarrow \qquad T(\zeta) = \frac{c}{24\pi} \frac{\pi^{2}}{2\tau^{2}}.
  \label{eqn:T-zeta}
\end{equation}

\begin{figure}[h]
  \centering
  \includegraphics[width=75mm]{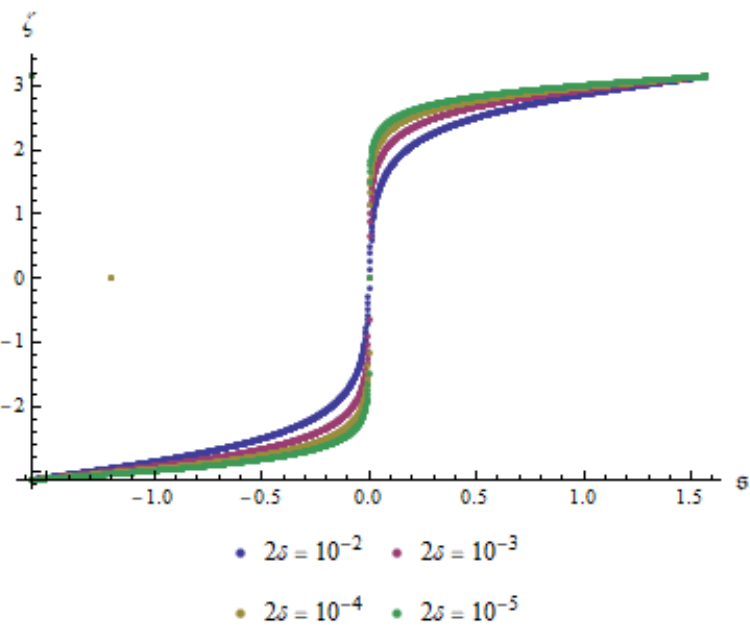}
  \includegraphics[width=75mm]{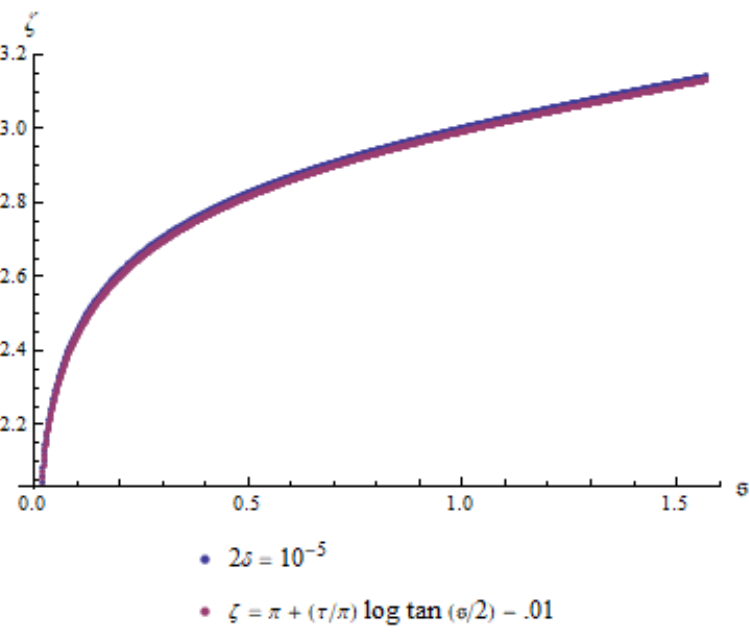}
  \caption{How the $\mfs \in \mathbb{R}$ circle maps to the $\zeta \in \mathbb{R}$ circle in the map \eqref{eqn:dsc-map}.\newline Left: We see that most of the $s$ circle gets compressed to a small angular width in $\zeta$, and that the concentration increases as $\delta \to 0$. This plot uses a different range for $\mfs,\zeta$ compared to the rest of the paper, to better show the concentration. \newline Right: Comparison of the numerical map and the analytic approximation \eqref{eqn:ffq-map} using parameter values from \eqref{eqn:hu-mu}. The approximation has been offset so that both graphs are visible.}
  \label{fig:conn-conf-concentration}
\end{figure}

Now we try to approximate the conformal transformation $\zeta(\mfs)$.
First, we notice that because every manifold in figure \ref{fig:conn-ket-bra} has the same time-reflection symmetry, we conclude that $\Im \mfs = 0$ and $\Im \zeta = 0$ map to each other.
Then, we notice that we don't need the approximation everywhere but only at the moment of time-reflection symmetry, since the conformal transformation for all real time can then be found from the fact that $\partial_{\bar{\mfs}} \zeta = \partial_\mfs \bar{\zeta} = 0$. 
We divide the $\Im \mfs = 0$ circle into two regions: 
\begin{align}
    \text{near-quench (nq) region} &= \{\mfs \in (0,\delta) \} \bigcup \{ \mfs \in  (\pi-\delta,\pi)\} \nonumber\\
    \text{far-from-quench (ffq) region} &= \{ \mfs \in (\delta, \pi-\delta) \}.
    \label{eqn:conn-regions}
\end{align}
Since correlation functions in the far-from-quench region take the same values as the strip \emph{vacuum} and the state at $\Im \zeta = 0$ is an excited state on the circle, the map must compress the far-from-quench region into a small interval $\zeta \in (\pi - \delta\zeta, \pi + \delta\zeta)$.
This is so because in the UV all correlation functions flow to their vacuum value.
This is similar to the limit $w_{0} \to 0$ in \eqref{eqn:disc-conf-1} that ensures factorisation between the two half-lines.
A numerical check with the parameters in \eqref{eqn:hu-mu} confirms this intuition, as seen in figure \ref{fig:conn-conf-concentration}.

The stress tensor on the $\mfs$ circle is, by the standard CFT transformation rules,
\begin{equation}
  T(\mfs) = \frac{c}{24 \pi} \left[ \frac{\pi^{2}}{2 \tau^{2}} \zeta'(\mfs)^{2} -  \{ \zeta, \mfs \} \right].
  \label{eqn:quench-T}
\end{equation} 
We will use this equation in two different ways in the two regions:  in the near-quench region we will use it to estimate $T(\mfs)$, whereas in the far-from-quench region we will use consistency with the global vacuum value of $T(\mfs)$ to approximate $\zeta(\mfs)$.

In the far-from-quench region, using the fact that the stress energy is the same as the global vacuum value
\begin{equation}
  T_{ffq} (\mfs) = - \frac{c}{24\pi} \frac{1}{2},
  \label{eqn:ffq-T}
\end{equation}
we find the equation
\begin{equation}
  \{\zeta,\mfs\} - \frac{\pi^{2}}{2 \tau^{2}} (\zeta')^{2} - \frac{1}{2} = 0.
  \label{eqn:ffq-diff-eqn}
\end{equation}
This differential equation can be solved by the observation that it is equivalent to the quadratic equation
\begin{equation}
  \left\{ \frac{4 \pi^{2}}{\tau^{2}} \tan^{2} \left( \mfs - \mfs_{0} \right) - \frac{4\pi^{2}}{\tau^{2} \mfc^{2}} \right\} (\zeta')^{2} + 4 \sec^{2} \left( \mfs - \mfs_{0} \right) \left( \frac{2\pi}{\tau \mfc} \zeta' - 1 \right) = 0
  \label{eqn:ffq-quad-eqn}
\end{equation}
$\mfc,\mfs_{0}$ are integration constants.
We use the $\mathbb{Z}_{2}$ symmetry to set $\mfs_{0} = 0$.
Integrating again and defining the quantity $\delta\zeta$ as the size of the image of the $\mfs$ cylinder on the $\zeta$ cylinder, we find
\begin{align}
  \zeta = \pi + \frac{\delta \zeta}{2} - \frac{2\tau}{\pi} \coth^{-1} \frac{\cosh \frac{\pi \delta\zeta}{2\tau} + \tan \frac{\mfs}{2}}{\sinh \frac{\pi \delta\zeta}{2\tau}}, \quad \delta \zeta = \frac{2\tau}{\pi} \sinh^{-1} \mfc.
  \label{eqn:ffq-map-0}
\end{align}
To fix the integration constant $\delta \zeta$, we impose a second consistency condition, namely that it reproduce the strip vacuum value for the one-point function of a primary operator at a point space-like to the quench.
We find that it is only reproduced in the limit $\frac{\tau}{\delta \zeta} \to 0$,\footnote{Take the bulk tension $\mathbb{T} = 0$. In the strip vacuum, the length of a geodesic from $\mfs = \bar{\mfs} = \sigma$ to the brane is $\log (2 \sin \sigma)$ up to a regularisation constant. Using equations \eqref{eqn:3d-g} and \eqref{eqn:3d-brane-pos} along with conformal transformation factors from \eqref{eqn:ffq-map-0}, we find this length to be
\begin{equation}
    \log \left[ 2 \frac{1 + \cosh \frac{\pi \delta \zeta}{2\tau} \sin \sigma}{\sinh \frac{\pi \delta \zeta}{2\tau}} \right] \xrightarrow{\frac{\tau}{\delta \zeta} \to 0} \log (2 \sin \sigma).
\end{equation}} and so we find 
\begin{equation}
    \zeta = \pi + \frac{\tau}{\pi} \log \tan \frac{\mfs}{2}
    \label{eqn:ffq-map}
\end{equation}
The expression \eqref{eqn:ffq-map} for $\zeta(\mfs)$ is one of the main results of this section, and will be used extensively in section \ref{ssec:conn-ent}. The same formula is also reproduced directly from the Schwarz-Christoffel map \eqref{eqn:dsc-map} in appendix \ref{app:dcSC}.

\begin{figure}[h]
  \centering
  \includegraphics[width=0.7\textwidth]{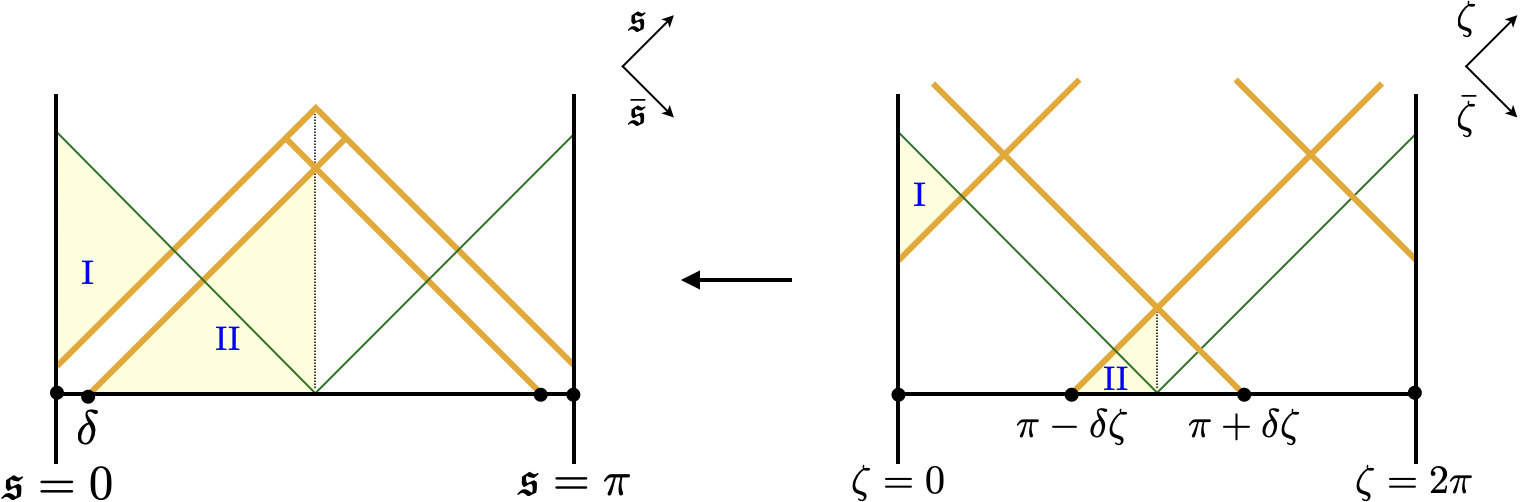}
  \caption{The map between the Lorentzian $\zeta$ and $\mfs$ cylinders. The regions marked $I,II$ are the far-from-quench regions that will be relevant to us; they map to each other and their coordinate extents are given  in equation \eqref{eqn:regions}. The regions between the orange lines are the near-quench regions; as can be seen the near-quench region is the majority of the $\zeta$ cylinder.}
  \label{fig:conn-regions}
\end{figure}

Because these are cylinders, there is some ambiguity in the choice of coordinate range; the expression \eqref{eqn:ffq-map} is consistent if we make the choices that regions $I,II$ in figure \ref{fig:conn-regions} have coordinate ranges
\begin{align}
    \text{region } I: &\; 
    \mfs=\mfs^+\in \left(0,\frac{\pi}{2}\right)
    \qquad 
    \bar{\mfs}=\pi-\mfs^-\in \left( \frac{\pi}{2},\pi \right)
    \nonumber\\
    \text{region }II: &\; 
    \mfs=\mfs^+\in \left(0,\pi\right)
    \qquad\ \  
    \bar{\mfs}=-\mfs^-\in \left(0, \frac{\pi}{2} \right)
    \label{eqn:regions}
\end{align}
where $\mfs^\pm = t_{gl}\pm \sigma$.
The extension to other regions is straightforward, but these two regions will be relevant below.

\begin{figure}[h]
  \centering
  \includegraphics[width=70mm]{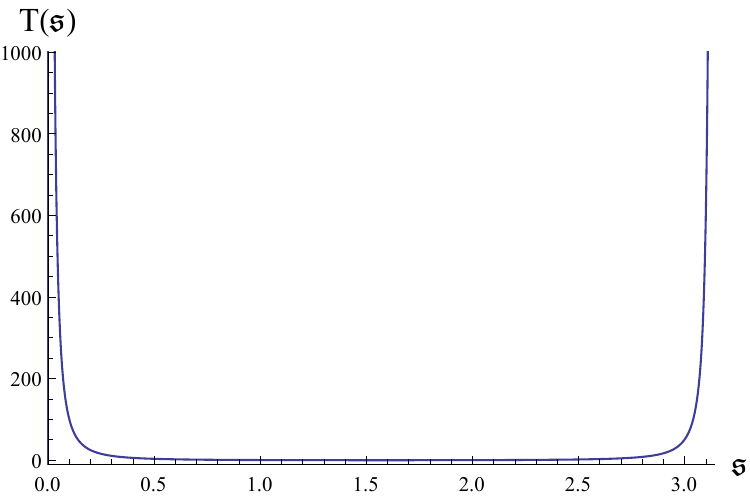}
  \caption{The numerical result for $T(\mfs)$ with $\delta = 5*10^{-6}$, see \cite{conn-code} for the code. We use the expression for the Schwarzian in \cite{dep-sc}. The result corroborates the expectation \eqref{eqn:nq-T} that the stress-energy in the near-quench region is positive and large. The result in the far-from-quench region is not trustworthy, because $\tau \approx .7$ in this case and the relevant limit is $\tau \to 0$.}
  \label{fig:conn-energy}
\end{figure}

In the near-quench region, $\zeta' \sim \delta^{-1}$ since it maps the $\mO(\delta)$ angular width in $\mfs$ to an $\mO(1)$ angular width in $\zeta$.
So, we expect the first term to dominate \eqref{eqn:quench-T} and find, since $\mathfrak{T} > 0$ in the black hole phase,
\begin{equation}
  T_{nq} (\mfs) \propto +\frac{c}{\delta^{2}}.
  \label{eqn:nq-T}
\end{equation}
This result is further corroborated by numerics \cite{conn-code}, see figure \ref{fig:conn-energy}, as well as an analysis of the Schwarz-Christoffel map, see equation \eqref{eqn:nq-T-answer} where the constant of proportionality is found to be $\frac{1}{16\pi}$.

This means that, similarly to the disconnected history, the joining quench throws out a shock of large positive energy and then the two sides stay in equilibrium.
The bulk geometry is thus $AdS$-Vaidya and the boundary particle trajectory is given by
\begin{align}
  t_{gl} (u) = 2 \tan^{-1} \tanh (\pi \tilde{T}_{0} u) \theta(-u) &+ 2 \tan^{-1} \left\{ \frac{\tanh (\pi \tilde{T}_{1} u)}{\sqrt{1 + 2 \alpha}}  \right\} \theta(u), \nonumber\\
  & \alpha \equiv \frac{1}{2} \frac{\tilde{T}_{1}^{2} - \tilde{T}_{0}^{2}}{\tilde{T}_{0}^{2}} = \mO \left( \frac{c \tilde{\beta}_{0}}{\phi_{r}} \right).
  \label{eqn:vaidya-bdp-gl}
\end{align}
See appendix \ref{app:JT} for a derivation.
Despite the fact that $\alpha \ll 1$, we will defer expanding in $\alpha$ to final expressions, since $e^{- 2\pi \tilde{T}_{1} u} \ll \alpha$ for large enough $u$.

We use \eqref{eqn:ffq-map} and \eqref{eqn:vaidya-bdp-gl} to calculate the conformal factor between the physical metric and the $\zeta$ cylinder.
\begin{align}
  ds^{2} = \frac{1}{\epsilon^{2}} dy d \bar{y} &= e^{2 \Omega(u)} d\zeta d \bar{\zeta} \nonumber\\
  e^{\Omega} = \frac{1}{\epsilon t_{gl}'(u) | \zeta' |}  &= \frac{\tanh (\pi \tilde{T}_1 u)}{\epsilon \pi \tilde{T}_1} \cosh^2 (\pi \tilde{T}_1 u) \nonumber\\[10pt]
  \Rightarrow \quad \Omega &\xrightarrow{u \to \infty} 2 \pi \tilde{T}_1 u.
  \label{eqn:conf-factor-conn}
\end{align}
Note that it is important to be cognizant of the order of limits to get the right behaviour here.
Again, we see that the dominant contribution comes from $t_{gl}'(u)$, justifying the choice in \eqref{eqn:g-conn}, and verifying the main calculation in section \ref{sssec:bulk-dual}.

We again can calculate a one-point function of a primary operator of dimension $\Delta$ at $y = \bar{y} = u$.
This point is in region $I$ of figure \ref{fig:conn-regions} and so we have $\mfs = t_{gl} (u), \bar{\mfs} = \pi - t_{gl} (u)$.
The one-point function is
\begin{align}
    \langle O(u) \rangle &= e^{-\Delta \Omega} \left\langle O \left( \zeta = 2\pi - \bar{\zeta} = \pi + \frac{\tau}{\pi} \log \frac{\tanh (\pi \tilde{T}_1 u)}{\sqrt{1+2\alpha}}  \right) \right\rangle_{d\zeta d \bar{\zeta}} \nonumber\\
    &\xrightarrow{u \to \infty} \left\langle O \left( \zeta = 2\pi - \bar{\zeta} = \pi - \tau \alpha  \right) \right\rangle_{d\zeta d \bar{\zeta}} e^{-2 \pi \tilde{T}_1 \Delta u}.
    \label{eqn:conn-opf}
\end{align}
The first factor here is neither too large nor too small, as can be shown by using the holographic description of the $\zeta$ cylinder.
Thus, we find that this one-point function decays exponentially with time.

\subsection{Entropies} \label{ssec:conn-ent}
To calculate the CFT entropies, we need to calculate the entanglement entropies on the $\zeta$ cylinder.
Again, we use the doubly holographic description for this.
The single interval bulk entropy in the 2d CFT can be calculated using the HRT formula \cite{Cooper:2018cmb}, 
\begin{align}
  S_{bulk} (\zeta_{1}, \zeta_{2}) &= \frac{c}{6} \log \left\{ \frac{4}{r_{h}^{2}} \min \left( \sinh^{2} \frac{r_{h} R}{2}\ ,\  \frac{1+\mathbb{T}}{1-\mathbb{T}} \prod_{i=1,2} \cosh (r_{h} t_{i}) \right) \right\} \nonumber\\
  &\qquad\qquad R^{2} \equiv (\zeta_{1} - \zeta_{2}) ( \bar{\zeta}_{1} - \bar{\zeta}_{2} ), \qquad t_{i} \equiv \frac{\zeta_{i} - \bar{\zeta}_{i}}{2}.
  \label{eqn:conn-zeta-ee}
\end{align}
We remind the reader of our notation in table \ref{tab:ents}.
The first possiblity is when the 3d HRT surface misses the ETW brane, whereas the second possibility is when the 3d HRT surface goes from each end-point to the brane.
We have thrown away a divergent constant, since we are only interested in generalised entropies --- in which case the constant only serves to renormalise $\phi_{r}$.
While \cite{Cooper:2018cmb} only does the calculation for the case where the two end-points are on the same time slice, the formula applies for points also at different times.
This is clear in the case where the HRT surface hits the brane since the answer is factorised; in the other case, this was argued to be the case in \cite{Castro:2014}.

An important point to note is that, in writing the expression for $R$ in \eqref{eqn:conn-zeta-ee}, we have assumed that both points are in region $I$ or $II$ of figure \ref{fig:conn-regions} and that we are using the labelling for $\zeta,\bar{\zeta}$ laid out implicitly in \eqref{eqn:regions} and \eqref{eqn:ffq-map}.
To see why this is important, consider placing one point at $\zeta = \bar{\zeta} = 0$ and one point  at $\zeta = \bar{\zeta} = \frac{3\pi}{2}$ (not in regions $I,II$).
A naive application of \eqref{eqn:conn-zeta-ee} gives $R = \frac{3\pi}{2}$ whereas a moment's thought tells us that the true distance is only $\frac{\pi}{2}$. 

To calculate entanglement entropy in $AdS_{2}$, we need to introduce factors for the conformal transformation from $\zeta$ to $\mfs$ and also a factor for the Weyl transformation from flat space to $AdS_{2}$.
We take one end-point to be in the flat space strip between the QM systems, meaning that instead of that from the Weyl transformation to $AdS_{2}$ we have to include a factor from the conformal transformation to $y$ coordinates.
Thus, the full CFT entanglement entropy is
\begin{align}
  S_{bulk} \left(u, (\mfs, \bar{\mfs}) \right) &= S_{bulk} \left(\zeta(t_{gl}(u)), \zeta(\mfs)\right) - \frac{c}{6} \log \left( \epsilon t_{gl}'(u) \sqrt{\zeta' (t_{gl}(u)) \bar{\zeta}' (\pi - t_{gl}(u))} \right) \nonumber\\
  &\qquad\qquad - \frac{c}{6} \log \left( \sqrt{\zeta'(\mfs) \bar{\zeta}'(\bar{\mfs})} \sin \frac{\mfs + \bar{\mfs}}{2} \right).
  \label{eqn:conn-bulk-ee}
\end{align}

Using \eqref{eqn:dilaton-soln},\eqref{eqn:tfd-charges} and \eqref{eqn:vaidya-charge}, we find that the dilaton in this geometry is
\begin{equation}
  \phi =\begin{cases}
    2\pi \tilde{T}_{0} \phi_{r} \frac{\cos t_{gl}}{\sin \sigma}  &\quad \text{to the past of the shocks} \\
    2\pi \tilde{T}_{0} \phi_{r} \left[ \frac{\cos t_{gl}}{\sin \sigma} + \alpha \frac{\sin t_{gl} - \cos \sigma}{\sin \sigma} \right] &\quad \text{to the future of the shocks}
  \end{cases}.
  \label{eqn:conn-phi}
\end{equation}
To calculate the entanglement between the two QM systems, we need to find the extrema of
\begin{equation}
  S_{gen,nE} (u,(\mfs,\bar{\mfs})) = S_{0} + \phi + S_{bulk} (u,(\mfs,\bar{\mfs})).
  \label{eqn:gen-ent}
\end{equation}

The first step in the calculation is to figure out when each of the two 3d HRT surfaces dominates in \eqref{eqn:conn-zeta-ee}.
We will place one point in region $I$ and the other in either region $I$ or $II$.
Since one point is in region $I$, the HRT surface that hits the brane has length $\sim \log \cosh r_{h} \pi \gg 1$.
Meanwhile the HRT surface that misses the brane has length $\sim \tau$ but the combination $r_{h} R \sim \tau^{0}$.
So this latter HRT surface always dominates in the cases of interest.
Thus, we find that
\begin{align}
  S_{bulk} (u, (\mfs, \bar{\mfs})) &= \frac{c}{6} \log \left\{ \frac{4}{\epsilon t_{gl}'(u) \sin \frac{\mfs + \bar{\mfs}}{2}} \frac{\pi^{2}/\tau^{2}}{\zeta'(t_{gl}(u)) \sqrt{\zeta'(\mfs) \bar{\zeta}'(\mfs)}} \sinh^{2} \frac{\pi R}{2\tau} \right\}.
  \label{eqn:conn-bulk-ee-2}
\end{align}
A surprising thing about this formula is that all $\tau$ factors cancel, and so there is no dependence on $\delta$ at all, meaning that the shocks don't seem to carry any entanglement.

Another implication of the dominance of this brane-missing HRT surface after the quench ($t_{gl} > \delta$) is that the original bifurcation surface $\mfs = \bar{\mfs} = \frac{\pi}{2}$ immediately stops being a QES.
This is because the distance $R$ is not extremal here for $u \neq 0$, whereas the dilaton and the other factors in \eqref{eqn:conn-bulk-ee-2} are extremal.
However, since the bulk entropy is a subleading term in the generalised entropy, we can look for the early-time QES close to the bifurcation surface.
This expansion is organised in powers of
\begin{equation}
  k \equiv \frac{c \tilde{\beta}_{0}}{12 \pi \phi_{r}}.
  \label{eqn:k}
\end{equation}
There are two things to keep in mind while doing this expansion,
\begin{equation}
  \alpha \sim \frac{k}{\delta} \quad \text{ and } \quad \text{when } \tilde{T}_{1} u > \log k,\ e^{- \pi \tilde{T}_{1} u} \ll k,\alpha.
  \label{eqn:k-exp-caveats}
\end{equation}
The transition between the $\alpha \ll e^{- 2\pi \tilde{T}_{1} u}$ and $\alpha \gg e^{- 2\pi \tilde{T}_{1} u}$ happens at around $u = \tilde{\beta}_{1} \log k^{-1}$, which is the scrambling time.

With these caveats in mind, we find by plugging \eqref{eqn:conn-bulk-ee-2} into \eqref{eqn:gen-ent} for the extremal surface equations
\begin{align}
  0 = \frac{6}{c} \partial_{\mfs} S_{gen,nE} &=  - \frac{1}{2k} \frac{\cos \bar{\mfs}}{\sin^{2} \sigma} - \frac{1}{2} \cot \sigma + \frac{1}{2} \cot \mfs + \frac{1}{2 \sin \mfs} \sqrt{\frac{\log \frac{ \cot \frac{\bar{\mfs}}{2}}{ \zeta_{u}}}{\log \frac{\tan \frac{\mfs}{2}}{\zeta_{u}}}} \coth \left[ \frac{1}{2} \sqrt{\log \frac{\tan \frac{\mfs}{2}}{\zeta_{u}}\log \frac{\cot \frac{\bar{\mfs}}{2}}{\zeta_{u}}} \right]  \nonumber\\
  0 = \frac{6}{c} \partial_{\bar{\mfs}} S_{gen,nE} &=  - \frac{1}{2k} \frac{\cos \mfs}{\sin^{2} \sigma} - \frac{1}{2} \cot \sigma + \frac{1}{2} \cot \bar{\mfs} - \frac{1}{2 \sin \bar{\mfs}} \sqrt{\frac{\log \frac{\tan \frac{\mfs}{2}}{\zeta_{u}}}{\log \frac{\cot \frac{\bar{\mfs}}{2}}{\zeta_{u}}}} \coth \left[ \frac{1}{2} \sqrt{\log \frac{\tan \frac{\mfs}{2}}{\zeta_{u}}\log \frac{\cot \frac{\bar{\mfs}}{2}}{\zeta_{u}}} \right].
  \label{eqn:conn-pre-page-qes-eqn}
\end{align}
Here, we have defined
\begin{align}
  \zeta_{u} \equiv \frac{\tanh \pi \tilde{T}_{1} u}{\sqrt{1+2\alpha}}, \quad \zeta (t_{gl}(u)) = \pi + \frac{\tau}{\pi} \log \zeta_{u},\ \bar{\zeta} (t_{gl}(u)) = \pi - \frac{\tau}{\pi} \log \zeta_{u}.
  \label{eqn:zeta-u-defn}
\end{align}

\begin{figure}[h]
  \centering
  \includegraphics[height=50mm]{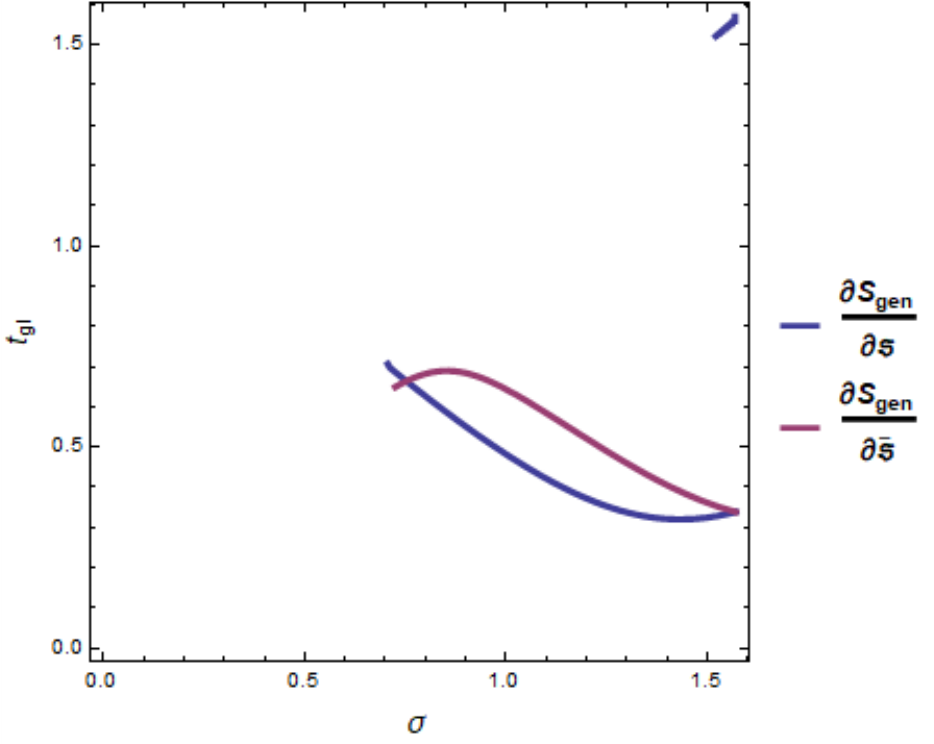}
  \caption{The position of all behind-shock QESs. The zeroes of both equations are plotted in the region behind the shock and spacelike to the boundary point $\mfs = \pi - \bar{\mfs} = 2 \tan^{-1} \zeta_u$.
  They intersect at the symmetric surface $\sigma = \pi/2$, and also at a point that is right behind the shock and nearly null-separated from the boundary point. The parameters have the values $k=.1,\zeta_{u} = .78$. }
  \label{fig:conn-scr-qes}
\end{figure}

There is always a QES behind the shock at the $\mathbb{Z}_{2}$-symmetric surface $\mfs + \bar{\mfs} = \pi$, as can be seen by the fact that the sum of the two QES equations with $\mfs + \bar{\mfs} = \pi$ vanishes identically.
An important thing to note is that because of the symmetry there are \emph{two} 3d HRT surfaces that go between the two points; this means that, because of the results of \cite{Murthy:2019,DW,MWW}, the HRT formula is expected to have an $\mO(\sqrt{c})$ error.
We use a numeric plot to look for any other QESs behind the shock, see figure \ref{fig:conn-scr-qes}, and find that there is also one right behind the shock.
In this case, there is another degenerate QES just behind the other shock as well; and the HRT formula is expected to have an $\mO(\sqrt{\phi_r})$ error.

\begin{figure}[h]
  \centering
  \includegraphics[width=0.7\textwidth]{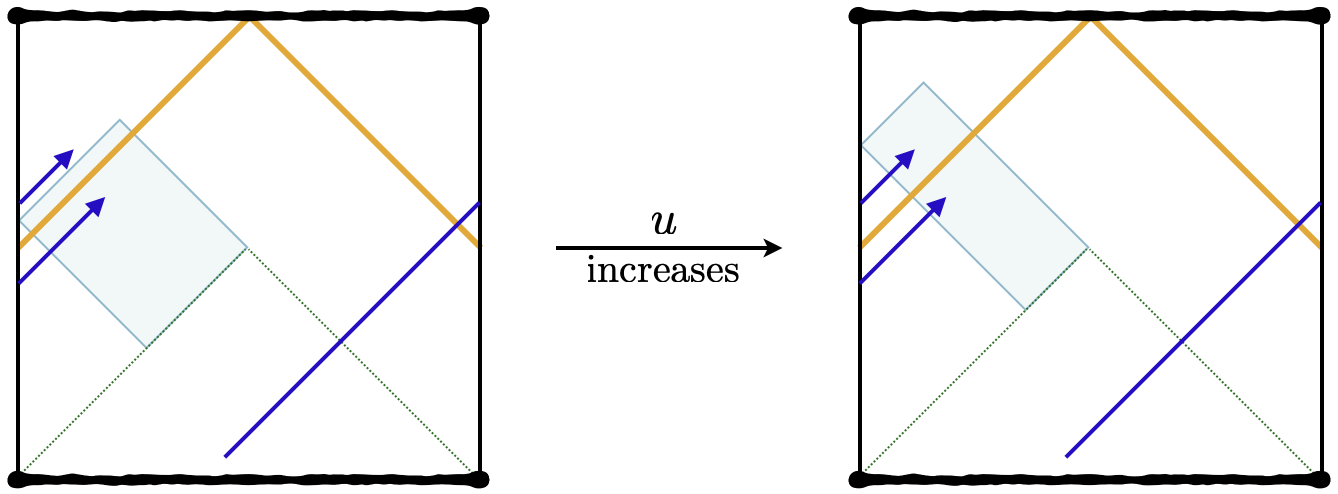}
  \caption{A simple quasiparticle explanation for the early-time decrease in the generalised entropy. The regions bounded by solid green lines are the quantum extremal wedges at the two times. The arrows are mirror quasi-particles that are entangled in the eternal black hole; as time goes on, more and more entangled pairs end up on the same side and the entanglement decreases. The turn-around at scrambling time is not so easily visualised.}
  \label{fig:conn-qp}
\end{figure}

We first deal with the QES on the symmetric surface $\sigma = \pi/2$ or $\mfs + \bar{\mfs} = \pi$.
Here, the equations simplify to
\begin{equation}
  \frac{\cos \mfs}{2k} + \cot \mfs + \frac{1}{\sin \mfs} \frac{\tan \frac{\mfs}{2} + \zeta_{u}}{\tan \frac{\mfs}{2} - \zeta_{u}} = 0.
  \label{eqn:conn-pre-page-symm-qes-eqn}
\end{equation}
It is easy to solve for $\tan \frac{\mfs}{2}$, but the exact expression is too long to reproduce here.
In a small $k$ expansion, we find
\begin{equation}
    \mfs_* - \frac{\pi}{2} = \begin{cases} \frac{1+\zeta_u}{1-\zeta_u} k + \mO(k^2) & \pi \tilde{T}_1 u \ll \log \alpha^{-1} \\ \sqrt{2k} + \left( 1 - \frac{1 - \zeta_u}{2k} \right) k + \mO(k^{3/2}) & \pi \tilde{T}_1 u \gg \log \alpha^{-1} \end{cases}.
\end{equation}
The generalised entropy at this surface is
\begin{align}
  S_{gen} (\text{connected, symmetric}) = S_0 &+ \frac{c}{6} \left. \begin{cases} \frac{1}{k} & \pi \tilde{T}_1 u \ll \log \alpha^{-1} \\ \frac{1}{k} - 2 & \pi \tilde{T}_1 u \gg \log \alpha^{-1} \end{cases} \right\} \nonumber\\ 
  & + \frac{c}{6} \log \left[ \frac{(1 + \alpha - \tanh \pi \tilde{T}_1 u)^{2}}{\epsilon \pi \tilde{T}_1} \cosh^2 \pi \tilde{T}_1 u \right] - \mO\left(\sqrt{c} \right) \nonumber\\
  &\xrightarrow{u \gg \phi_r/c} S_0 + \frac{c}{6k} - \frac{c}{3} + \frac{c}{6} \log \frac{\alpha^2}{\epsilon \pi \tilde{T}_1} + \frac{c}{3} \pi \tilde{T}_1 u - \mO \left( \sqrt{c} \right).
  \label{eqn:pre-shock-gen-ent}
\end{align}
It decreases till the scrambling time and then increases linearly after.
The early time decrease can be easily seen in a quasiparticle picture, as shown in figure \ref{fig:conn-qp}; the linear increase after scrambling time isn't as clear.

The QES right behind the shock can be found by taking the ansatz
\begin{equation}
    \mfs = 2 \tan^{-1} \zeta_u + \delta \mfs, \qquad \delta \mfs = \mO(k^2), \quad \bar{\mfs} = \mO(k).
    \label{eqn:qes3-scaling}
\end{equation}
With this scaling, the leading order QES equations become
\begin{align}
    \frac{6}{c} \partial_{\bar{\mfs}} S_{gen,nE} &\approx - \frac{1}{2k} \frac{1-\zeta_u^2}{\zeta_u} + \frac{1}{2 \bar{\mfs}} \nonumber\\
    \frac{6}{c} \partial_{\mfs} S_{gen,nE} &\approx \frac{1+\zeta_u^2}{2 \zeta_u} \left[ - \frac{1}{k} + \sqrt{ \frac{\log \frac{2}{\bar{\mfs}}}{2 \left( \zeta_u + \frac{1}{\zeta_u} \right) \delta \mfs} } \right]
    \label{eqn:qes3-eqns}
\end{align}
The solutions are
\begin{align}
    \bar{\mfs}_* &= \frac{2 \zeta_u^2}{1 - \zeta_u^2} k \nonumber\\
    \mfs_*       &= 2 \tan^{-1} \zeta_u - \zeta_u^2 \frac{\log \frac{1 - \zeta_u^2}{\zeta_u^2 k}}{2 \left( \zeta_u + \frac{1}{\zeta_u} \right)} k^2
    \label{eqn:qes3-pos}
\end{align}
and the generalised entropy is
\begin{align}
    S_{gen} (\text{connected, at-shock}) = S_0 &+ \frac{c}{6} \left( \frac{1}{\zeta_u k} + 2 \right) \nonumber\\
    &+ \frac{c}{6} \log \left\{ \frac{k^2 \tanh^{7/2} \pi \tilde{T}_1 u}{2\epsilon} \sqrt{\frac{k}{1-\zeta_u^2}} \left( \log \frac{1-\zeta_u}{k \zeta_u} \right)^2 \cosh^2 \pi \tilde{T}_1 u \right\} \nonumber\\
    &- \mO \left( \sqrt{\phi_r} \right) \nonumber\\
    &\xrightarrow{u \gg \phi_r/c} S_0 + \frac{c}{6k} + \frac{c}{3} \pi \tilde{T}_1 u - \mO \left( \sqrt{\phi_r} \right).
    \label{eqn:qes3-sgen}
\end{align}
where the last term appears because there is a degenerate QES behind the other shock.

At later times, $u \sim \phi_{r}/c$ there is a new QES that develops and dominates in region $I$ of the geometry.
It turns out that it is outside the horizon, as in the disconnected history.
Expanding the expression at large $u$, we find that the expression for the bulk entropy for a general point in the exterior of the black hole and in region $I$ is remarkably simple in terms of the $y,\bar{y}$ coordinates,
\begin{equation}
  S_{bulk} (u,y,\bar{y}) = \frac{c}{6} \log \left\{ \frac{2}{\epsilon \pi \tilde{T}_1} \frac{\sinh \left[ \pi \tilde{T}_{1} (y-u) \right] \sinh \left[ \pi \tilde{T}_{1} (\bar{y} + u) \right]}{ \sinh \left[ \pi \tilde{T}_{1} (y+\bar{y}) \right] } \right\}.
  \label{eqn:conn-late-time-bulk-ent}
\end{equation}
The dilaton takes the usual $AdS$-Vaidya value,\footnote{This follows from the fact that in Schwarzchild coordinates the exterior of the shock is identical to a black hole of temperature $\tilde{T}_{1}$.}
\begin{equation}
  \phi = 2\pi \phi_{r} \tilde{T}_{1} \coth \left[ 2 \pi \tilde{T}_{1} (y + \bar{y}) \right].
  \label{eqn:conn-ext-dilaton}
\end{equation}
Since the dilaton is independent of $y - \bar{y}$ and the bulk entropy has a time-reflection invariance about $y - \bar{y} = 2u$, it is clear that the QES is \emph{at} $y - \bar{y} = 2u$.
Extremising the spatial coordinate, we find
\begin{equation}
  u_{QES} = u, \quad \frac{y_{QES} + \bar{y}_{QES}}{2} = \frac{1}{2\pi \tilde{T}_{1}} \sinh^{-1} \frac{\sqrt{1+2\alpha}}{k}.
  \label{eqn:conn-ext-qes-1}
\end{equation}
This is logarithmically close to the horizon, similar to \cite{Almheiri:2019,Chen:2020jvn}.
Finally, we find that the late time generalised entropy takes its equilibrium value
\begin{align}
  S_{gen} (\text{connect,late-time}) &= S_{0} + 2\pi \tilde{T}_{1} \phi_{r} \sqrt{1 + \frac{k^{2}}{1+2\alpha}} + \frac{c}{6} \log \tanh \frac{\sinh^{-1} \frac{\sqrt{1+2\alpha}}{k}}{2} - \mO\left(\sqrt{\phi_r}\right) \nonumber\\
  &\approx S_{0} + 2\pi \tilde{T}_{1} \phi_{r}  - \mO\left(\sqrt{\phi_r}\right).
  \label{eqn:conn-ee-page}
\end{align}

\begin{figure}[h]
  \centering
  \includegraphics[width=.97\textwidth]{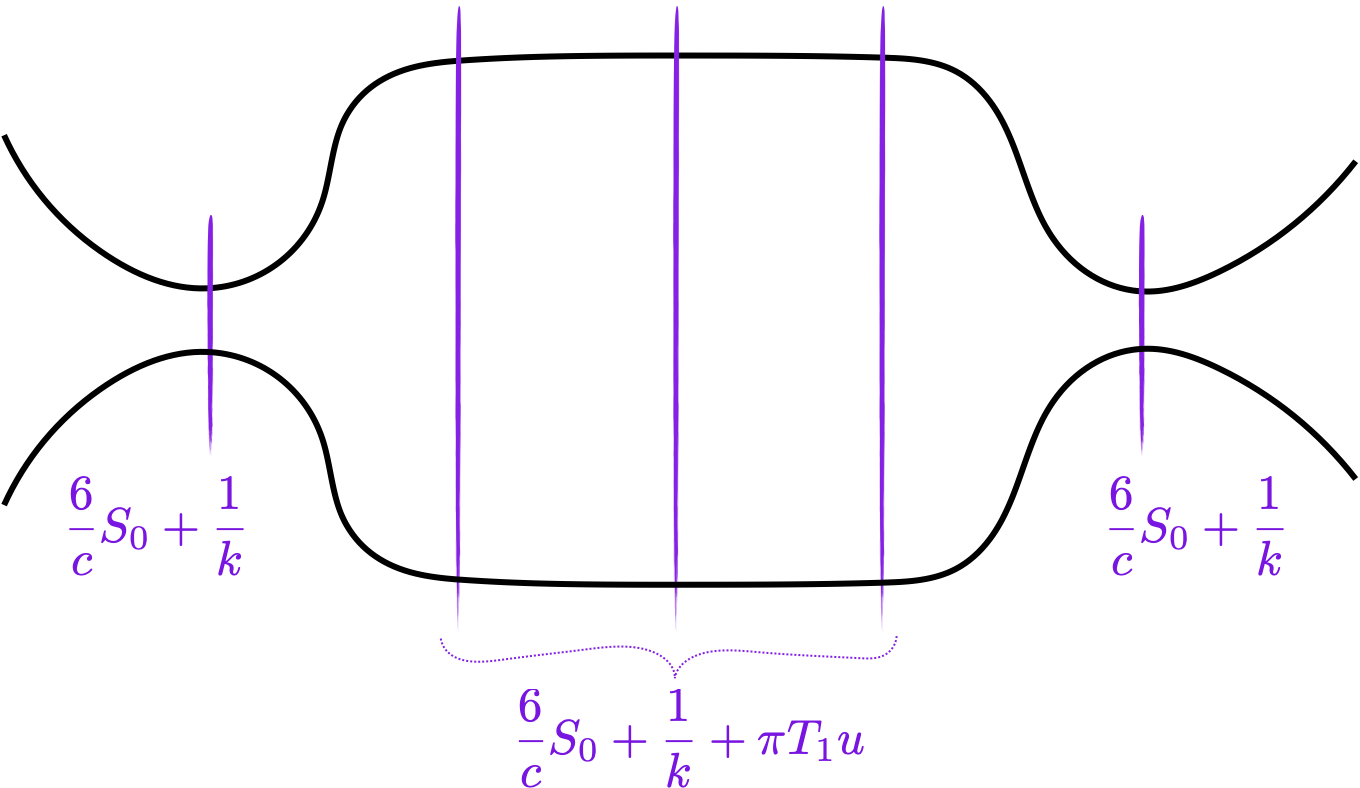}
  \caption{All the QESs. Dashed blue lines denote entanglement DOFs, and purple lines mark the location of QESs, and they are labelled by the value of $\frac{6}{c} S_{gen}$. $u \gg u_{Page}$. \newline The QESs marked with the same generalised entropies aren't actually degenerate, but our analysis is insufficient to distinguish them. \newline We have defined the quantity $k = \frac{c}{12\pi \phi_{r} \tilde{T}_{0}}$.}
  \label{fig:conn-lunch}
\end{figure}

\begin{figure}
    \centering
    \includegraphics[width=120mm]{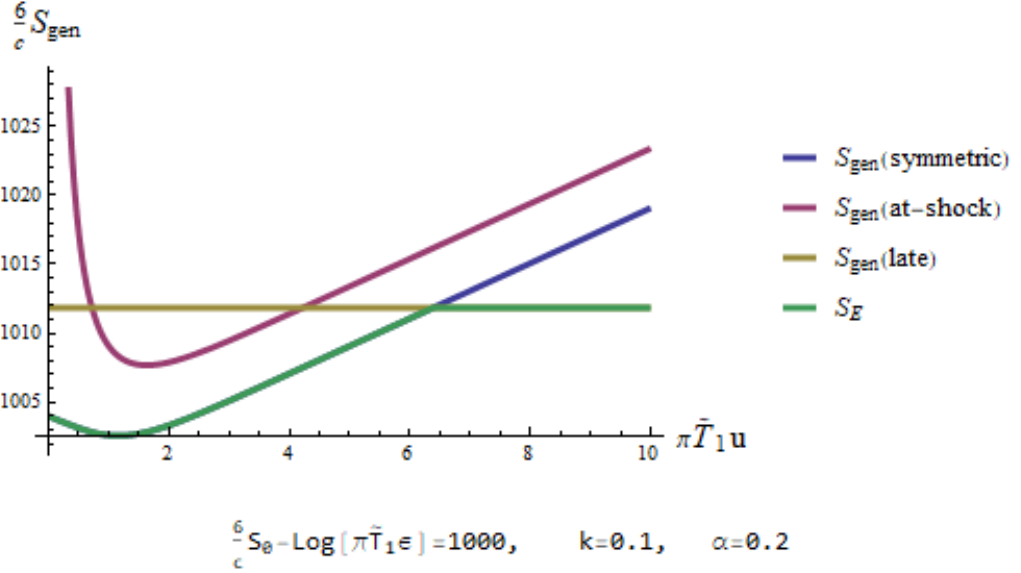}
    \caption{The generalised entropies of the various QESs and the UV EE. We have ignored uncertainties inherent to the HRT prescription for simplicity.}
    \label{fig:conn-ee}
\end{figure}

We have thus found that, apart from a pre-scrambling-time decrease in the entropy, the Page curve here is substantially similar to the model in section \ref{sec:evap}.
At early times, there are two pairs of QESs whose generalised entropies are the same up to the uncertainties inherent in the HRT prescription.
Ignoring these uncertainties, we plot the shape of the Python's lunch in figure \ref{fig:conn-lunch} and the Page curve in figure \ref{fig:conn-ee}.

\section{Discussion} \label{sec:conc}
We have studied, in three models, the effect of entangling two $AdS_{2}$ black holes with each other, with an eye towards understanding whether there is any topology change.
We have calculated the Page curve and found the expected behaviour in all three cases.
We find that the occurrence of a Hawking-Page-like transition to a connected topology depends on the model, i.e., on the details of how the black holes are entangled.

In the non-dynamical dPSSY model, there is an unambiguous change of bulk topology and the relative enhancement occurs because of `entanglement loops.'
The mechanism is not a topology-changing process, i.e. a history in which the topology differs between Cauchy slices.
It is only a Hawking-Page-like transition caused by the relative enhancement of the contribution of a subleading saddle with a `ket-ket' wormhole between the two boundaries.\footnote{These `ket-ket' wormholes are always present in the Hartle-Hawking state \cite{Anous:2020}.}

In the dynamical case, in which we allow the two spacetimes to radiate into each other in real time, the situation is less clear.
While the norm path integral itself does not seem to exhibit any phase transitions, the norm path integral with the insertion of a set of operators that do not have the identity in their OPE does.
Other correlation functions and the entropy, on the other hand, do \emph{not} seem to transition.
We are unable to offer a clean and unified explanation of these disparate facts.

The other notable aspect of the dynamical example is that, in the path integral that transitions, the transition happens not because of a separation of scales but because of a simple factor of $2$ between the temperatures of the black holes in the disconnected vs. connected histories.
The relative enhancement in the dynamical example arises from the large boosts of the sort common in black hole mechanics, which are encoded in two bulk dimensions in an exponentially growing conformal factor.
The connected history with the wormhole is a factor of two colder than the disconnected history and so suffers smaller large boosts.
One could have hoped to find an eternal traversable wormhole as in \cite{MQa}, in which case it would have been a separation of scales, but this is not obtained because of the energy released by the coupling, and we in fact find a Maldacena-Qi geometry by explicitly projecting onto the ground state of the interacting Hamiltonian.

Another possibility that we have not explored is that neither the disconnected nor the connected history is the dominant saddle.
One that we should expect to find is a baby-universe-emitting geometry like the ones in \cite{Saad:2019pqd} in which the length of the ER bridge stops growing.
Since the growth of this length is related to the growth of the conformal factor that gives the main contribution in the transition studied above, we might expect these baby-universe-emitting histories to dominate over the connected history at a time-scale $u = \mO(\text{poly}(S_0))$, where the polynomial can also be linear. 
An interesting point is that the connected geometries we consider can be seen as a baby-universe exchange in the Euclidean past.

It would also be interesting to get a better understanding of the dynamical case.
Apart from the lack of clarity in the existence of such a transition in this case there is also the question of the role of entanglement.
While the transition in the dPSSY model was unambiguously driven by entanglement, the role of entanglement in the dynamical case is somewhat obscure.
We have also not explored the possibility of finding a transition by `cooling' down the coupled system by coupling it to a bath \cite{Maldacena:2019ufo}.
Further, it would be interesting to understand if there is some unitary operation that causes a phase transition.
It would also be useful to study these questions with SYK techniques \cite{Lensky:2020}.

Apart from answering the above questions, it might also be useful to repeat our analysis for some other set-ups, two of which seem especially notable.
The first is that of black holes formed by collapse, where presumably the saddle is not a ket-ket wormhole but a more general baby-universe exchange.
The second case is that of higher dimensions, especially odd bulk dimensions.
For example, in three dimensions there is no potential analog of $S_0$, i.e. a topological term giving a controlled sum over topologies.
The smallness of the overlap with the thermofield double is taken care of by a normalisation factor in three dimensions, but the norm path integral and related quantities are harder to understand. Further, in higher dimensions there are also additional saddles as compared to the two we encountered in two bulk dimensions, and it would be interesting to explore the interplay of these various saddles. 

Let us conclude with two interesting observations about our set-up itself.
First, this set-up allows us to put spatial and spacetime wormholes on the same footing; since the ket-ket wormhole in this story is the same Euclidean geometry as the Euclidean wormholes of \cite{Penington:2019kki}.
This is in a sense not surprising, since both spatial \cite{Marolf:2013,Harlow:2015lma,Harlow:2018tqv} as well as spacetime \cite{Saad2018, Penington:2019kki,Stanford2020,Belin:2020hea} wormholes have been settings for framing factorisation puzzles in gravity. Nevertheless, our analysis brings into focus the intricate relation between the ER=EPR scenario and Euclidean wormholes. 

Secondly, this set-up somehow bridges recent discussions of averaging \cite{Saad:2019lba,Penington:2019kki,Stanford2020,Belin:2020hea,Altland:2020ccq}, quantum error correction \cite{Almheiri:2015,Harlow:2016vwg,Faulkner:2020} and `third-quantized' gravity \cite{Berenstein:2017abm,Jafferis:2017tiu,Marolf:2020,Marolf:2020a,Anous:2020}.
The most obvious connection is with averaging, since the set-up of two black holes with real-time coupling was the one that was found to give the clearest demonstration of the averaged nature of semiclassical gravity \cite{Stanford2020}.
The one advantage of going from bra-ket to ket-ket wormholes is that now the wormhole has a Hamiltonian interpretation and so is easily related to discussions of quantum error correction.

In the discussions of quantum error correction, it has been argued that the map from boundary states to bulk states involves a projection (or composition with a conditional expectation \cite{Faulkner:2020}) into the code subspace.
The code subspace of states with ER bridges (and no matter) in JT gravity has the property that the ADM energies of the two boundaries agree as semi-classical operators.
So, one might expect that among the myriad things that the projection or conditional expectation does, one must be that it implements the above equality.

\begin{figure}[h]
    \centering
    \includegraphics[width=140mm]{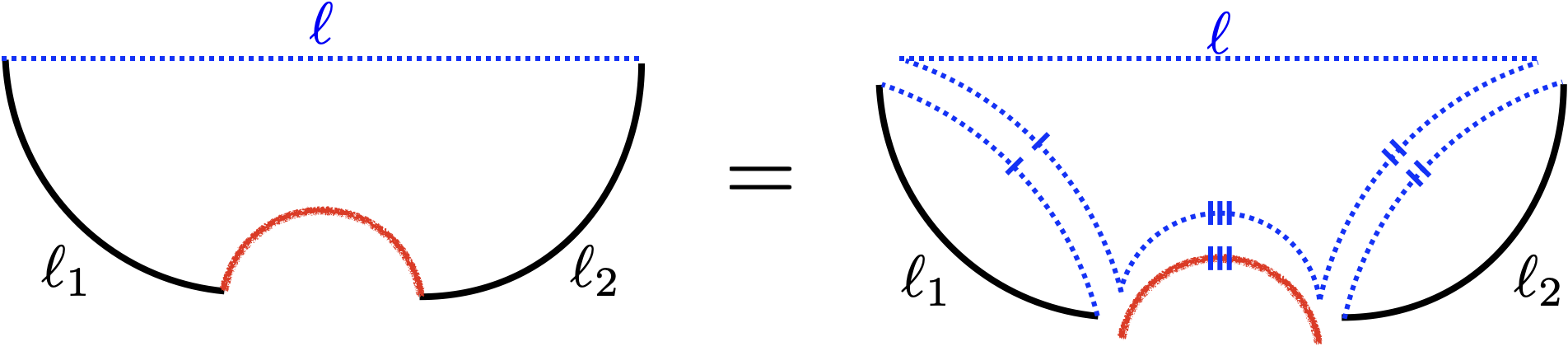}
    \caption{The calculation of the connected component of the Wheeler-de Witt function can be decomposed into an integral over these intermediate lengths.}
    \label{fig:npd}
\end{figure}

The way this equality is manifested in our set-up is shown in figure \ref{fig:npd}.
The path integral that calculates the connected component in the WdW wavefunction can be broken into an integral over three intermediate lengths as in \cite{Penington:2019kki}.
The crucial object in the centre that `sews' up the boundaries was introduced in \cite{Yang:2018gdb}; its wavefunction matches a GHZ state $\sim \sum_E \ket{E}^{\otimes 4}$.
This central object is precisely what implements the projection onto equal energies for the two boundaries.
On the other hand, this object (rather, a three-dimensional analog) was also found to be a generator of non-perturbative diffeomorphisms \cite{Jafferis:2017tiu} 

In other words, in a particularly sharp setting for showing the averaged nature of gravity, we find that the non-perturbative diffeomorphism that connects the connected and disconnected spacetimes is precisely the projector that ensures that the two boundaries have the same energy! 

%%%%%%%%%%%
\section*{Acknowledgements}
We would like to thank Vijay Balasubramanian, John Cardy, Arjun Kar, Cathy Li, Adam Levine, Raghu Mahajan, Edward Mazenc, Shiraz Minwalla, Suvrat Raju, Daniel Ranard, Eva Silverstein, Arvin Shahbazi-Moghaddam, Jonathan Sorce, Douglas Stanford, Leonard Susskind, Sandip Trivedi, Gabriel Wong and Zhenbin Yang for discussions.
We thank Douglas Stanford for comments on a draft.
We also thank Chenglie Hu for publishing their code \cite{hu1998algorithm}.
RMS would also like to thank AE (of the immortal wave), without whom this paper wouldn't have been possible.

Finally, LA would like to thank everyone in the theoretical physics community for making her time there both immensely enjoyable as well as educational. Thank you all.

\appendix
\section{Solutions of Semi-Classical Lorentzian JT Gravity} \label{app:JT}
The basic idea is that the entire information in the JT solution is the trajectory of the boundary particle and the position of the boundary particle is decided by a Dirichlet boundary conditions on the dilaton, so the info is in the parameters for the dilaton solution and those parameters are an $SL(2,\mathbb{R})$ charge.
Useful references for this formalism are \cite{Maldacena:2016upp,Maldacena2017,Bulycheva:2019naf}.

We think of AdS${}_{2}$ in embedding space, defined by
\begin{equation}
  Y^{2} = Y^{a} \cdot Y^{a} \equiv - (Y^{-1})^{2} - (Y^{0})^{2} + (Y^{1})^{2} = -1.
  \label{eqn:embedding-eqn}
\end{equation}
The embedding coordinates are related to global and Poincare coordinates by
\begin{align}
  Y^{a} &= \left( \frac{\cos t_{gl}}{\sin \sigma} , \frac{\sin t_{gl}}{\sin \sigma} , - \cot \sigma \right) = \left( \frac{1-t_P^2+z^2}{z}, \frac{t_P}{z}, \frac{1+t_P^2-z^2}{z}\right)
  \label{eqn:embedding-coords-relns}
\end{align}

In the absence of matter, the solution for the dilaton can be written as
\begin{equation}
  \phi = - Q \cdot Y.
  \label{eqn:dilaton-soln}
\end{equation}
$Q^{a}$ is an $SL(2,\mathbb{R})$ charge that measures the charge of the cutout under the three isometries of $AdS_{2}$.
Because the boundary of the cutout is given by a fixed value of the dilaton, see \eqref{eqn:bd-conds}, the trajectory of the boundary satisfies
\begin{align}
  Q \cdot X &= - \phi_r, \qquad  X \equiv \lim_{\epsilon \to 0} \epsilon Y, \text{ and } X^{2} = 0 \nonumber\\
  Q^{2} &= - 4 \phi_{r} M,
  \label{eqn:bd-particle-trajectories}
\end{align}
where $M$ is the ADM energy.
For a Schwarzchild black hole of temperature $T$, $M = (\pi T)^2 \phi_r$.

The real power of this formalism is that it is also a way to package the solutions of semi-classical JT gravity.
To bulk matter, we assign a charge
\begin{equation}
    Q_m^a = \int (\zeta^a)^\mu n^\nu T_{\mu\nu},
    \label{eqn:matter-charge}
\end{equation}
where the $\zeta^a$s are the Killing vectors of $AdS_2$.
Then, invariance under the $SL(2,\mathbb{R})$ gauge symmetry means that
\begin{equation}
 (Q_l + Q_r + Q_{m})^{a} = 0,
 \label{eqn:charge-gauge}
\end{equation}
where $Q_{l,r}$ are the charges for the left and right boundary particles respectively.
The choice of gauge is the freedom in different ways of solving this equation.
So, by calculating $Q_m$ and choosing an appropriate gauge, we can solve for $Q_{l,r}$, which using \eqref{eqn:bd-particle-trajectories} allows us a full solution.

For the eternal black hole, we fix gauge so that
\begin{equation}
  Q_{r}^{a} = - Q_{l}^{a} = - \frac{2\pi \phi_r}{\beta} (1, 0, 0).
  \label{eqn:tfd-charges}
\end{equation}

The calculation of $Q_m$ simplifies for point particles of mass $m$.
The trajectory of a massive particle $Y^{a} (s)$ satisfies
\begin{align}
  d_{s}^{2} Y + m^{2} Y &= 0.
  \label{eqn:geod-eqn}
\end{align}
Its $SL(2,\mathbb{R})$ charge is also an integral of motion,
\begin{align}
  Q_{a} = \varepsilon_{abc} Y^{b} d_{s} Y^{c}, \quad &Q \cdot Y = 0 \text{ by antisymmetry, } \nonumber\\
  Q^{2} &= m^{2}.
  \label{eqn:particle-charge}
\end{align}
Massless particles satisfy the $m \to 0$ limit of this equation.
For a massless particle passing through the point
\begin{equation}
  Y^{a} = (\sqrt{1 + y^{2}}, 0, s_1 y), \quad y > 0, s_1 = \pm 1,
  \label{eqn:null-particle-pos}
\end{equation}
we find in the limit in which the initial position goes towards the boundary,
\begin{equation}
  \lim_{y\to \infty} Q^{a}_{s_{1}} = \Delta E (s_{1}, 0, 1).
  \label{eqn:null-particle}
\end{equation}

Finally, let us use the above equations to calculate a boundary particle trajectory in AdS-Vaidya.
We will do this in global coordinates explicitly, since this is the one in which the answer is hard to guess.
The strategy is to take the relation between global and embedding coordinates in \eqref{eqn:embedding-coords-relns} and plug it into \eqref{eqn:bd-particle-trajectories} to find a differential equation for the trajectory.
It is a differential equation because the definition \eqref{eqn:bd-conds} of $u$ means that
\begin{equation}
  \sin \sigma_{bd} (u) = \epsilon t_{gl}'(u) \quad \Rightarrow \quad X = \frac{ ( \cos t_{gl}, \sin t_{gl}, -1 ) }{t_{gl}'},
  \label{eqn:sigma-bd}
\end{equation}
The first equation of \eqref{eqn:bd-particle-trajectories} is a linear equation in $X$ and therefore a differential equation for $t_{gl} (u)$.

For a TFD with a shock on the right, we have from \eqref{eqn:tfd-charges}, \eqref{eqn:null-particle} and the $SL(2,\mathbb{R})$ gauge condition $Q_{r} + Q_{l} + Q_{m} = 0$,
\begin{equation}
  Q_{r,V}^{a} = - \mu \left( 1 + \alpha, 0, \alpha \right), \quad \mu = \frac{2\pi \phi_{r}}{\beta},\ \alpha = \frac{\Delta E}{\mu}.
  \label{eqn:vaidya-charge}
\end{equation}
As a check, we can calculate the ADM mass after the shock using \eqref{eqn:bd-particle-trajectories}; we get $[\pi T \sqrt{1+2\alpha}]^2 \phi_r$, which is a well-known answer.
Plugging the value \eqref{eqn:vaidya-charge} of the charge and the expression \eqref{eqn:sigma-bd} for $X$ into \eqref{eqn:bd-particle-trajectories} gives for the post-shock solution
\begin{align}
  t_{gl} (u) &= 2 \tan^{-1} \frac{\tanh \left( \pi T \sqrt{1+2 \alpha}\; u \right)}{\sqrt{1+2 \alpha}}.
  \label{eqn:t-gl-vaidya-soln}
\end{align}
Following a similar procedure in Poincare coordinates gives
\begin{equation}
    t_P (u) = \frac{1}{\pi T \sqrt{1+2\alpha}} \tanh (\pi T \sqrt{1+2\alpha}\; u)
    \label{eqn:tP-vaidya-soln}
\end{equation}
after the shock.

\section{Exact Analysis of the Microcanonical dPSSY Model}\label{ssec:resolvent}
In section \ref{sec:dPSSY}, we focused on replica-symmetric saddle point geometries. But following \cite{Penington:2019kki}, we can in fact also perform the full gravitational path integral by summing over geometries which contribute in the planar limit. In this section only, we will set $\phi_{r} = 2\pi$, and we will also restrict to the microcanonical ensemble in this section for simplicity.

\begin{figure}[h!]
    \centering
    \includegraphics[height=6cm]{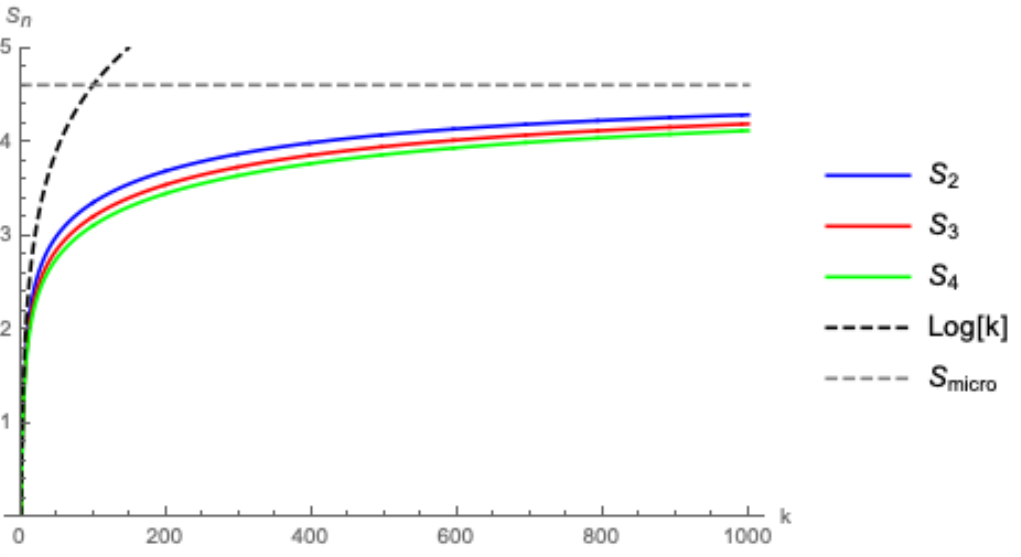}
    \caption{The first few R\'enyi entropies as a function of $k$.}
    \label{fig:Renyi}
\end{figure}

We start with the state \sout{of} \eqref{eqn:dPSSY-state}, that is:
\beq
|\Psi\rangle = \frac{1}{\sqrt{\cN}}\sum_{i,j=1}^D M_{ij}|\ell, i\rangle_1 \otimes |\ell,j\rangle_2^*,
\eeq
where we have taken $\ell_1=\ell_2$ for simplicity, and once more
$$\cN = \mathrm{Tr}(M^{\dagger}M)\,Z_1^2 + \mathrm{Tr}(M)\mathrm{Tr}(M^{\dagger})\, Z_2.$$
The reduced density matrix on, say, the first factor is given by
\beqn
\rho &=& \frac{1}{\cN}\sum_{i,i'=1}^D\sum_{j,j'=1}^DM_{ij}M^*_{i'j'}\langle \ell,j'| \ell,j\rangle^*_2\;|\ell, i\rangle\langle \ell, i'|_1  \nonumber\\
&=&\frac{1}{\cN}\sum_{i,i'=1}^D\sum_{j,j'=1}^DM_{ij}\,M^{\dagger}_{j'i'}\langle \ell,j| \ell,j'\rangle_2\;|\ell, i\rangle\langle \ell, i'|_1  .
\eeqn
From here, we get 
\beq
\mathrm{Tr}\,\rho^n = \frac{1}{\cN^n}\sum_{i_1,j_1}\sum_{i'_1,j'_1}\cdots \sum_{i_n,j_n}\sum_{i'_n,j'_n} M_{i_1j_1}\langle i_2'|i_1\rangle M^{\dagger}_{j_2',i_2'}\langle j_2|j_2'\rangle M_{i_3,j_2}\cdots \langle j_1|j_n'\rangle.
\eeq
To further simplify our analysis, we will now restrict the matrix $M_{ij}$ to be an orthogonal projection, i.e., we require that $M^2 = M$, and $M^{\dagger} = M$, with $k= \mathrm{Tr}\,M$. In order to proceed, we introduce the resolvent following \cite{Penington:2019kki}:
\beq
R(x) = \mathrm{Tr}\frac{1}{x^2 - \rho} =  \sum_{n=0}^{\infty}\frac{1}{x^{2(n+1)}} \mathrm{Tr}\,\rho^n.
\eeq
We can equivalently write
\beq
R(x) = \int d\gl \;D(\gl)\frac{1}{x^2 - \gl}|\gl\rangle\langle\gl |=\int d\gl\,\frac{D((\gl)}{2\sqrt{\gl}}\left[\frac{1}{x-\sqrt{\gl}} - \frac{1}{x+\sqrt{\gl}}\right] |\gl\rangle\langle\gl |. 
\eeq
This resolvent will thus have two branch cuts, one along the positive real $x$ axis and one along the negative real $x$ axis, symmetric under reflection about the imaginary axis. The eigenvalue density can be obtained from the resolvent as
\beq \label{D}
D(\gl) = \frac{2\sqrt{\gl}}{2\pi i}\; \text{disc}_{x=+\sqrt{\gl}}R(x) .
\eeq
The resolvent defined in this way is closely related to the trace of the resolvent defined and computed in \cite{Penington:2019kki}, which we will call $R_{(0)}(\lambda)$. The main difference is that our resolvent only involves gravitational amplitudes with an even number of boundaries, and in addition, the normalization factors are different. But we can restrict the PSSY resolvent to an even number of boundaries by simply taking its anti-symmetric part in $\lambda$. So, we find
\beq \label{R}
R(x) = \frac{\mathrm{Tr}\,1 - k}{x^2} + \frac{\alpha}{2x}\left[R_{(0)}(\alpha x) - R_{(0)}(-\alpha x)\right],
\eeq
where 
$$\alpha  =\left(\frac{1}{k} + \frac{Z_2}{Z_1^2}\right)^{1/2}.$$
We now restrict to the microcanonical ensemble. In this case, the PSSY resolvent is given by
\beq \label{R0}
R_{(0)}(\gl) = -\frac{(\gb-\gc)-\gb\gc\gl}{2\gl}\pm \frac{\gb\gc}{2\gl}\sqrt{\left(\gl - \left(\frac{1}{\sqrt{\gb}}-\frac{1}{\sqrt{\gc}}\right)^2\right)\left(\gl - \left(\frac{1}{\sqrt{\gb}}+\frac{1}{\sqrt{\gc}}\right)^2\right)}
\eeq
where 
$$ \gb = e^{\mathbf{S}},\;\;\;\;\ga= \left(\frac{1}{k} + \frac{1}{\gb}\right)^{1/2}.$$
Here $\mathbf{S}$ is the microcanonical entropy at chosen energy $E$ and energy window $\Delta E$:
\beq 
\gb=e^{\mathbf{S}} = e^{S_0}\rho(E)\Delta E,\;\;\;\rho(E) = \frac{E}{2\pi^2}\sinh(2\pi E),
\eeq
and note that the dependence on the brane tension $\mu$ drops out in the microcanonical ensemble. The PSSY resolvent $R_{(0)}$ only has a branch cut along the positive real axis, but the combination in equation \eqref{R} has the right structure to give us branch cuts along both the positive and negative real axes, as we expected. Using equations \eqref{R}, \eqref{D} and \eqref{R0}, we thus obtain the eigenvalue density
\beq
D(\gl) = \frac{\ga\gb\gc}{4\pi}\sqrt{\left[\sqrt{\gl} - \left(\frac{1}{\sqrt{\ga\gb}}-\frac{1}{\sqrt{\ga\gc}}\right)^2\right]\left[ \left(\frac{1}{\sqrt{\ga\gb}}+\frac{1}{\sqrt{\ga\gc}}\right)^2-\sqrt{\gl}\right]},
\eeq
for 
$$\left(\frac{1}{\sqrt{\ga\gb}}-\frac{1}{\sqrt{\ga\gc}}\right)^4 \leq \gl \leq \left(\frac{1}{\sqrt{\ga\gb}}+\frac{1}{\sqrt{\ga\gc}}\right)^4.$$
When $k \ll \gb$, the density is highly peaked around $\gl = \frac{1}{k}$ with a width $\frac{\Delta \gl}{\gl} \sim \sqrt{\frac{k}{\gb}}$. Thus, the entropy is approximately given by $\log\,k$. On the other hand when $k \gg \gb$, the density is highly peaked around $\gl = \frac{1}{\gb}$ with a width $\frac{\Delta \gl}{\gl} \sim \sqrt{\frac{\gb}{k}}$. Finally, we can also obtain the R\'enyi entropies by expanding the resolvent in equation \eqref{R} in $x$. The first few R\'enyi entropies are shown in figure \ref{fig:Renyi}. 

\section{Details of the Schwarz-Christoffel map} \label{app:dcSC}
In this appendix, we want to derive equation \eqref{eqn:ffq-map} directly using the doubly-connected Schwarz-Christoffel (SC) map. Recall from equation \eqref{eqn:dsc-map}, that the derivative of the relevant SC map is given by
\beq
f'(z)= C \frac{\Theta(\mu,z)\Theta(\mu,-z)\Theta(\mu,\frac{z}{\mu})}{\Theta^3(\mu,-\frac{z}{\mu})},
\eeq
where $C$ is a constant, and recall that $\mu=e^{-\tau}$ is the conformal modulus. The theta functions are
\beq
\Theta(\mu,z)= \frac{1}{\prod_{k=1}^{\infty}(1-\mu^{2k})}\sum_{n=-\infty}^{\infty}\mu^{2n}(-z)^n.
\eeq
This form is related to that in \cite{dep-sc,driscoll2002schwarz} by the so-called Jacobi triple product identity.
In the present case, we can conveniently re-write $f'$ in terms of the more standard Jacobi theta function:
\beq
f'(z)= C \frac{\EF(\frac{1}{2i}\log(z);\mu)\EF(\frac{1}{2i}\log(-z);\mu)\EF(\frac{1}{2i}\log(-z/\mu);\mu)}{\EF^3(\frac{1}{2i}\log(z/\mu);\mu)},
\eeq
where
\beq
\EF(u;q) =\sum_{n=-\infty}^{\infty}q^{n^2}e^{2inu}.
\eeq
We are interested in the limit $\mu \to 1$; we will later show that this is indeed the correct limit by relating $\mu$ to the Euclidean cutoff parameter $\delta$. In the $\mu\to 1$ limit, it is helpful to use the following modular transformation property for the elliptic function:
\beq
\EF(u;q=e^{-\tau}) = \frac{1}{(\tau/\pi)^{1/2}\exp(\frac{u^2}{\tau})}\EF\left(u'=\frac{\pi u}{i\tau};q' = e^{-\pi^2/\tau}\right).
\eeq
Writing $z=re^{i\theta}$, we find that one of the relevant theta function becomes:
\beqn
\EF\left(\frac{1}{2i}\log\left(z\right);\mu\right)&=&\EF\left(\frac{1}{2i}\left(\log\left(r\right)+i\theta\right);\mu\right)\nonumber\\
&=&\frac{1}{\left(\tau/\pi\right)^{1/2}\exp\left(-\frac{\left(\log\,r+i\theta\right)^2}{4\tau}\right)}\EF\left(u'=-\frac{\pi\left(\log\,r+i\theta\right)}{2\tau};q' = e^{-\pi^2/\tau}\right)\nonumber\\
&=&\frac{\exp\left(\frac{\left(\log\,r\right)^2-\theta^2+2i \theta \log\,r}{4\tau}\right)}{\left(\tau/\pi\right)^{1/2}}\sum_{n=-\infty}^{\infty}e^{-\frac{\pi^2n^2}{\tau}+\frac{n\pi \theta}{\tau}-\frac{i\pi n}{\tau}\log r}\nonumber\\
&=&\frac{\exp\left(\frac{\left(\log\,r\right)^2+2i \theta \log\,r}{4\tau}\right)}{\left(\tau/\pi\right)^{1/2}}\sum_{n=-\infty}^{\infty}e^{-\frac{\pi^2}{\tau}\left(n-\frac{\theta}{2\pi}\right)^2-\frac{i\pi n}{\tau}\log r}.
\eeqn
We will always work in the domain $0\leq \theta < 2\pi$. Note from above that when $\theta < \pi,$ the $n=0$ term dominates in the sum, when $\theta = \pi$ the $n=0$ and $n=1$ terms are degenrate, and when $\pi < \theta < 2\pi$, the $n=1$ term dominates. So as long as we stick to the domain $0\leq \theta < 2\pi$, we need only keep the $n=0$ and $n=1$ terms in our analysis:
\beqn
\EF\left(\frac{1}{2i}\log\left(z=re^{i\theta}\right);\mu\right)&=&\frac{\exp\left(\frac{\left(\log\,z\right)^2}{4\tau}\right)}{\left(\tau/\pi\right)^{1/2}}\left(1+e^{-\frac{\pi^2}{\tau}-\frac{i\pi }{\tau}\log z}\right)\nonumber\\
&=&\frac{\exp\left(\frac{\left(\log\,r\right)^2-\theta^2+2i \theta \log\,r}{4\tau}\right)}{\left(\tau/\pi\right)^{1/2}}\left(1+e^{-\frac{\pi\left(\pi-\theta\right)}{\tau}-\frac{i\pi }{\tau}\log r}\right). 
\eeqn

In order to further understand the SC map, we will analyse its derivative in the complex $z$-plane in two coordinate patches: (i) the far from quench region $-\varepsilon< (\pi-\theta)<\varepsilon$, where $\tau\ll\varepsilon\ll1$, and (ii) the near quench region, which is the complement of patch (i). (The terminology far-from-quench or near-quench will become clear shortly.) 

\subsection{Far from quench}
When $z$ is in patch (i), i.e., the far-from-quench region, we get
\beq
\EF\left(\frac{1}{2i}\log\left(z=re^{i\theta}\right);\mu\right)=\frac{\exp\left(\frac{\left(\log\,r\right)^2-\theta^2+2i \theta \log\,r}{4\tau}\right)}{\left(\tau/\pi\right)^{1/2}}\left(1+e^{-\frac{\pi\left(\pi-\theta\right)}{\tau}-\frac{i\pi }{\tau}\log r}+\mO\left(e^{-\pi^2/\tau}\right)\right). 
\eeq
The other theta function (with $z \to -z$) relevant for us in this patch is then given by
\beq
\EF\left(\frac{1}{2i}\log\left(-z\right);\mu\right)=\frac{\exp\left(\frac{\left(\log\,r\right)^2-\left(\pi-\theta\right)^2-2i \left(\pi-\theta\right) \log\,r}{4\tau}\right)}{\left(\tau/\pi\right)^{1/2}}\left(1+\mO\left(e^{-\pi^2/\tau}\right)\right). 
\eeq
Putting things together, we get
\beqn
f'\left(z\right) &=& - iC\,e^{-\frac{\pi}{\tau}\left(\frac{\pi}{2}-\theta\right)-\frac{i\pi \log r}{\tau}-\log\,r-i\theta-\tau/2}\frac{\left(1+e^{-\frac{\pi}{\tau}\left(\pi-\theta\right)-\frac{i\pi \log r}{\tau}}\right)}{\left(1-e^{-\frac{\pi}{\tau}\left(\pi-\theta\right)-\frac{i\pi \log r}{\tau}}\right)^3}\nonumber\\
&=&-\frac{ iCe^{\frac{\pi^2}{2\tau}-\frac{\tau}{2}}}{4z}\;\frac{\cosh\left(\frac{\pi}{2\tau}\left(\pi-\theta\right)+\frac{i\pi \log r}{2\tau}\right)}{\sinh^3\left(\frac{\pi}{2\tau}\left(\pi-\theta\right)+\frac{i\pi \log r}{2\tau}\right)}.
\eeqn
We can integrate this to obtain
\beq
f\left(z\right) =A + C\frac{\tau e^{\frac{\pi^2}{2\tau}-\frac{\tau}{2}}}{4\pi}\frac{1}{\sinh^2\left(\frac{\pi^2}{2\tau}+\frac{i\pi}{2\tau}\log\,z\right)},
\eeq
where $A$ and $C$ are both constants. Now, we impose the boundary conditions that $f\left(z=-\mu\right) = 0$, and $f\left(z=-\sqrt{\mu}\right) = -1$. These conditions fix the constants of integration to be  $$A =1, \;\;\;C=\frac{4\pi}{\tau}e^{-\frac{\pi^2}{2\tau}+\frac{\tau}{2}},$$
and thus the function becomes
\beq
f\left(z\right) =1 + \frac{1}{\sinh^2\left(\frac{\pi^2}{2\tau}+\frac{i\pi}{2\tau}\log\,z\right)}=\frac{\cosh^2\left(\frac{\pi^2}{2\tau}+\frac{i\pi}{2\tau}\log\,z\right)}{\sinh^2\left(\frac{\pi^2}{2\tau}+\frac{i\pi}{2\tau}\log\,z\right)}.
\eeq
When $r=\sqrt{\mu}$, then it is a simple matter to check that $|f|=1$, i.e., the time-reflection symmetric slice in the complex $z$ plane maps to the time-reflection symmetric slice in the $f$-plane, as expected. Writing $z= e^{-\tau/2-i\zeta}$ and $f= e^{2i\mathfrak{s}}$, we can now straightforwardly solve for $\zeta\left(\mathfrak{s}\right)$ from the above formula, and we find
\beq
\zeta = \pi +\frac{\tau}{\pi}\log\,\tan\frac{\mathfrak{s}}{2},
\eeq
which agrees with equation \eqref{eqn:ffq-map} derived in the main text using the Schwarzian method. 

\subsection{Near quench}
Next, we wish to understand the SC map $f$ in the near quench region. Our main goal is to relate the conformal modulus $\mu=e^{-\tau}$ to the regulator $\delta$. In order to do so, it is sufficient to focus on the real axis $\theta=0$ and $\mu \leq r \leq 1$. In this case, the relevant theta functions are given by
\beq
\EF\left(\frac{1}{2i}\log\left(r\right);\mu\right)= \frac{\exp\left(\frac{\left(\log\,r\right)^2}{4\tau}\right)}{\left(\tau/\pi\right)^{1/2}}\left(1+ \mO\left(e^{-\frac{\pi^2}{\tau}}\right)\right),
\eeq
\beq
\EF\left(\frac{1}{2i}\log\left(-r\right);\mu\right)=\frac{2\exp\left(-\frac{\left(\pi^2-\left(\log\,r\right)^2\right)}{4\tau}\right)}{\left(\tau/\pi\right)^{1/2}}\left[\cos\left(\frac{\pi}{2\tau}\log\,r\right)+\mO\left(e^{-\frac{2\pi^2}{\tau}}\right) \right].
\eeq
Putting everything together, we get on the real axis:
\beq
f'\left(r\right)=-2Ce^{-\frac{\pi^2}{2\tau}-\frac{1}{2}\left(\tau+2\log\,r\right)}\sin\left(\frac{\pi}{\tau}\log\,r\right)\left[1+\mO\left( e^{-\frac{\pi^2}{\tau}}\right)\right].
\eeq
Therefore, we get
\beqn
f\left(r\right) &=& f\left(r_0\right)+\int_{r_0}^rdr'\,f'\left(r'\right)\nonumber\\
&=&f\left(r_0\right)+\frac{2\tau}{\pi}Ce^{-\frac{\pi^2}{2\tau}-\frac{\tau}{2}}\left[\cos\left(\frac{\pi}{\tau}\log r\right)-\cos\left(\frac{\pi}{\tau}\log r_0\right)\right]+\cdots.
\eeqn
Now we impose boundary conditions. Taking $r_0 = \mu$ and $f\left(r_0\right) = e^{-2\delta}$, we get
\beq
f\left(r\right) = e^{-2\delta}+\frac{4\tau}{\pi}Ce^{-\frac{\pi^2}{2\tau}-\frac{\tau}{2}}\cos^2\left(\frac{\pi}{2\tau}\log r\right)+\cdots.
\eeq
Next, setting $f\left(\sqrt{\mu}\right)=1$ and $f\left(1\right) = e^{2\delta}$ gives:
\beq
1 = e^{-2\delta}+\frac{2\tau}{\pi}Ce^{-\frac{\pi^2}{2\tau}-\frac{\tau}{2}}+\cdots
\eeq
\beq
e^{2\delta} = e^{-2\delta}+\frac{4\tau}{\pi}Ce^{-\frac{\pi^2}{2\tau}-\frac{\tau}{2}}+\cdots
\eeq
Using the value of $C$ obtained previously, both the above equations are solved if we make the identification 
\beq
\delta = 4 e^{-\frac{\pi^2}{\tau}}+\cdots,
\eeq
where the $\cdots$ denote terms suppressed by more powers of $e^{-\frac{\pi^2}{\tau}}$. Note that as $\tau \to 0$, we find $\delta \to 0$. This justifies working in the $\tau \to 0$ limit. 

Finally, we can now also obtain the function $f$ on the time-reflection symmetric slice in the near quench region. As before, setting $f=e^{2i\mathfrak{s}}$ and $z= e^{-\frac{\tau}{2}-i\zeta}$, we find
\beq
(1-e^{2i\mathfrak{s}}) = 8e^{-\frac{\pi^2}{\tau}}\sin\frac{i\pi \zeta}{\tau},\;\;\Rightarrow \zeta(\mathfrak{s}) = -i\frac{\tau}{\pi}\sin^{-1}\left(\frac{1-e^{2i\mathfrak{s}}}{8e^{-\frac{\pi^2}{\tau}}}\right).
\eeq
Note that a finite interval around $\zeta=0$ gets mapped to an infinitesimal neighbourhood of $\mathfrak{s}=0$, which is the near quench region. We can compute the stress tensor on the $\mathfrak{s}$-plane from here, and we find
\beq
T(\mathfrak{s}) = \frac{c}{24\pi}\left[-2\frac{e^{\frac{4\pi^2}{\tau}}(e^{2i\mathfrak{s}}-1)^4-64e^{\frac{2\pi^2}{\tau}}(5e^{4i\mathfrak{s}}-4e^{2i\mathfrak{s}}+2)+4096}{(e^{\frac{2\pi^2}{\tau}}(e^{2i\mathfrak{s}}-1)^2-64)^2}\right].
\eeq
We see that the stress tensor is highly peaked at $\mathfrak{s}=0$, with
\beq
T(\mathfrak{s}=0) \sim \frac{c}{24\pi}\frac{3}{2\delta^2}.
\label{eqn:nq-T-answer}
\eeq

\bibliographystyle{JHEP}
\bibliography{refs.bib}
\end{document}